\documentclass[pre,reprint,superscriptaddress,showpacs]{revtex4-1}
\usepackage{epsfig}
\usepackage{natbib}
\usepackage{color}
\usepackage{graphics}
\usepackage{amssymb,amsmath}
\usepackage{amsmath}
\usepackage{psfrag}
\usepackage{graphicx}

\usepackage{float}
\usepackage[caption = false]{subfig}

\usepackage{color}
\usepackage{graphics}
\usepackage{amssymb,amsmath}
\usepackage{amsmath}
\usepackage{psfrag}
\usepackage{graphicx}

\newcommand{\be}{\begin{equation}}      
\newcommand{\ee}{\end{equation}}      
      
\newcommand{\bef}{\begin{figure}}      
\newcommand{\eef}{\end{figure}}      
\newcommand{\bea}{\begin{eqnarray}}    
\newcommand{\eea}{\end{eqnarray}}

\begin{document}

\title {Long-lived transient structure in collisionless self-gravitating 
systems}
       
\author{David Benhaiem} \affiliation{Istituto dei Sistemi
  Complessi, Consiglio Nazionale delle Ricerche, Via dei Taurini 19,
  00185 Roma, Italia} 
  \author{Francesco Sylos Labini} \affiliation{Centro Studi e
  Ricerche Enrico Fermi,  Roma, Italia }
\affiliation{Istituto dei Sistemi
  Complessi, Consiglio Nazionale delle Ricerche, Via dei Taurini 19,
  00185 Roma, Italia} 
\affiliation{INFN Unit Rome 1, Dipartimento di Fisica, Universit\'a di
  Roma Sapienza, Piazzale Aldo Moro 2, 00185 Roma, Italia }
  \author{Michael Joyce} \affiliation{Laboratoire de Physique
  Nucl\'eaire et de Hautes \'Energies, UPMC IN2P3 CNRS UMR 7585,
  Sorbonne Universit\'e, 4, place Jussieu, 75252 Paris Cedex 05,
  France}

\date{\today}

\begin{abstract}

The evolution of self-gravitating systems, and long-range interacting systems 
more generally, from initial configurations far from dynamical equilibrium is often described
as a simple two phase process: a first phase of violent relaxation bringing it to
a quasi-stationary state  in a few dynamical times, followed by a slow adiabatic
evolution driven by collisional processes.  In this context the complex spatial 
structure evident, for example, in spiral galaxies is understood either in terms
of instabilities of quasi-stationary states, or a result of dissipative non-gravitational
interactions. We illustrate here, using numerical simulations, that purely  self-gravitating 
systems evolving from quite simple initial configurations can in fact give rise 
easily to structures of this kind of which the lifetime can be large compared to
the dynamical characteristic time, but short compared to the collisional relaxation 
time scale. More specifically, for a broad range of non-spherical and non-uniform 
rotating initial conditions,  gravitational relaxation gives rise quite generically 
to long-lived non-stationary structures of a rich variety, characterized by 
spiral-like arms, bars and even ring-like structures in special cases.  These structures 
are a feature of the intrinsically out-of-equilibrium nature of the system's collapse, 
associated with a part of the system's mass while the bulk is well virialized. 
They are characterized by  predominantly radial motions in their outermost parts, but also
incorporate an extended flattened region which rotates coherently about a well virialized 
core of triaxial shape with an approximately  isotropic velocity dispersion.  
  We characterize the kinematical and dynamical properties of these complex velocity fields 
  and  we briefly discuss the possible relevance of these simple
  toy models to the observed structure of real galaxies
  emphasizing the difference between dissipative and dissipationless
  disc formation.  
\end{abstract}

\pacs{05.10-a,05.90.+m,04.40.-b,98.62.-g,98.62.Hr} 

\maketitle 



\section{Introduction}
The dynamical evolution of many particles solely
interacting  by Newtonian gravity is a fundamental paradigmatic 
problem in physics, which is essential  for the modeling and interpretation 
of astrophysical structures. The study of this problem 
can also be placed in the broader framework of long-range interactions,
which, from the point of view of statistical mechanics, share essential
features (for a review, see, e.g., \cite{Campa_etal_2014,Chavanis_2013,Levin_etal_2014}).
An  approach  based on equilibrium statistical mechanics 
(for the case of gravity, see, e.g., \cite{Padmanabhan_1989,Chavanis_2002})
is physically relevant to such systems only on  time scales that are  very
long compared to the those characteristic of the mean-field
dynamics driven by the collective force fields sourced by many particles,
and described by Vlasov equations. This dynamics leads typically to
dynamical equilibria referred to variably as ``virial equilibria'', or
``collisionless equilibria" or ``quasi stationary states" (QSS) ---
we will adopt the latter term here. These states are interpreted
as stationary states of the appropriate Vlasov equation.
Such states can also manifest instabilities leading to 
further evolution, giving rise, for example, to changes in symmetry
(through ``radial orbit instability") and/or  to the development
of complex spatial structure (through, e.g., spiral wave instabilities).
On longer time scales, diverging in particle number when expressed 
in terms of the ``dynamical time'' characteristic of the mean-field time,
and described theoretically by a  broader framework than the
Vlasov equations incorporating ``collisional" terms,  such systems  
then typically evolve adiabatically through QSS, finally
evolving to thermal equilibrium if such a state is well defined
(see, e.g., \cite{Dauxois_etal_2002,Yamaguchi_etal_2004,Campa_etal_2009,Joyce+Worrakitpoonpon_2010,Marcos_2013,Benetti_etal_2014,Campa_etal_2014}. )

For the case of gravity in three dimensions, numerical study indicate
(see, e.g., \cite{Marcos_etal_2017} and references therein)
that collisional evolution is driven primarily by two-body collisions 
on a time scale of the order of  $\tau_{coll} \sim (N/ \ln(N)) \tau_{dyn}$ 
where $\tau_{dyn} \sim 1/\sqrt{G \rho}$ (where $\rho$ is the system's 
average density)  as originally argued by \cite{Chandrasekhar_1943}.
In most astrophysical systems the timescale for such relaxation
is this much longer than the Hubble time and their dynamics, on relevant
timescales, is thus expected to be accurately described by the 
collisionless (mean-field) dynamics. The framework for the
study of  stellar and galactic dynamics (see, e.g., \cite{Binney_Tremine}) 
is thus that of the collisionless Boltzmann equation 
 (i.e,. the Vlasov equation coupled to the Poisson equation).
Because of the extremely long time scale of collisional relaxation, 
the assumption of  stationarity is often used as a 
working hypothesis for studying the structure of galaxies,
with possible secular evolution through collisionless dynamics.
In this context the striking spatial structuration of galaxies
--- most evidently spiral structure --- has been described
analytically in terms of instabilities of such QSS
 (see, e.g., \cite{Dobbs_Baba_2014} and references
therein). With modern numerical studies of galaxy formation
 (see, e.g.,
   \cite{Gott_1977,Lake+Carlberg_1988,Katz_1991,Katz+Gunn_1991,Katz_1992,Steinmetz+Mueller_1994,Steinmetz+Mueller_1995,Marinacci_2014,Naab_2017}
   and references therein)  there has been extensive
study of the origin of such structure, but an essential
role in its generation is then played by dissipative
non-gravitational processes. In this paper we report results  
showing that structures of this kind can arise purely
within the restricted framework of self-gravitating 
systems. Further these structures, despite the fact that
they are  generated by a collisionless dynamics, are the
result of a far from equilibrium state which persists for 
times which are very long compared to the dynamical
time on which the initial approximate virialization
of the system occurs. 

To study these transient configurations we evolve numerically,
using gravitational N-body simulations, a set of  toy models; this  
allows us to understand the  basic physical processes which
give rise to such out-of equilibrium structures. 
The relaxation of isolated self-gravitating systems from simple
initial conditions (IC), of the kind we consider here, has been extensively
studied in numerical simulations over several decades
(see, e.g., \cite{Henon_1973,vanalbada_1982,aarseth_etal_1988,david+theuns_1989,aguilar+merritt_1990,theuns+david_1990,boily_etal_2002,barnes_etal_2009}). The
focus of such studies has, almost exclusively, been on the formation
and properties of the virialized structures which are very efficiently
produced by the collapse's dynamics. In particular, in the context of the theory
of galaxy formation, there was much interest in the capacity of such
IC  to produce structures resembling elliptical
galaxies.  The focus of our study here is, instead, on an aspect of
these systems which has been overlooked: { the production of rich
  transient structures from the small, but non-negligible, fraction of
  loosely bound and ejected mass which is very generically also
  produced by the relaxation process}
\citep{Joyce+Marcos+SylosLabini_2009,syloslabini_2012,syloslabini_2013}.  This
phenomenon is, we believe, of basic theoretical interest and
potentially of considerable relevance to understanding astrophysical
structures. In a recent article
\citep{Benhaiem+Joyce+SylosLabini_2017}, we have shown that, starting
from a specific class of simple idealized IC  ---
uniform rotating ellipsoidal configurations --- the relaxation of the
system under its self-gravity typically leads to extended transient
structures resembling qualitatively that of the outer parts of spiral
galaxies.  In the study reported in this paper, we
investigate a much broader range of IC, and also with a greater
range of particle number, whether these phenomena occur more
generically.  Our principle finding is that the generation of such
structures --- which, although transient in nature, may be very
long-lived (in units of the system's characteristic dynamical time)
--- is indeed a quite robust and generic feature of violently
relaxing systems.  Further the morphologies of the structures produced
are even more rich and diverse than we had anticipated.

The spiral-like structures generated far out-of-equilibrium in the 
systems we study arise in a manner very different to that usually 
envisaged in the modeling of such structures in the astrophysical 
literature. The mechanisms proposed are perturbative in nature,
envisaging the spiral-like structure as the result of instability of
an equilibrium disc configuration  (see, e.g., \cite{Dobbs_Baba_2014,
 Binney_Tremine}). In addition, it is well known that the 
 formation of flat disc-like configurations can be  obtained with 
 simulations of collapsing spheroids where, in addition 
 to gravity, the dissipative effects of several astrophysical processes 
 are taken into account
 (see, e.g.,
   \cite{Gott_1977,Lake+Carlberg_1988,Katz_1991,Katz+Gunn_1991,Katz_1992,Steinmetz+Mueller_1994,Steinmetz+Mueller_1995}
   and references therein). 
 Indeed,  in this context, an  indispensable element for the appearance 
 of such structures is  the inclusion of gas with a dissipative dynamics 
 (e.g., cooling).   The principle motivation for including this dynamics 
   was initially to allow such structures to be produced. The central finding of 
   our study, in contrast,  is that disc-like configurations
with transient spiral arms and with bars and/or rings may be
produced by a purely ({  i.e., non dissipative}) gravitational dynamics.  We
will discuss below in further detail this essential difference with respect 
to previous works in the literature.  We also stress the peculiar features of 
the complex velocity fields generated by the gravitational dynamics we have 
investigated, which are different from the structures of this kind generated  
by dissipative dynamics. Indeed,  while in the latter case the velocity fields 
are essentially dominated  by rotational motions, in the former case these 
show different  regimes and are essentially radial in the systems' outermost regions.

The article is organized as follows.  In Sect.\ref{numerical} we 
describe the implementation of the numerical simulations,
discussing  our choice of numerical parameters, 
of IC, and the specific quantities we have measured. 
In Sect.\ref{results} we present our results, focusing in 
particular on the spiral-like, bars and rings transients produced.
We describe in considerable detail the mechanism producing them,
which varies in detail from one kind of IC  to the other.
We  turn in Sect.\ref{observations} to a 
brief discussion of the possible relevance to real galactic structures 
of our simple models.
The differences between  dissipative and {  non-dissipative} dynamics
in the process of the formation of a disc-like structure is discussed 
in Sect.\ref{galform}. 
Finally in Sect.\ref{conclusions} we summarize our findings
and outline some of the many further questions raised by
them.


\section{Numerical Simulations}
\label{numerical} 

Our {  numerical experiments  consist of} 
molecular dynamics simulations of an 
Hamiltonian system of $N$ point particles 
in an infinite (three dimensional) domain, interacting by 
a pair potential $\phi$ which is that of Newtonian gravity 
with a short distance regularization. 
They have been performed using the
publicly available (and widely used)  
code {\tt Gadget-2} \citep{Springel_2005},
in the appropriate version (non-expanding universe,
open boundary conditions). The regularization
of the potential  --- the so-called ``gravitational softening" ---
in this code is implemented by solving the Poisson 
equation for spherical continuous clouds, 
with compact support of characteristic 
radius $\varepsilon$ (and total mass equal to
that of the particle).  The force thus takes
exactly its Newtonian value at separations
greater than $\varepsilon$. The functional
form of the regularized potential for $r < \varepsilon$, 
of which the exact expression can be found in \citep{Springel_2005}, 
is a cubic spline interpolating between 
the exact Newtonian potential at $r = \varepsilon$
and a constant value at $r=0$, with vanishing
gravitational force at this point.

\subsection{Precision and softening}

For simulations of the dynamics from the IC  we consider,
which give rise typically to very significant contraction of the system
--- leading to very high densities and short characteristic time-scales ---
the choice of the force softening length and numerical parameters
controlling the accuracy of the time-stepping and force calculation
requires particular care.  

We have performed simulations, which we call ``low precision" (LP),
with the numerical parameters of the code set at the values suggested by
the {\tt Gadget-2} user guide \footnote{See
  https://wwwmpa.mpa-garching.mpg.de/gadget/users-guide.pdf.}.  We
monitor the total energy and find that these runs typically conserve
it to within about $0.5\%$. We have then also run ``high precision"
(HP) simulations using values of the relevant parameters
 which lead to
energy conservation to within one part in $10^{5}$
\footnote{Specifically the parameter {\tt ErrTolTheta} is $10^{-10}$
  for HP compared to $0.7$ for LP, while {\tt ErrTolForceAcc} is
 $10^{-10}$ for HP and $5\times 10^{-4}$ for LP.}.  We also monitor
conservation of total linear and angular momentum conservation and
observe similar results. 

We have { run both LP  (with $N$ in the range
$[10^4,5\times 10^5]$) and HP (with $N$ in the range
$10^6$) simulations of realizations of our
IC (described in detail below), and found no apparent significant differences 
for the macroscopic quantities which we study. In what follows
we will report only results for LP simulations,
which likewise give results completely consistent with the 
lower $N$ simulations at HP and LP. Thus it appears that
our results are $N$ independent --- and represent those of
an $N \rightarrow \infty$ limit. However, as we discuss further
below, such a conclusion should be treated with caution.}  

Our default value of the force softening length (the code's parameter
$\varepsilon$), and that used in the specific simulations reported
below, is $10^{-3}$ in the units of length we define further below. In
practice this means it is $10^{-3}$ of the smallest characteristic
length in the IC  (e.g., shortest semi-principal axis
in the case of an initial ellipsoid), and is such that it is always
considerably smaller than the typical size of the system when it is
most contracted. We have also performed extensive tests to control the
effect of varying the force-softening parameter, and found that we
indeed obtain very stable results, provided this condition is
respected on the comparison of $\varepsilon$ and the minimal size
reached by the whole structure during the collapse.

We have also run some test simulations in which the particles have two
different masses.  Specifically in a reference simulation in which all
the particles have mass $m$, we have re-sampled randomly with
particles of mass $\alpha \times m$ and $m/\alpha$, with
$\alpha=2,5,10$, determining their number so that the total mass is
fixed.  We have then studied the spatial and velocity properties of
the two species separately, and found them to be in good agreement
both with one another and with the properties found in the original
single mass simulation of the IC.  This provides strong
evidence that the dynamics producing the distributions we have
described are indeed representative of the mean-field (or
collisionless) dynamics of the continuum IC we have sampled. Indeed
this is consistent with what one would expect as, even for the longest
times we simulate (at very most $200$ dynamical times), collisional
two body effects would be expected to be negligible for the particle
numbers we consider even in the virialized core of the system.

\subsection{Initial conditions}
\label{Initial_conditions} 

 { We wish to study evolution of self-gravitating $N$ body systems 
 starting from IC which are sampled from spatial mass distributions 
 which break rotational symmetry, and also mass distributions 
 which are non-uniform (i.e. for which the mass density is not 
 constant in the compact region where it is non-zero). Clearly 
 the space of parameters that describe a generic IC  of this type 
 is infinite. We have chosen three very simple few parameter
 families of such IC. In order of increasing complexity, we
 consider mass distributions which are 
 (i) uniform ellipsoids,  (ii) uniform ellipsoids with a central 
 spherical region of higher density,   and, (iii)  a collection
 of uniform spheres of varying radii with centers randomly
 placed in a spherical region.  For the initial velocities we consider only two
 very simple cases:  coherent rigid body type rotation of the whole 
 structure, and in a few cases only, some additional random uncorrelated
 motion. The precise details of our choices {  for} how the different
 IC are then parametrized, are given below.
  
With these choices of IC  we aim to single out the effect of the initial 
shape of an isolated mass distribution on its collapse and subsequent 
evolution. This aspect appears to have been largely overlooked in the 
literature, which  has focused instead mostly on the effect of internal correlations 
 between density fluctuations, for the case of a spherical cloud
\citep{Gott_1977,Lake+Carlberg_1988,Katz_1991,Katz+Gunn_1991,Katz_1992,Steinmetz+Mueller_1994,Steinmetz+Mueller_1995}.
   As we discuss below, one of our main results is that
   the spherical IC  represent a very peculiar and
   pathological case that give rise a to dynamics that suppresses 
   a common characteristic feature of the gravitational collapse  of 
   isolated over-densities: the formation of a disc-like structure 
   with spiral type arms that occurs }{ when the IC  significantly 
   break spherical symmetry. }

\begin{table}
\begin{center}
\begin{tabular}{|c|c|c|c|c|c|c|c|c|}
\hline
Name & $\frac{a_1}{a_3}$ & $\frac{a_2}{a_3}$ & $\frac{R}{a_3}$ & $|b_{rot}|$ & $\frac{M_e}{M_s}$ & $N_c$ &  $\frac{\ell_c}{\Lambda_c}$ & {$\lambda$ }\\
\hline
A1 &  2 & 1        & --  & 0.2  & -- & -- & --&  0.19 \\  
A2 &  2 & 1.25   & --  & 1.0 & -- & -- & -- &  0.30\\  
\hline
B1 &  1.5  & 1      & 0.1    &  0.25 & 1 & -- & &0.12  \\  
B1a & 1.5 & 1      & 0.01  &  0.25 & 1 & -- & &0.29 \\  
B2 &  1.5  & 1.5   & 0.1    &  0.50 & 3 & -- & &0.29 \\  
\hline
C1 & -- & -- &  -- & 1  &  --  & 5 & 0.5                &0.28\\  
C2 & -- & -- &  -- & 0.8  &  --  & 10 & 0.5           &0.17\\ 
\hline
\end{tabular}
\end{center}
\caption{Values of the relevant parameters (see text) for the specific 
simulations we report here.}
\label{table-param}
\end{table}


\subsubsection{Uniform ellipsoidal clouds}

For this class { of systems} the particles are 
\begin{itemize}
\item  randomly distributed, without correlation,
with uniform probability  {\it inside an ellipsoid},  and
\item given a  {\it coherent rigid body rotation}
about the shortest semi-principal axis.
\end{itemize} 

The three parameters which we choose to characterize 
them are:
\begin{itemize}
\item  The ratios, $a_1/a_3$ and $a_2/a_3$,  
of the two longer semi-principal axes ($a_1 \geq a_2$)
to the shortest one $a_3$.
\item The virial ratio associated with the rotation
\be
b_{rot} =
\frac{2 K_{rot}}{W_0} \;,
\label{def-brot}
\ee
where $K_{rot}$ is the kinetic energy of the rotational
motion, and $W_0$ is the initial gravitational potential 
energy. {  As mentioned above, we have considered 
some numerical experiments in which we add a random motion 
in addition to the coherent rotational one. The amplitude 
of this motion has been taken to be a fraction of the order of $10-20\%$ 
of the rotational one. We have not found any significant difference 
on a macroscopic scale, but the transient features like the 
spiral arms are indeed more diffuse when random motions have 
the highest amplitude we have tested.}
\end{itemize} 

We have performed simulations for a large variety of prolate (i.e.,
$a_1>a_2=a_3$), oblate (i.e., $a_1=a_2>a_3$) and triaxial (i.e.,
$a_1>a_2>a_3$) shapes, as well as the special case when the cloud is
spherical (and thus rotational symmetry is broken only by the finite
$N$ Poissonian fluctuations).  {{We have explored the range of values
    defined by the ratio $a_1/a_3$, $a_2/a_3$ $\in [0.05,2]$.}  For
  the velocities we have explored the range $b_{rot} \in [-1, -0.05]$.
  We report in detail results for just two representative cases, A1
  and A2, for which the parameter values are given in the Table
  \ref{table-param}. We report also {  the value of the so-called 
  ``spin parameter'' as defined in \cite{Peebles_1969,Peebles_1971}: } %
  \be
\lambda= \frac{ J |E|^{1/2}} {G M^{5/2}}\,,
\ee
where $J$ is the total angular momentum of the system, $E$ the binding energy,
$M$ the total mass and $G$ Newton's constant. 
{  This parameter is widely used in the astrophysical context when characterizing the 
angular momentum of astrophysical systems (see also, e.g.,   \cite{Barnes+Efstathiou_1987,Bullock_etal_2001}). 
It simply provides a dimensionless measure of the  angular momentum in the natural units of a self-gravitating 
system (given by dimensional analysis by  the combination of $GM^{5/2}/E^{1/2}$). It thus provides an indication 
of the degree of rotational support of a self gravitating system provided by its angular momentum.}

\subsubsection{Non-uniform ellipsoidal clouds}

In this class  { of systems} we consider  a coherently rotating 
ellipsoidal configuration exactly as described above, 
but then {\it modify it in a spherical region around 
its center}, ascribing
\begin{itemize}
\item
a different number (and thus) mass density 
of the uniformly distributed particles, and
\item
only random uncorrelated velocities,  
sampled uniformly in velocity space
up to a maximal magnitude.
\end{itemize}

We consider only the case that the kinetic
energy $K_s$ of the particles in the spherical 
region is such that $2K_s=-W_s$  where 
$W_s$ is their potential energy, i.e., this
central region is initially approximately 
virialized. The initial conditions are thus
chosen to probe the dynamics of the collapse
of a rotating ellipsoidal  cloud which already has
a smaller virialized structure inside it. 

There are then {\it five} parameters characterizing this family of IC,
which we choose to be: $a_1/a_3$, $a_2/a_3$, $R/a_3$ (where $a_i$ are
again the lengths of the semi-principal axes, and $R$ the radius of
the sphere), the ratio $M_e/M_s$ of mass (i.e. particle densities) in
the ellipsoidal region to that in the sphere, and the ratio $b_{rot}$
as given in (\ref{def-brot}), with $W_0$ the total potential energy of
the configuration {of the ellipsoid alone}.

We have considered simulations in which $b_{rot} \in[-1,-0.1]$ which
means that in almost all cases all particles are initially bound. For
the other parameters we have explored 
{$a_1/a_3$, $a_2/a_3$ $\in [0.05,2]$, }} {$R/a_3 \in[0.01,0.1] $}, and
$M_e/M_s \in [1/3,3]$.

We note that the characteristic time for collapse of
the rotating cloud, in units of the dynamical time 
of the sphere, is  $\sim (a_3/R)^{3/2} \dot (M_s/M_e)^{1/2}$.
Thus we explore the range in which this ratio is
between about ten to thirty. 
We report in detail results for just three representative cases,
B1, B1a and B2, for which the parameter values are given
in Table \ref{table-param}.

\subsubsection{Non-spherical and non-uniform clouds}

The third class of models mass distributions which still
rotate coherently, but in which rotational symmetry
is broken in a more random and less idealized manner.

Specifically:
\begin{itemize} 
\item We choose $N_c$ points randomly with uniform
probability in a sphere of radius $R_0$.
\item Taking each of these $N_c$ points as centers, 
we distribute randomly $N_p$ points in spheres of
radius $\ell_c$ centered on them. 
\item We calculate the moment of inertia tensor
and use it to determine the direction of the 
principal axis with the largest eigenvalue. 
We give a coherent rigid body rotation to
the whole cloud about this axis.
\end{itemize} 

For a given total particle number, this is thus 
a three parameter family of configuration. We 
take these parameters to be $N_c$,
$\ell_c//\Lambda_c $ and $b_{rot}$,
where the latter is given again by
equation (\ref{def-brot}) with $W_0$
the total initial potential energy. 
$\Lambda_c$ here is the mean distance 
between neighboring clouds, 
$\Lambda_c \approx 0.55( 4 \pi R_0^3/3 N_c)^{1/3}\;$,
and the parameter $\ell_c//\Lambda_c$ thus
characterizes the degree of overlap of the
individual clouds. We are in practice interested
in values not so different from unity so that the 
initial density fluctuations are not so large,
and there is a global collapse of the whole 
structure (rather than separate collapses
for the sub-structures and the whole structure).  
We have studied the range of parameters
 $N_c \in [3,50]$, 
  $\eta=l_c/\Lambda_c \in [0.1,2]$and 
 $b_{rot} \in[-1,-0.1]$.  

We again report results for just two cases, C1 and C2, specified
in the Table \ref{table-param}, whose features are representative 
of this class of models. 

{  It is important to note that our random sampling with a finite number
$N$ of particles introduces mass density fluctuations, which are
additional to those intrinsic to our continuum IC. Such density fluctuations 
can of course play a role in the dynamics, and as their amplitude is
$N$-dependent (with $\frac{\delta \rho}{\rho} \sim 1/\sqrt{N}$), one might 
expect this to induce potentially an $N$-dependence in our results even
for macroscopic results. However, as detailed above, the parameters of 
our IC are in fact chosen so that  $\frac{\delta \rho}{\rho} \sim 1$, so one 
might expect the finite $N$ fluctuations to be negligible.
We discuss in more detail finite $N$ effects and the problem 
of taking the continuum limit in Sect.\ref{thermolimit}.
}

As noted above this is indeed consistent with our numerical 
findings over a range of $N$ between $10^4$ and $10^6$. 
For this reason we present in the paper only results for
the case $N=10^6$ and do not discuss any further 
the role of $N$ in our results. Nevertheless we caution that  
it is quite  possible that the  physical processes we simulate 
might have subtle dependencies on $N$ which do not show 
up clearly in the range we have simulated. Notably, as 
illustrated by the study in \citep{Benhaiem_etal_2018} of 
the special case of  spherical IC, such dependencies may arise 
because the finite $N$ fluctuations break symmetries of 
our continuum IC.

\subsection{Physical quantities measured}
\label{physquant} 


A useful basic quantity to monitor the global evolution of the 
system is the ``gravitational radius'' defined by 
\be
\label{Rg} 
R^g (t) = \frac{GM_b(t) ^2}{|W_b(t)|} \;,
\ee
where $W_b(t)$ is the gravitational potential energy of the bound
particles and $M_b(t)\le mN$ is their mass. 

To characterize the system's shape we compute the three eigenvalues $\lambda_i$ { (with $i=1,2,3$) }
of
its inertia tensor, which, in the case of an ellipsoid,
are related to the lengths $a_i$ of the three semi-principal axes by
\be
\label{lambda} 
\lambda_i = \frac{1}{5} M_b (a^2_j+a^2_k)
\ee
where $i \ne j \ne k$ and $i,j,k,=1,2,3$. It follows that
 $\lambda_1 \le \lambda_2 \le \lambda_3$.
It is standard then to introduce three different 
combinations of the $\lambda_i$: the {\it flatness parameter}, 
\be 
\label{iota} 
\iota = \frac{\lambda_3}{\lambda_1} -1 \;, 
\ee 
the {\it triaxiality index}, 
\be 
\label{tau} 
\tau = \frac{\lambda_3-\lambda_2}{\lambda_3 - \lambda_1} 
\ee
and the {\it disk parameter},  
\be 
\label{phi} 
\phi = \frac{\lambda_3-\lambda_1}{\lambda_2+ \lambda_1} \;.
\ee
These parameters
allow one to distinguish not
only between different type of ellipsoids (e.g., prolate, oblate and
triaxial) but also between other shapes ({ i.e.,} bars vs. disks). For instance
a sphere has (0, --, 0) a disk (1,0,0.5) and a narrow cylinder (i.e.,
a bar) ($\gg 1$,1, $\approx 1$). 




In addition to the radial component of a particle's velocity $\vec{v}$,  
\be
\label{vr}
v_r = \frac{ \vec{v}\cdot \vec{r}}{|\vec{r}|} 
\ee
we define the vectorial ``transverse velocity" as
\be
\label{vc}
\vec{v}_t (r)= \frac{\vec{r} \times \vec{v}(r)}{|\vec{r}|} \;,
\ee
i.e., the vector of which the magnitude is that of the non-radial component
of the velocity, but oriented parallel to the particle's angular momentum 
relative to the origin. 

We will denote the average of a quantity in a spherical shell about
radius $r$ by $\langle \cdots \rangle$. Coherent rotation of all the
particles in the shell about the same axis then corresponds to
$\langle |\vec{v}_t| \rangle$= $|\langle \vec{v}_t \rangle|$.
Furthermore we consider  
the anisotropy parameter
\be
\label{beta}
\beta(r)  = 1 - \frac{\langle v_t^2 \rangle}{2  \langle v_r^2 \rangle}  \;, 
\ee
where   $\langle v_t^2 \rangle$ and $\langle v_r^2 \rangle$ 
are respectively the average square value of the transversal
and radial velocity. The anisotropy parameter
has the following limiting behaviors: $\beta \rightarrow 0$ for an
isotropic velocity distribution, and $\beta \rightarrow 1$ when 
${\langle v_t^2 \rangle} \ll {2  \langle v_r^2 \rangle}$, i.e. when
the motion is predominately radial.

To characterize the kinematics further, we also consider the different 
components of the radial acceleration, which can be decomposed
as 
\be
a_r = \dot{v}_r - \frac{v_t^2}{r} = \dot{v}_r - a_c \;,
\ee
where $a_c$ is the magnitude of the centripetal acceleration
associated with the transverse component of the velocity.  To quantify
in a simple manner the degree of circular vs.  radial motion, we will
consider the ratio
\be
\label{zeta} 
\zeta = \frac{\langle \dot{v}_r\rangle }{\langle a_r \rangle} \;.
\ee
When particles' motion is purely circular we thus have 
$\zeta = 0$, while if it is purely radial we have $\zeta=1$:
thus  $\zeta$ captures  different properties
of the velocity field than $\beta$, although they both 
tend to unity when the motion is purely radial.


\section{Results}
\label{results}

The phenomenology of the gravitational 
collapse of the IC we study is,  in many respects,
very similar to that of non-rotating isolated clouds 
discussed at length in previous 
works. We first summarize these behaviors, and in particular recall the
mechanism by which particles may gain enough energy even to
be ejected from the system.  We then subsequently focus on the 
features of the evolving system which are specifically associated 
with the initial rotational motion, and in particular the emergence
of long-lived spiral-like structure, as well as transient bars and/or rings,
in the spatial configuration. { We recall that all the results presented 
explicitly in the article are for simulations with $N=10^6$ particles, 
but that all the quantities considered have shown no apparent $N$
dependence for simulations of the same IC with $N$ ranging from 
$10^4$ to $5 \times 10^5$. As noted above, this corresponds to
there being no apparent dependence on the fluctuations 
associated with the particle sampling of the continuous mass
distributions characterizing the initial conditions.} 

\subsection{Units}
As unit of length we take $a_3=1$ for our first two 
families of IC, and $R_0=1$ for the third one.
As unit of time we then take
\be
\label{taud}
\tau_d = \sqrt{ \frac{\pi^2} {8 G M}} \;, 
\ee
i.e., the characteristic time for the collapse of
a sphere of radius unity (where $M$ is the
total mass of the system). Finally particle
energies will be given in units of ${GMm}$
where $m$ is the particle mass (and $M=Nm$).

\subsection{Collapse and re-expansion}

\begin{figure}
\vspace{1cm} { \par\centering
\resizebox*{8cm}{6cm}{\includegraphics*{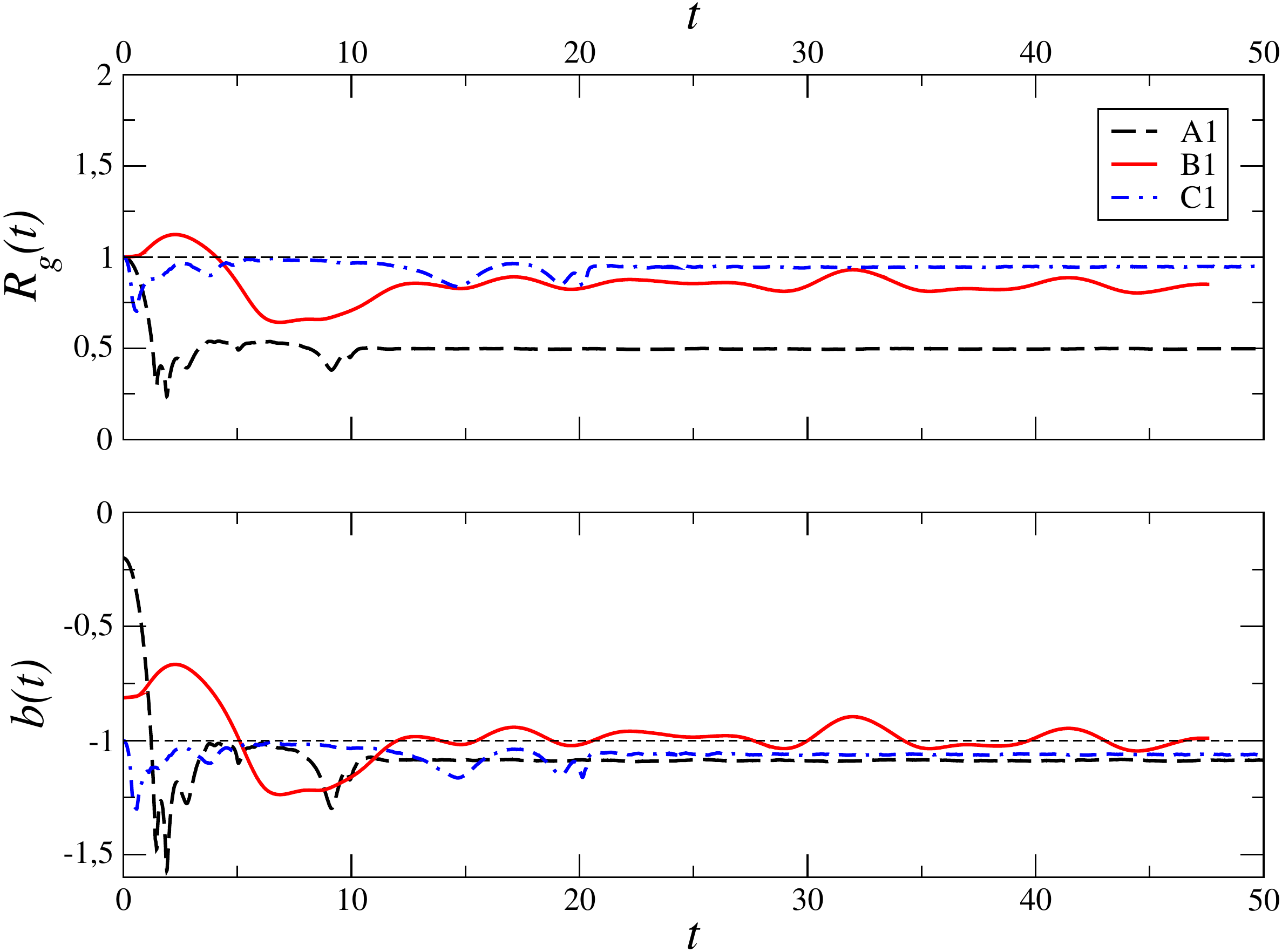}}
\par\centering }
\caption{Upper panel: gravitational radius as a function
  of time for different simulations. 
  Lower panel: global virial ratio. }
\label{Rgt} 
\end{figure}

Fig. \ref{Rgt} (upper panel) shows the evolution of the gravitational
radius (Eq.\ref{Rg}), for three different initial conditions. These
are those which show the largest and the smallest variation among the
ones we have selected. In the case of A1 and C1 the system, which is
initially far from equilibrium, contracts globally reaching a minimum
on a time scale of order $\tau_d$ ($\approx 10 \tau_d$ for the
case of B1 because of the lower density of the external ellipsoid),
then it re-expands and, after a number of {damped} oscillations which
varies, rapidly settles down to a fairly stable value. A similar
behavior is manifested by the virial ratio (lower panel of
Fig. \ref{Rgt}) and thus the stabilization of $R_g$ reflects the
relaxation of the system to a state close to virial equilibrium.  As
we will see below, this inference is only approximately true, as a
small fraction of the mass remains in a time-dependent configuration
on much longer time scales: indeed, it is this fraction of the mass
distribution which we will discuss at length below.

For the cases of A1 and C1 a fraction of the mass is
ejected after the collapse, and as a result the global virial ratio
stabilizes around a value smaller than -1; in the case of B1 
the collapse is less violent and there is no mass ejected. 
We note that for B1 and C1, as all other simulations of these classes,
the gravitational radius is reduced by a smaller factor than for the
case of homogeneous ellipsoids.  Compared to this case, the fluctuations 
of the gravitational field generated are thus much weaker. 


\subsection{Particle energies distributions: before and after collapse}

\begin{figure}
\vspace{1cm}
{ \par\centering
\includegraphics[width = 3.5in]{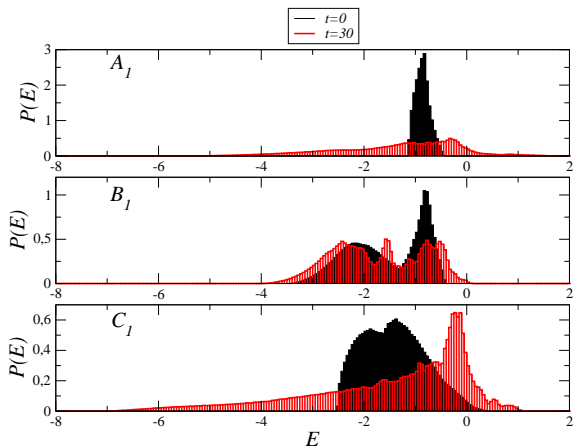} 
\par\centering }
\caption{Energy distribution at $t=0$ and $t=30$ 
 for the simulation A1 (upper panel), B1 (middle panel),
  and C1 (bottom panel) at different times (see
  labels). }
\label{Pe_A1B1C1}
\end{figure}

Fig. \ref{Pe_A1B1C1} shows the distribution $P(E)$ of particle
energies $E$ at two different times in the same three
simulations as in the previous figure.  Plotting these distributions
at longer times than the last one shown we find no noticeable
evolution, i.e., these distributions represent well in principle a
final stationary distribution.
We see that their qualitative behavior divides them clearly 
as in the previous figure: in the cold simulation, A1, the
change in the energy distribution brought by the dynamics is
much more marked, with (i) a much more widely spread energy
distribution compared to that in B1 and C1, and (ii) a significant
fraction of the mass with positive energy, while there is a much
smaller fraction of (or even no) such particles in the other cases. In
the case of B1 there is a small but non-negligible evolution of the
particle energy distribution while C1 represents an intermediate case
with respect to A1 and B1.

The correlation between the behavior in Fig. \ref{Rgt} and
Fig. \ref{Pe_A1B1C1} is simple to understand: changes in particle
energy are driven by the variation of the gravitational field, and
this is much more violent in the cold cases. In the warmer case the
variation of the field is relatively gentle, and mixing in phase space
has time to play a greater role in the relaxation process.
Nevertheless, as we will see, the generation of even a small number of
particles with positive energy or indeed a significant fraction of
bound mass with energy close to zero, is sufficient to produce a
considerable and non-trivial evolution in configuration space.

Let us recall another relevant feature of these energy
changes, which we have shown in previous study of IC without
rotation \citep{Joyce+Marcos+SylosLabini_2009,syloslabini_2012}: the particles 
which  have a large energy {\it gain}, and which thus constitute 
predominantly both the unbound and loosely bound mass, are 
those which {\it lie initially in the outer part of the structure}.
In the  spherical case, these are the particles initially in the 
outer shells, and in the ellipsoidal case, particles which are
at large radii and close to the longest semi-principal axis.  

The reason for this correlation between the energy gain/loss and
particle initial position, is related to the times at which particles
first pass through the center of the structure: particles which 
pass through the center after the bulk of the particles,
experience the intense gravitational field which changes their energy
at a time when it is weakening, simply because the bulk of mass
generating it is already re-expanding. Such particles thus fall into a
potential which is deeper than the one they subsequently climb out of,
and they thus have a greater net boost of their energy.  In the
ellipsoidal case it is quite evident why the initially outermost
particles arrive late: in the evolution from such an initial condition,
collapse occurs first along the shortest axis and last along the
longest axis \citep{Lin_Mestel_Shu_1965}.  In the quasi-spherical case
the reason for the average late arrival of particles in the outermost
shells is more subtle as it is due to a boundary effect: particles near
the boundary experience a lower average density and thus have a longer
fall time (see \cite{Joyce+Marcos+SylosLabini_2009} for a detailed discussion).


\subsection{Mean density profiles}

\begin{figure}
\vspace{1cm} { \par\centering
\includegraphics[width = 3.5in]{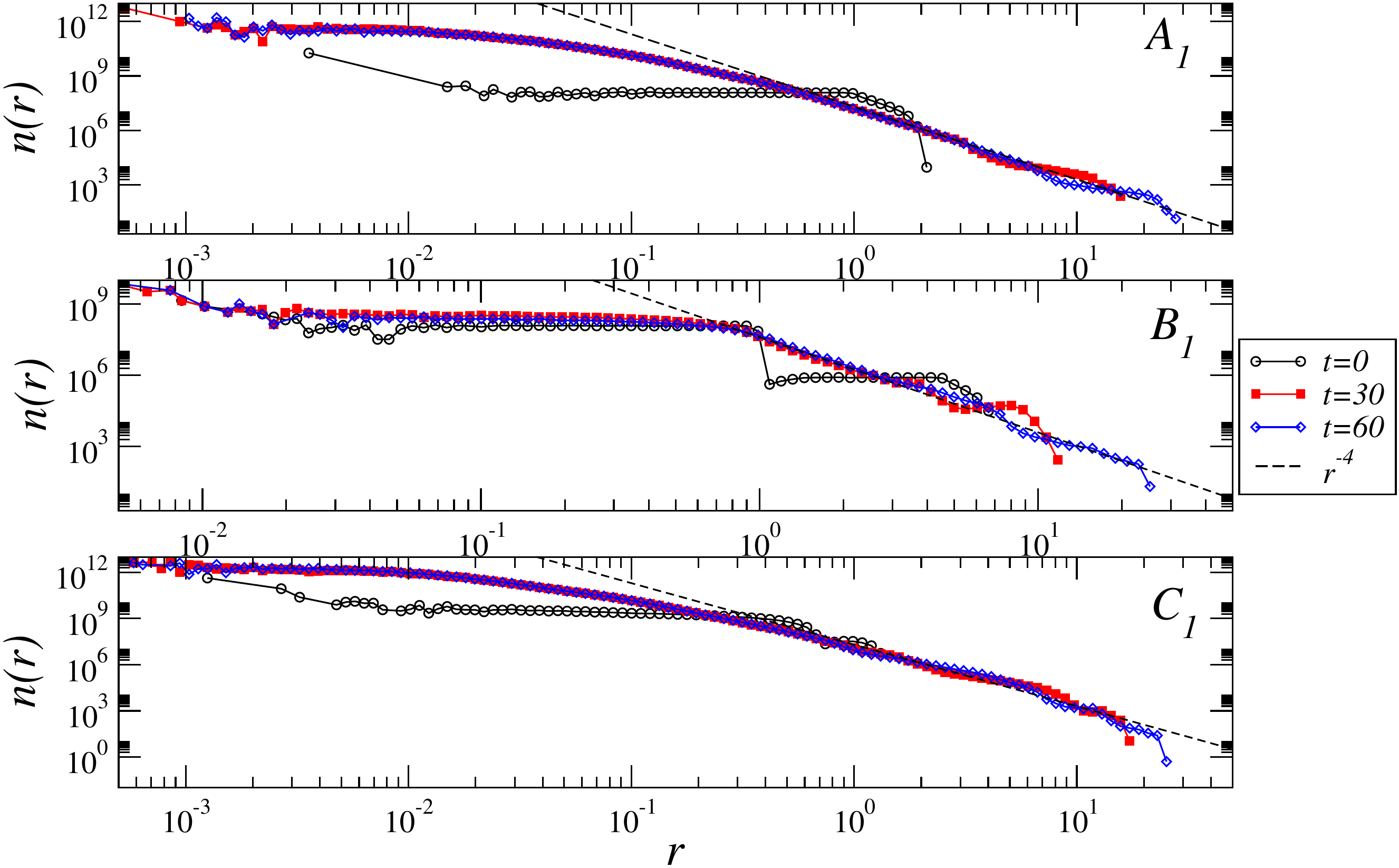} 
\par\centering }
\caption{Density profile for the simulation A1 (upper panel), B1 (middle panel),
and C1 (bottom panel) at different times (see labels). The solid line 
corresponds to a decrease proportional to $r^{-4}$.}
\label{nr_A1B1C1}
\end{figure}

Fig. \ref{nr_A1B1C1} shows the mean density averaged in radial shells
of equal logarithmic thickness $ \Delta\left(\ln(r)\right)$ as a function 
of radius, for the same simulations as in the previous figures \footnote{We take 
as center the particle which has the lowest gravitational potential.  
Note that we will use the same radial binning everywhere below
when we compute averages.}.
 We again 
obtain results very similar to that for non-rotating IC, with 
cold IC producing (i) a more
compact core than the warmer IC,  
%
%
and 
(ii) a characteristic 
$1/r^4$ decay of the density at large radii. As discussed
for example in \cite{syloslabini_2012},  this latter behavior is 
associated with the very loosely bound particles on highly 
radial orbits, and can be explained in a simple manner
by considering that the outermost particles
move approximately  in a  central and stationary potential.   
In these plots the system appears to settle down to a stationary form
on the same times scales as inferred from the plots in the previous
figures (and which are slightly longer for the warm cases).  We can
just make out the signature of some continuing evolution at the
largest radii. It is this which we now focus on.


\begin{figure*}
{\par\centering
\subfloat{\includegraphics[width =1.3in,height=1.1in]{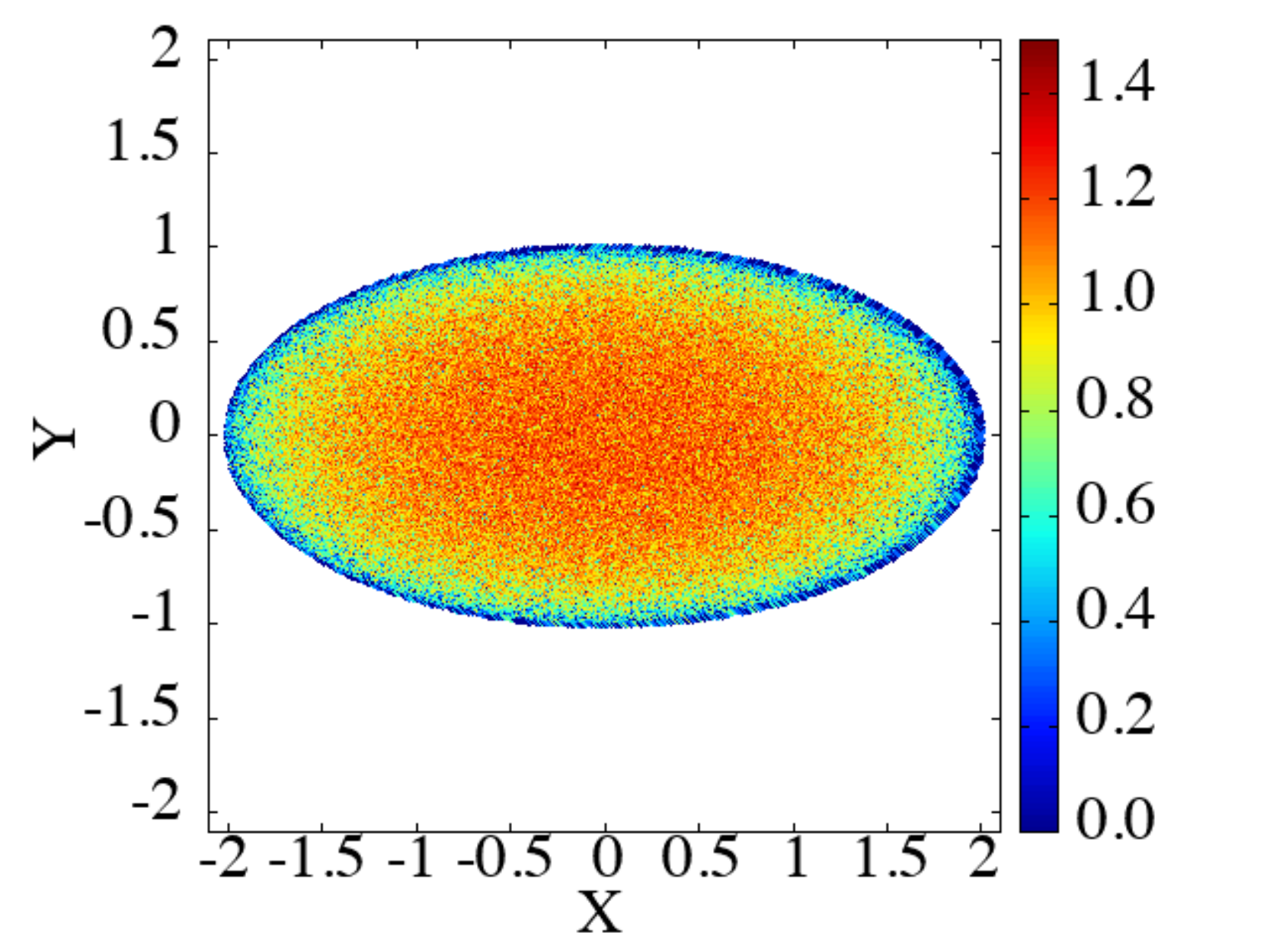}} 
\subfloat{\includegraphics[width =1.3in,height=1.1in]{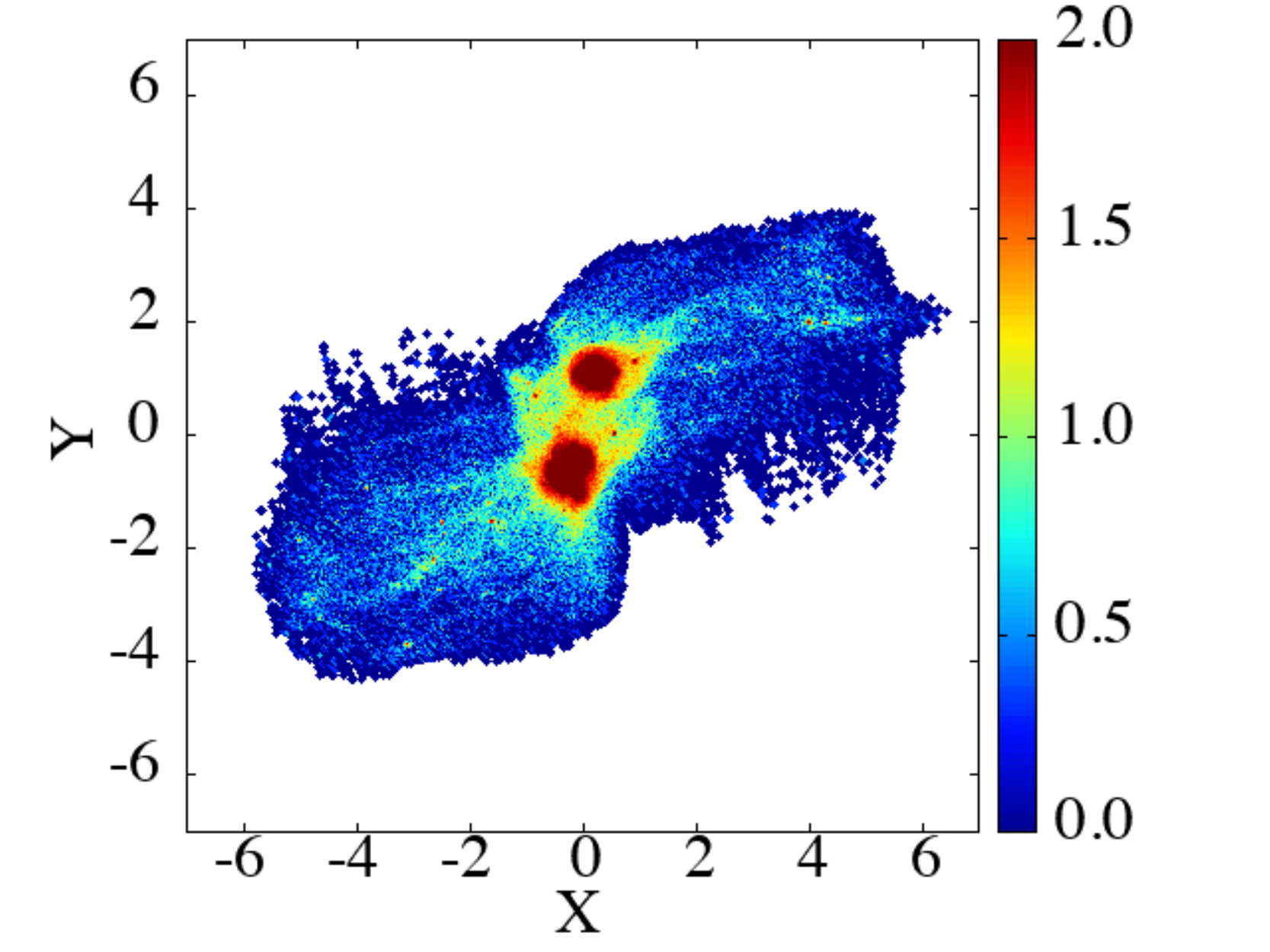}} 
\subfloat{\includegraphics[width =1.3in,height=1.1in]{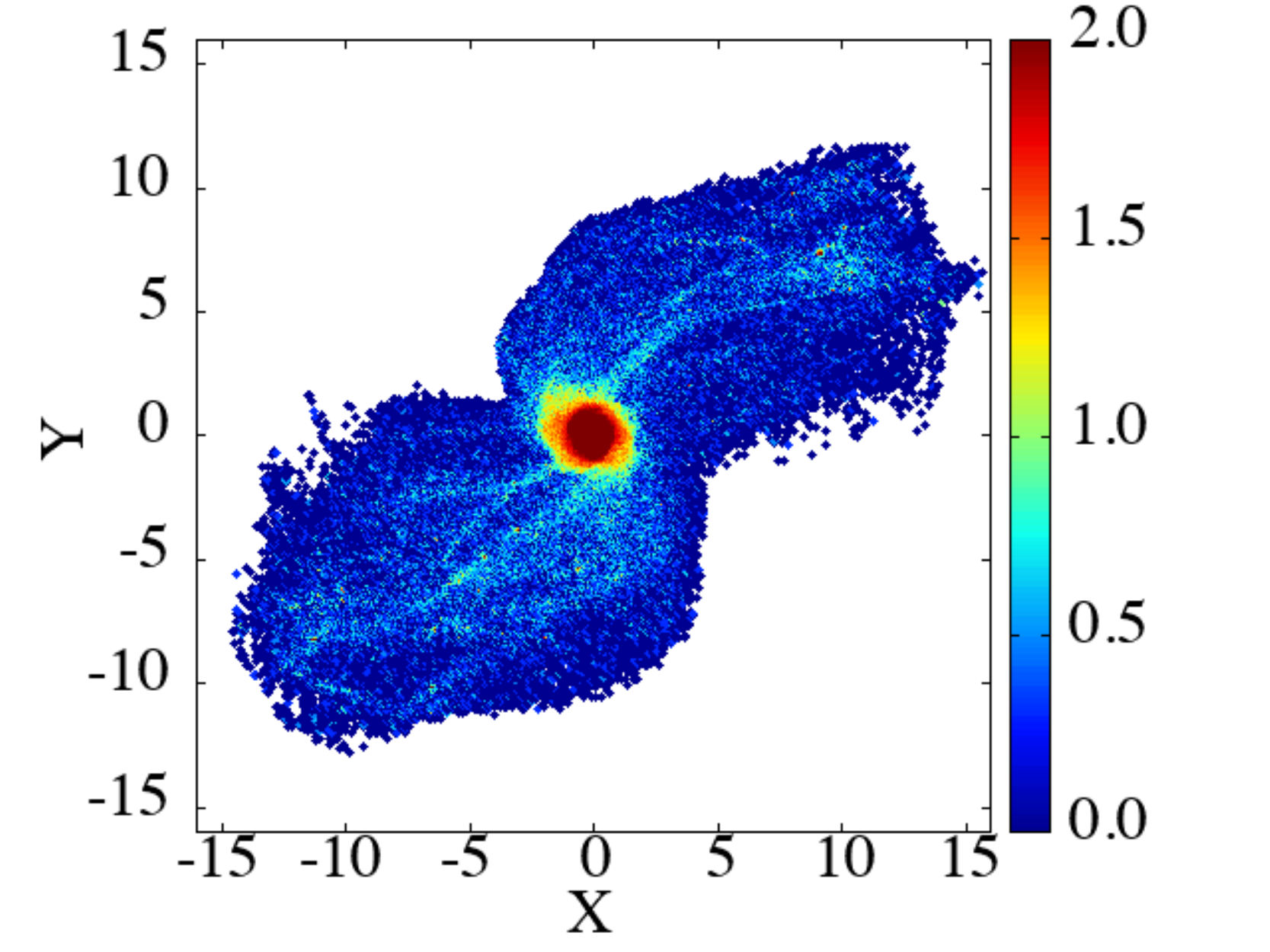}} 
\subfloat{\includegraphics[width =1.3in,height=1.1in]{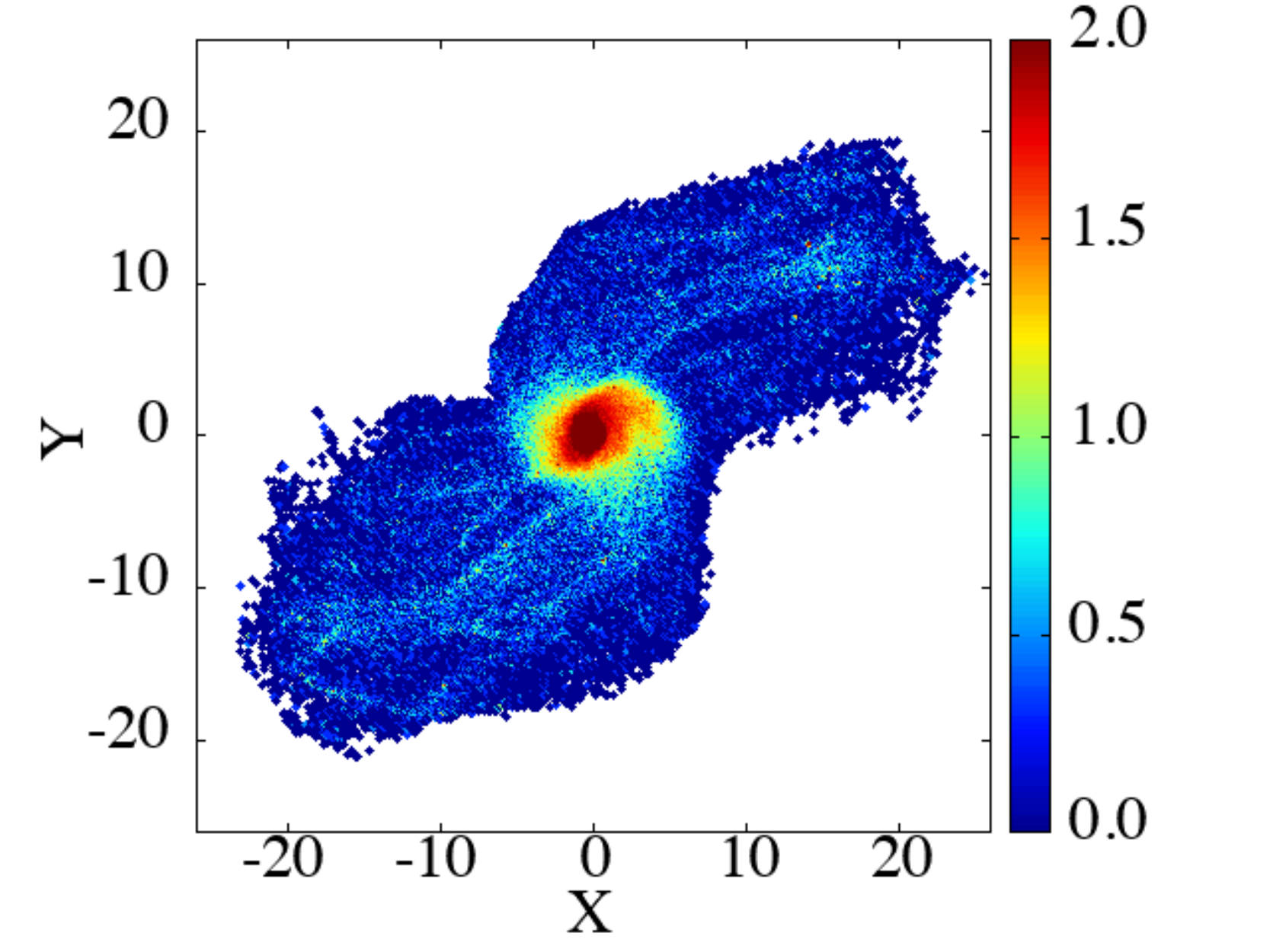}} 
\subfloat{\includegraphics[width =1.3in,height=1.1in]{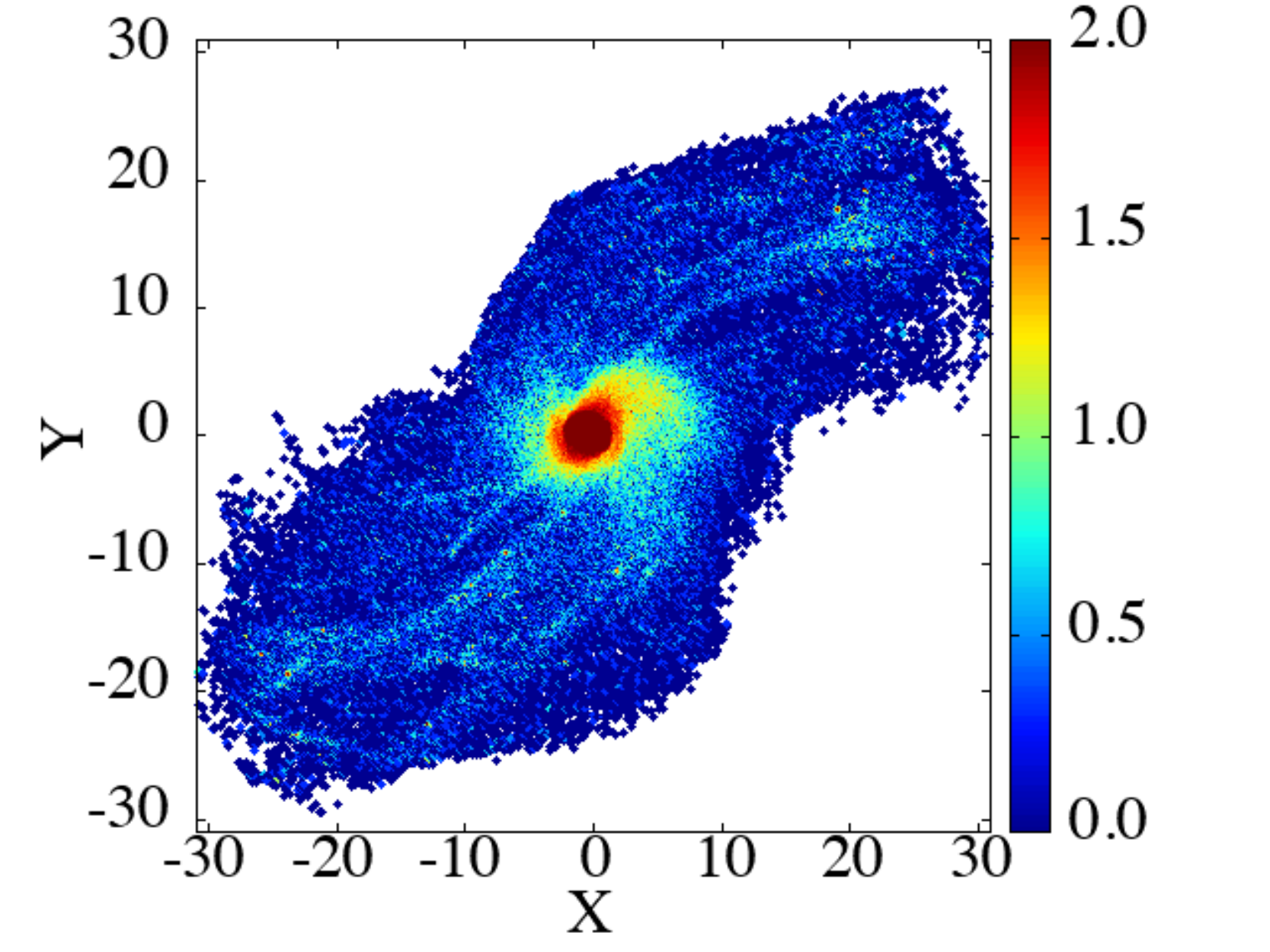}}\\ 
\subfloat{\includegraphics[width =1.3in,height=1.1in]{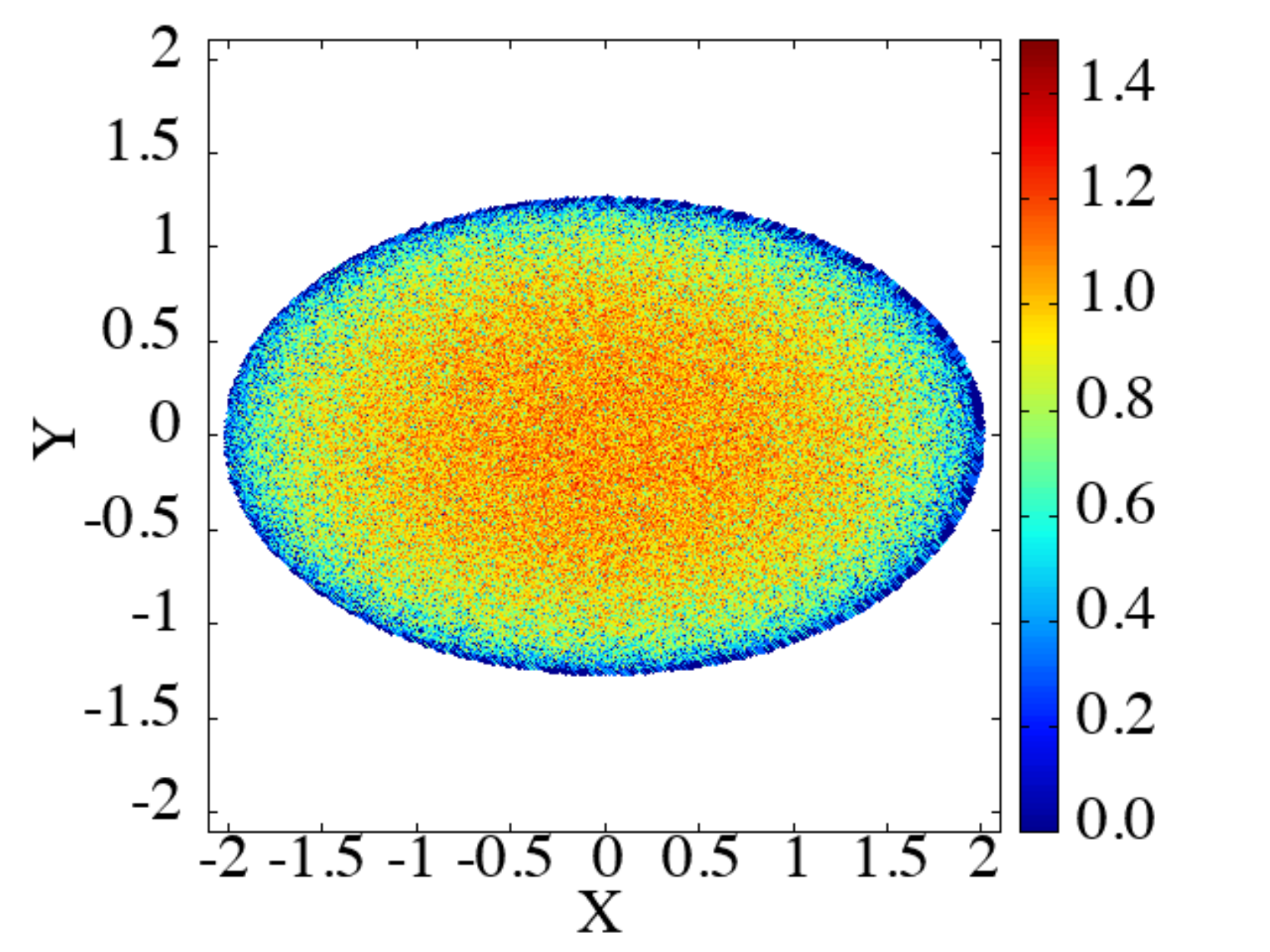}}  
\subfloat{\includegraphics[width =1.3in,height=1.1in]{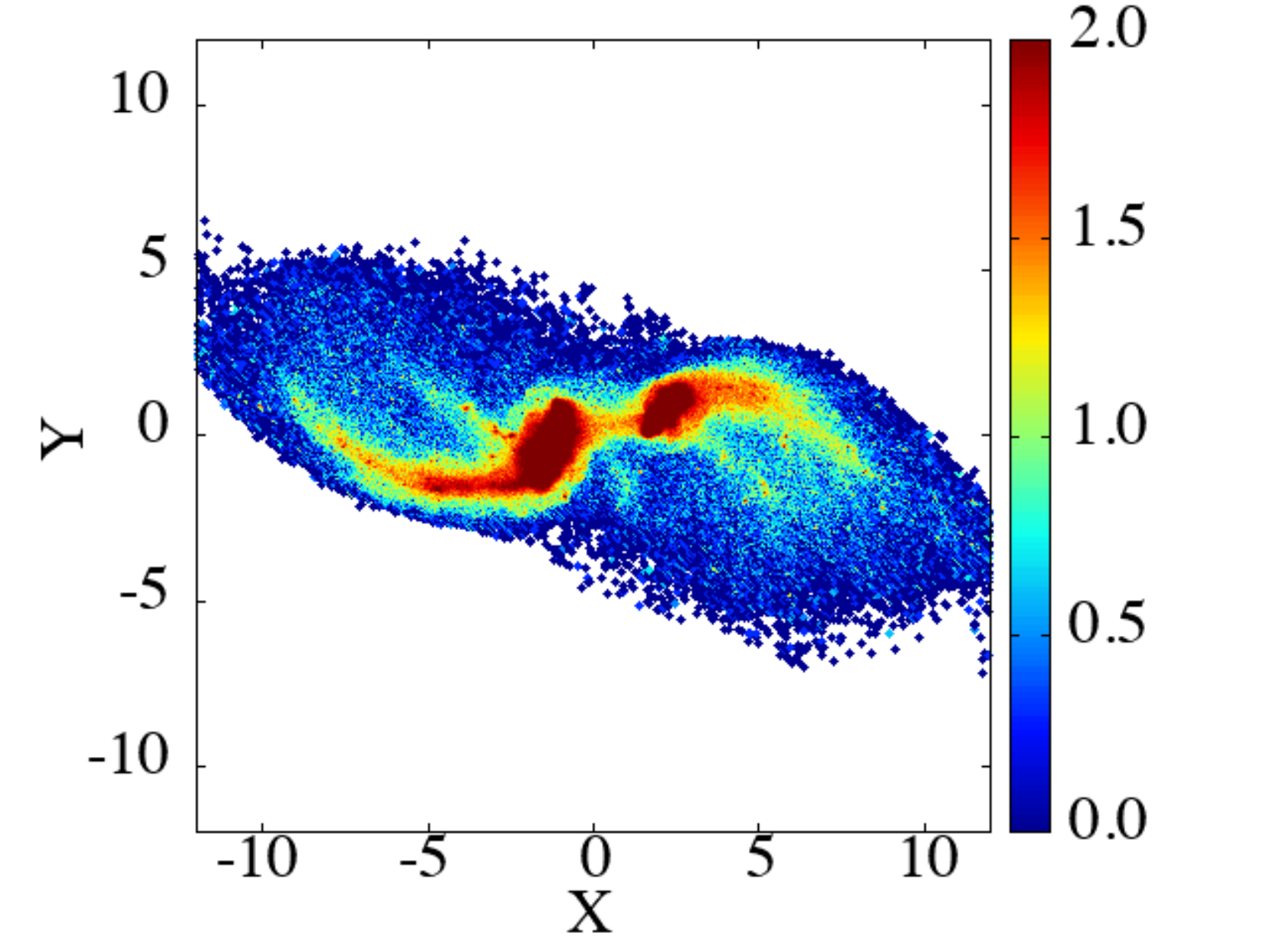}}  
\subfloat{\includegraphics[width =1.3in,height=1.1in]{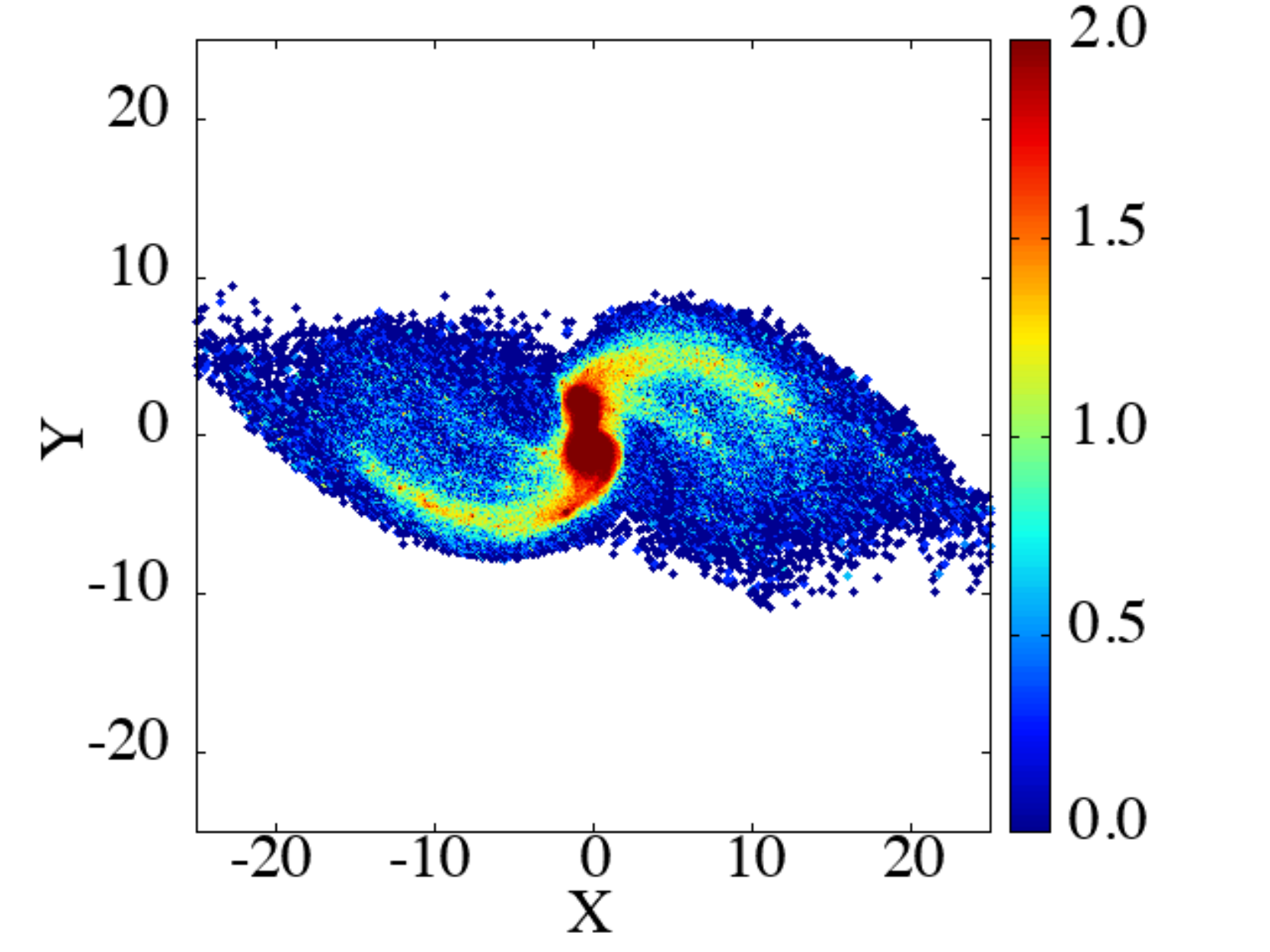}}  
\subfloat{\includegraphics[width =1.3in,height=1.1in]{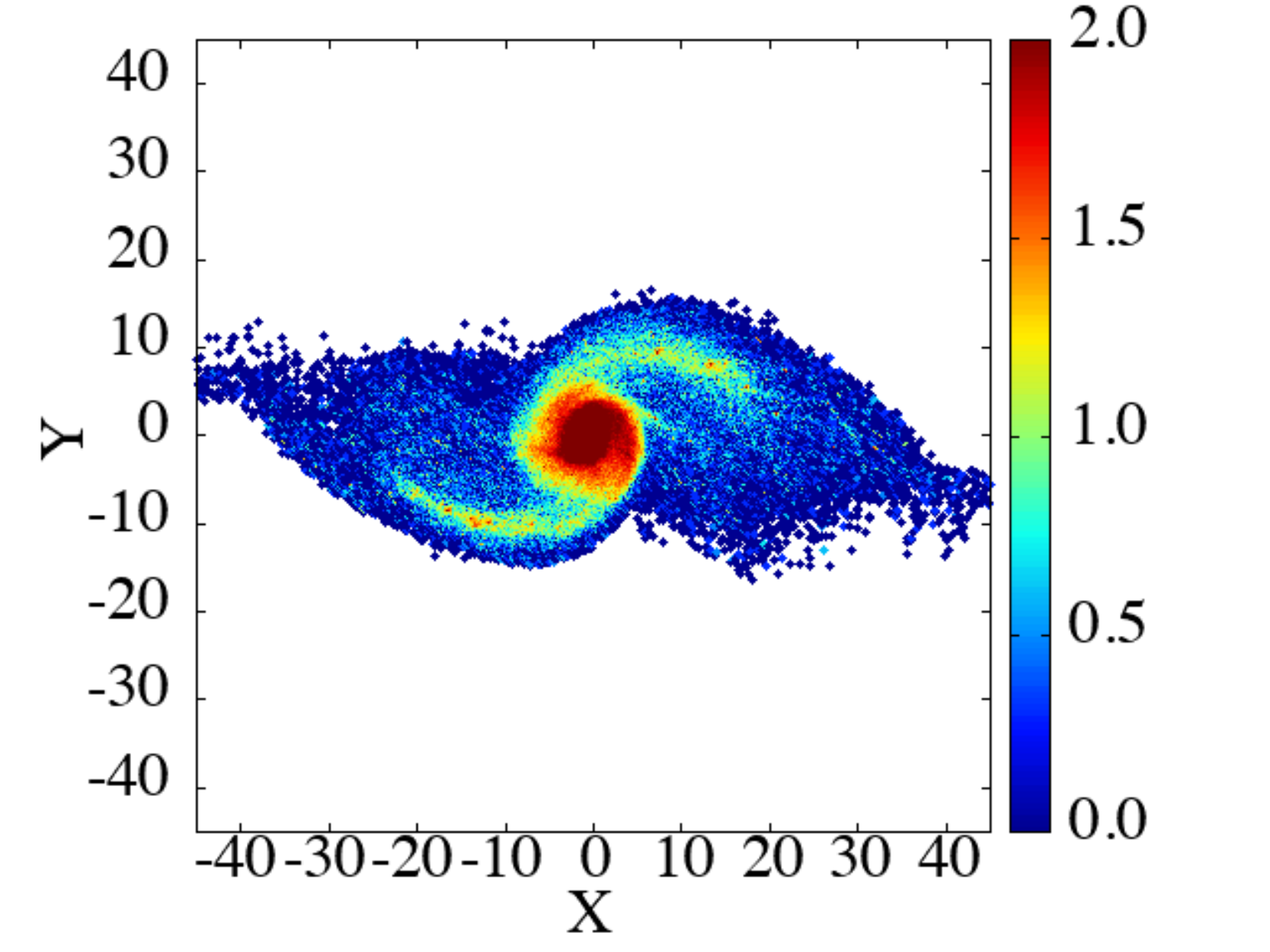}}  
\subfloat{\includegraphics[width =1.3in,height=1.1in]{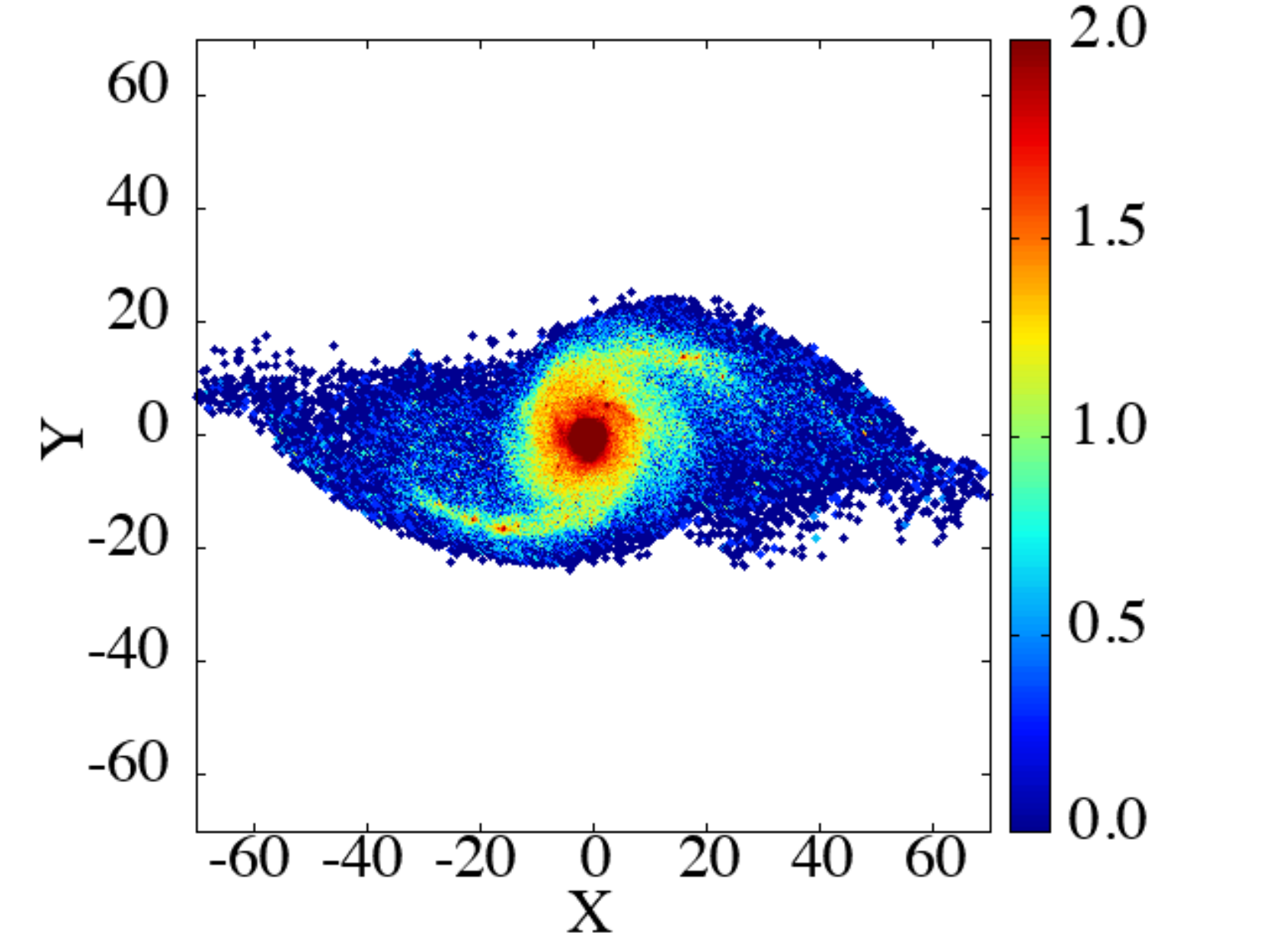}}\\  
\subfloat{\includegraphics[width =1.3in,height=1.1in]{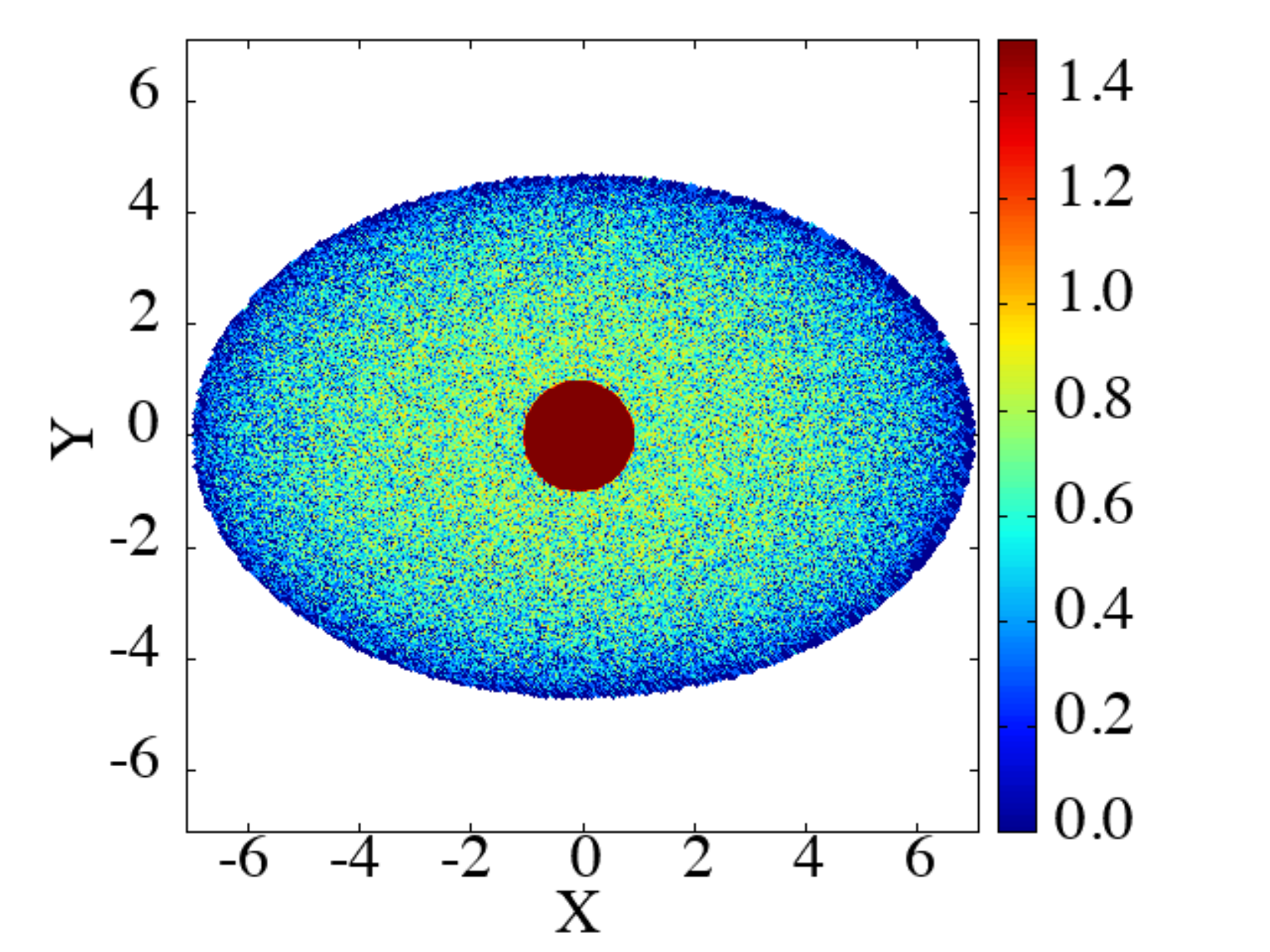}}  
\subfloat{\includegraphics[width =1.3in,height=1.1in]{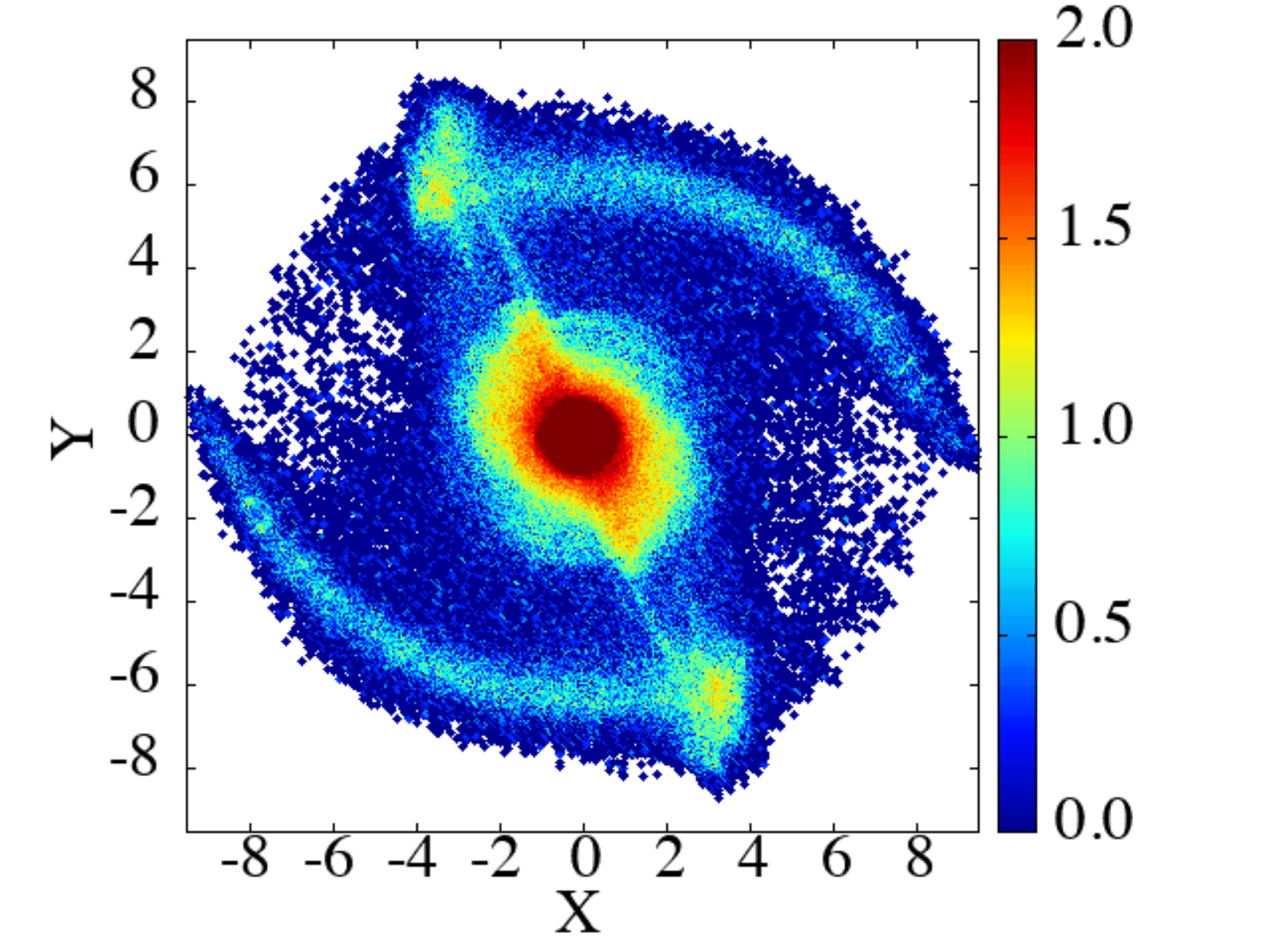}}  
\subfloat{\includegraphics[width =1.3in,height=1.1in]{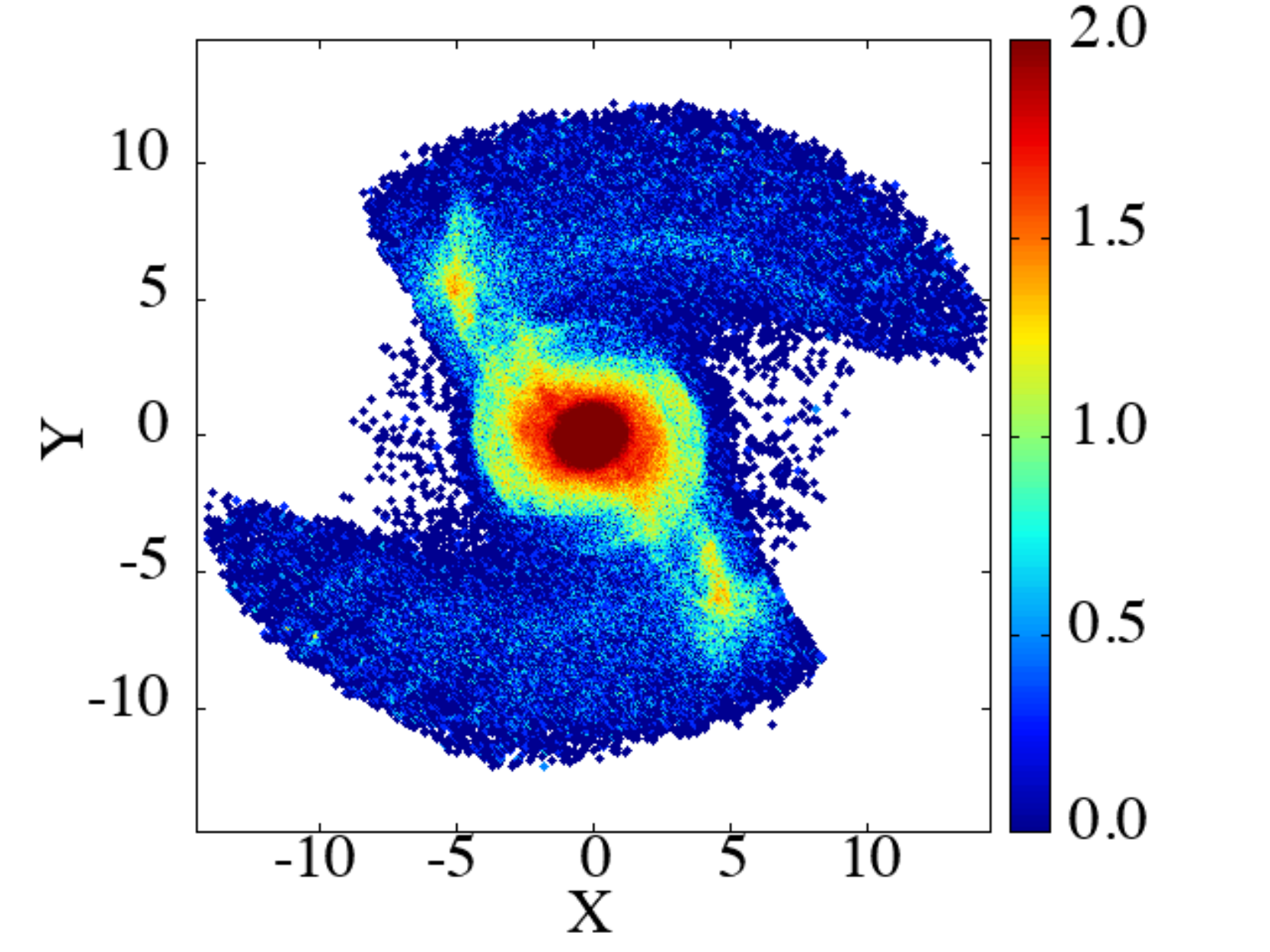}}  
\subfloat{\includegraphics[width =1.3in,height=1.1in]{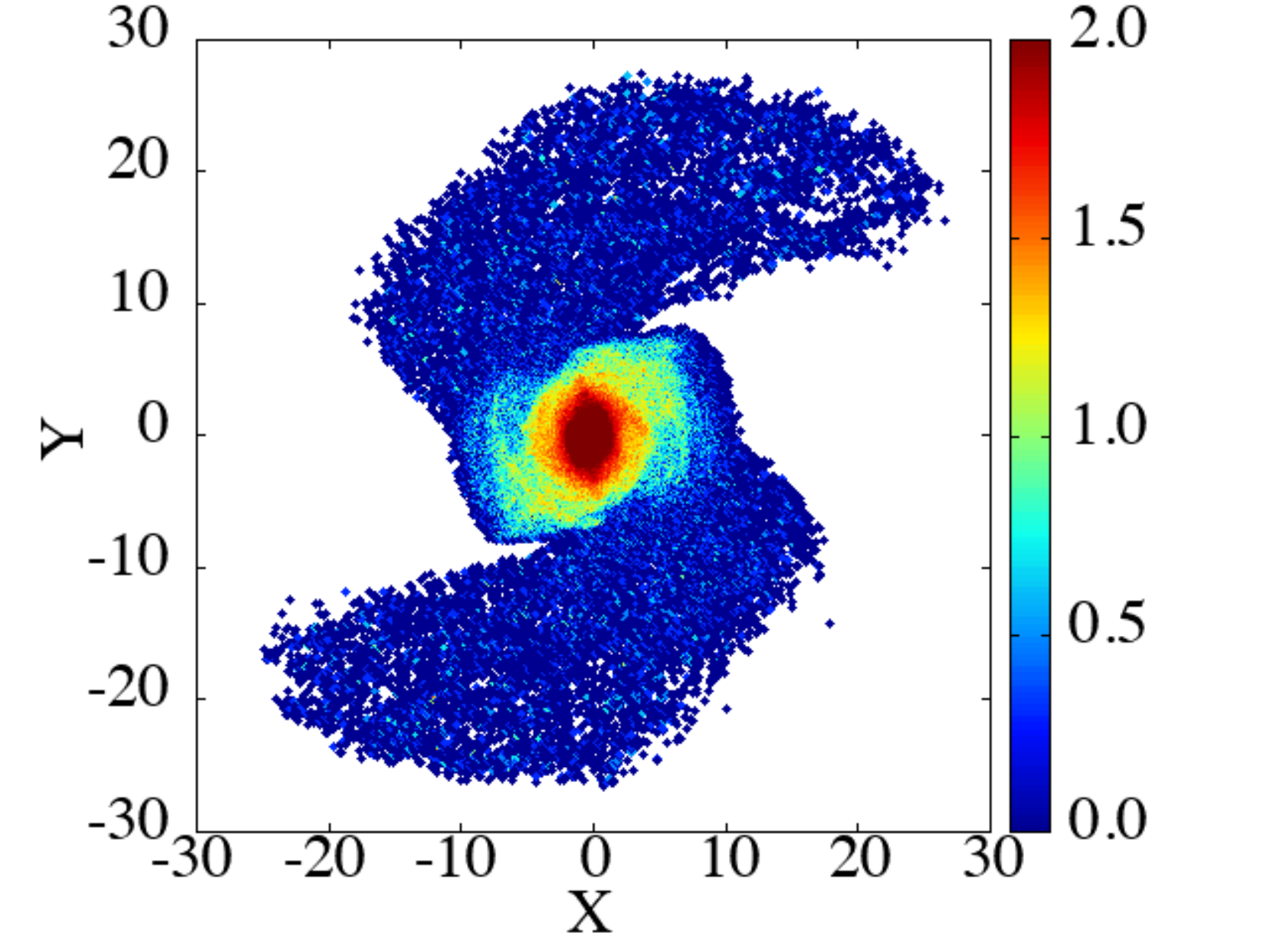}}  
\subfloat{\includegraphics[width =1.3in,height=1.1in]{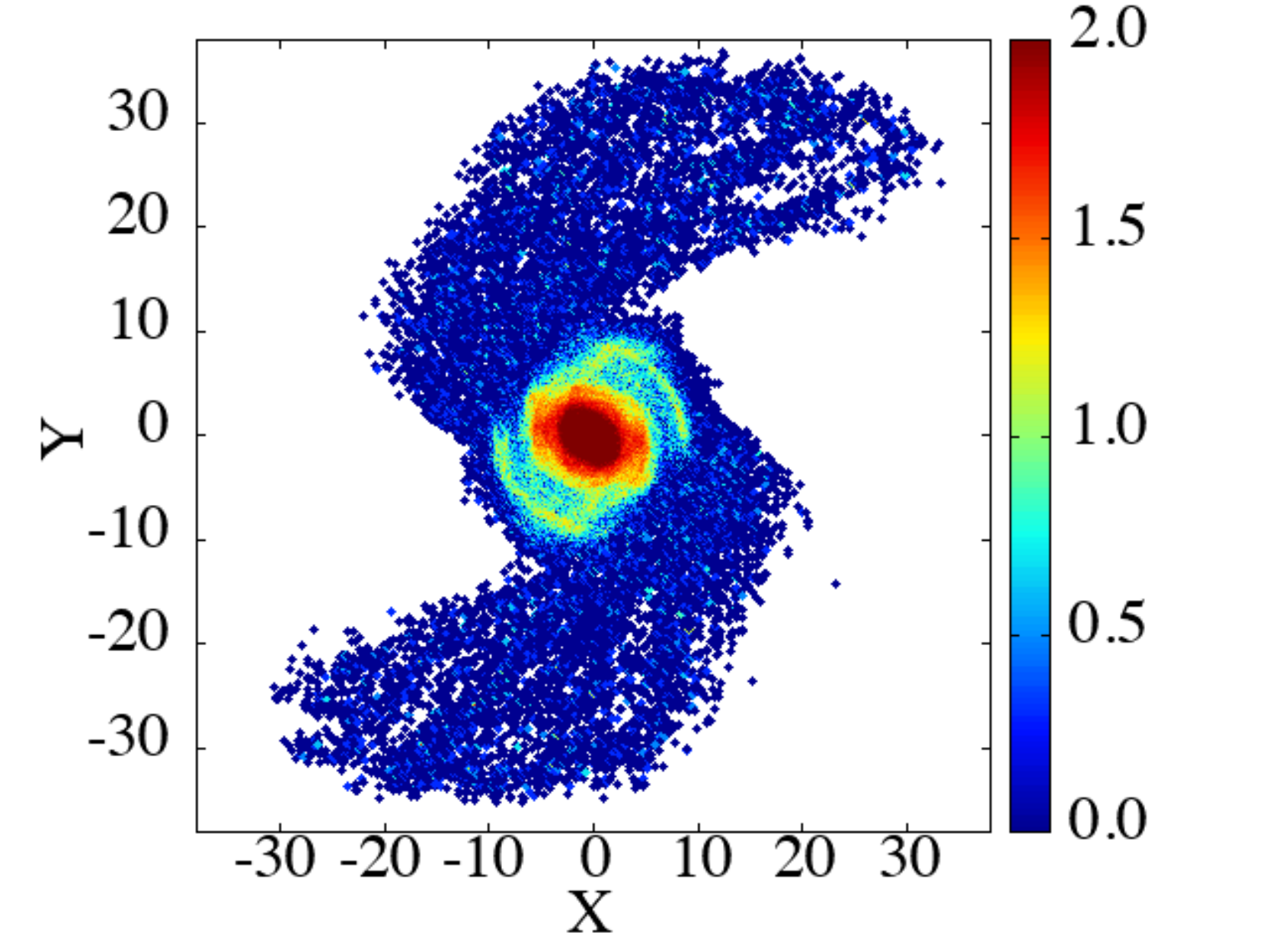}}\\  
\subfloat{\includegraphics[width =1.3in,height=1.1in]{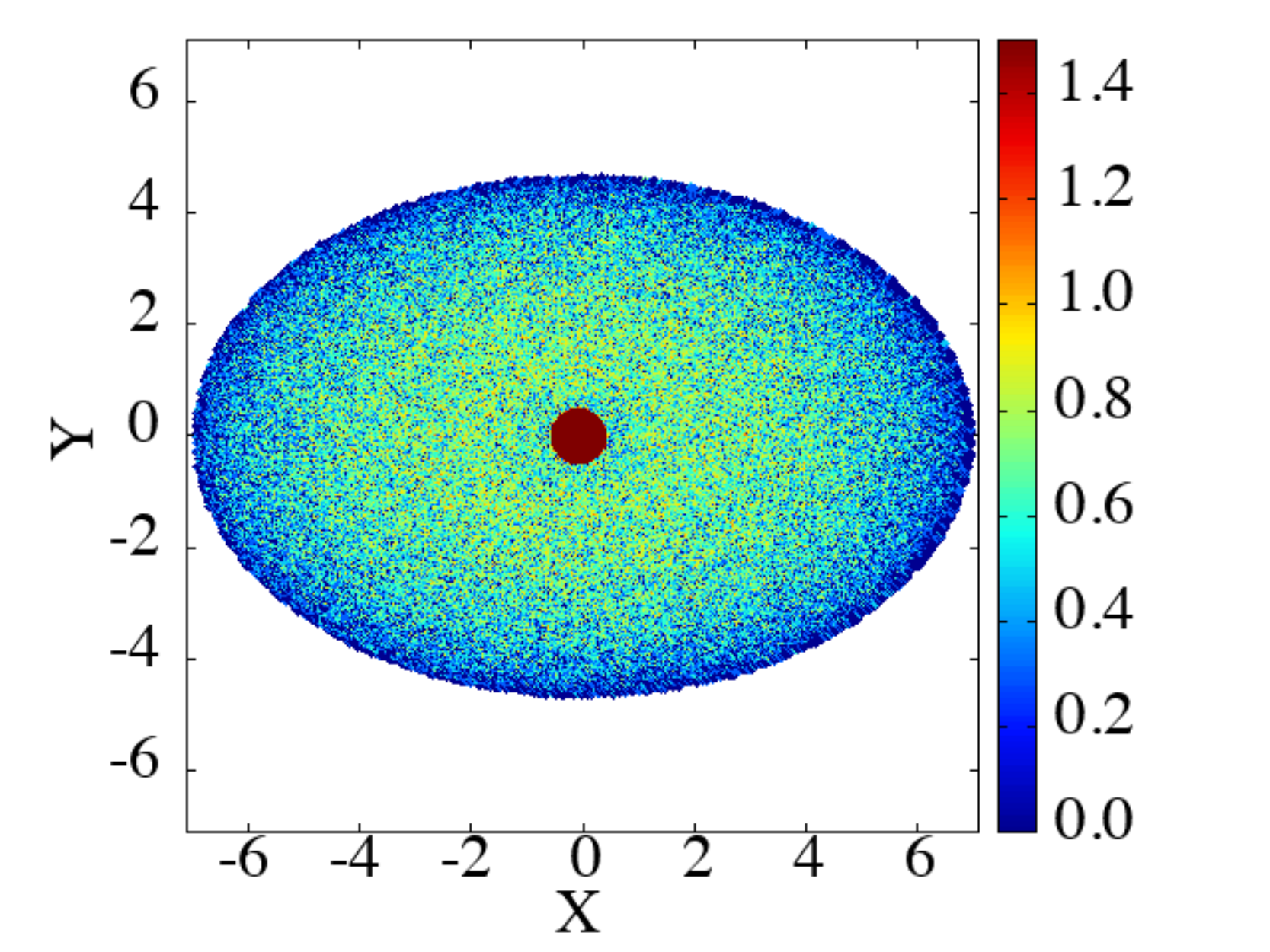}} 
\subfloat{\includegraphics[width =1.3in,height=1.1in]{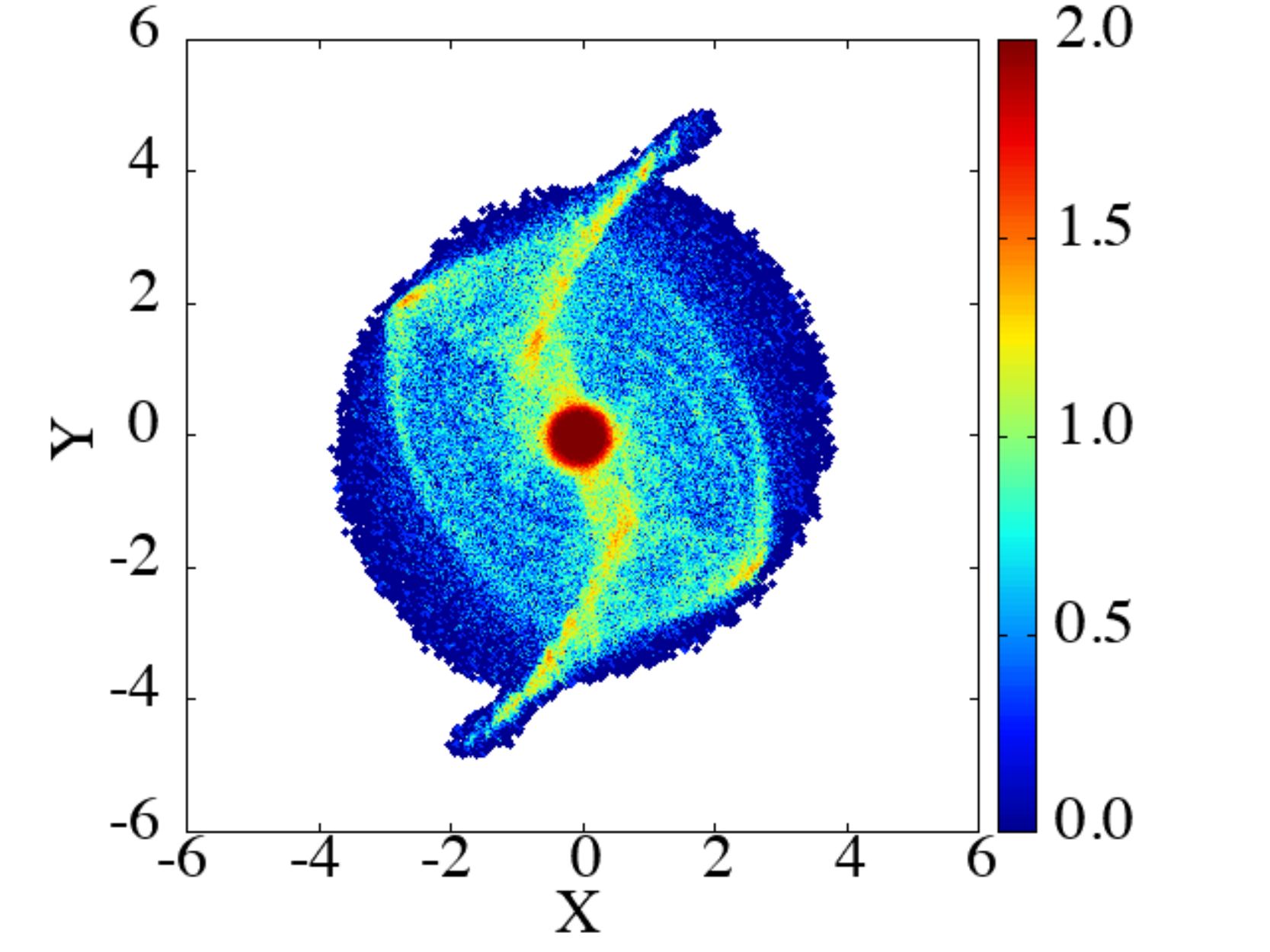}}  
\subfloat{\includegraphics[width =1.3in,height=1.1in]{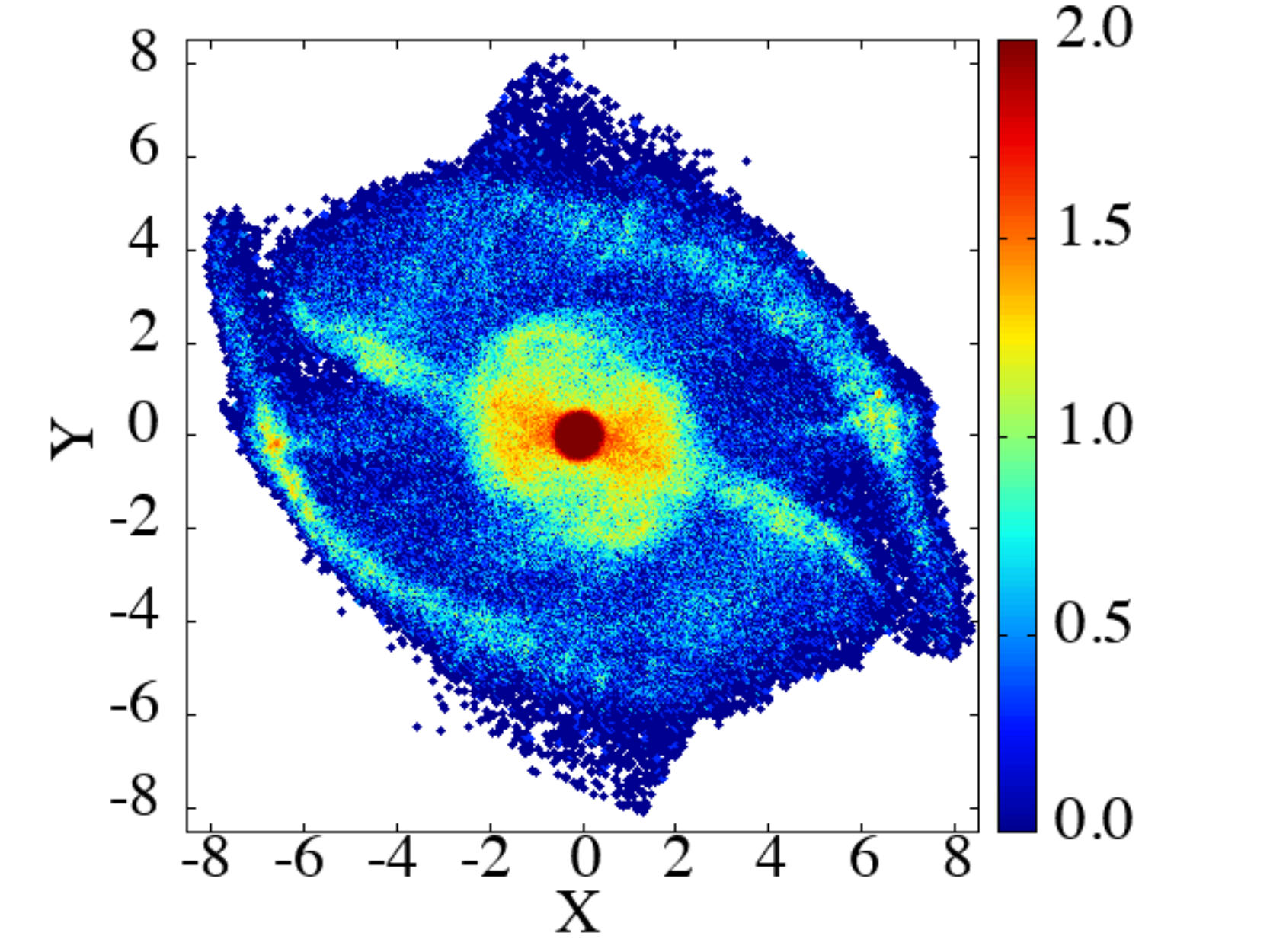}}  
\subfloat{\includegraphics[width =1.3in,height=1.1in]{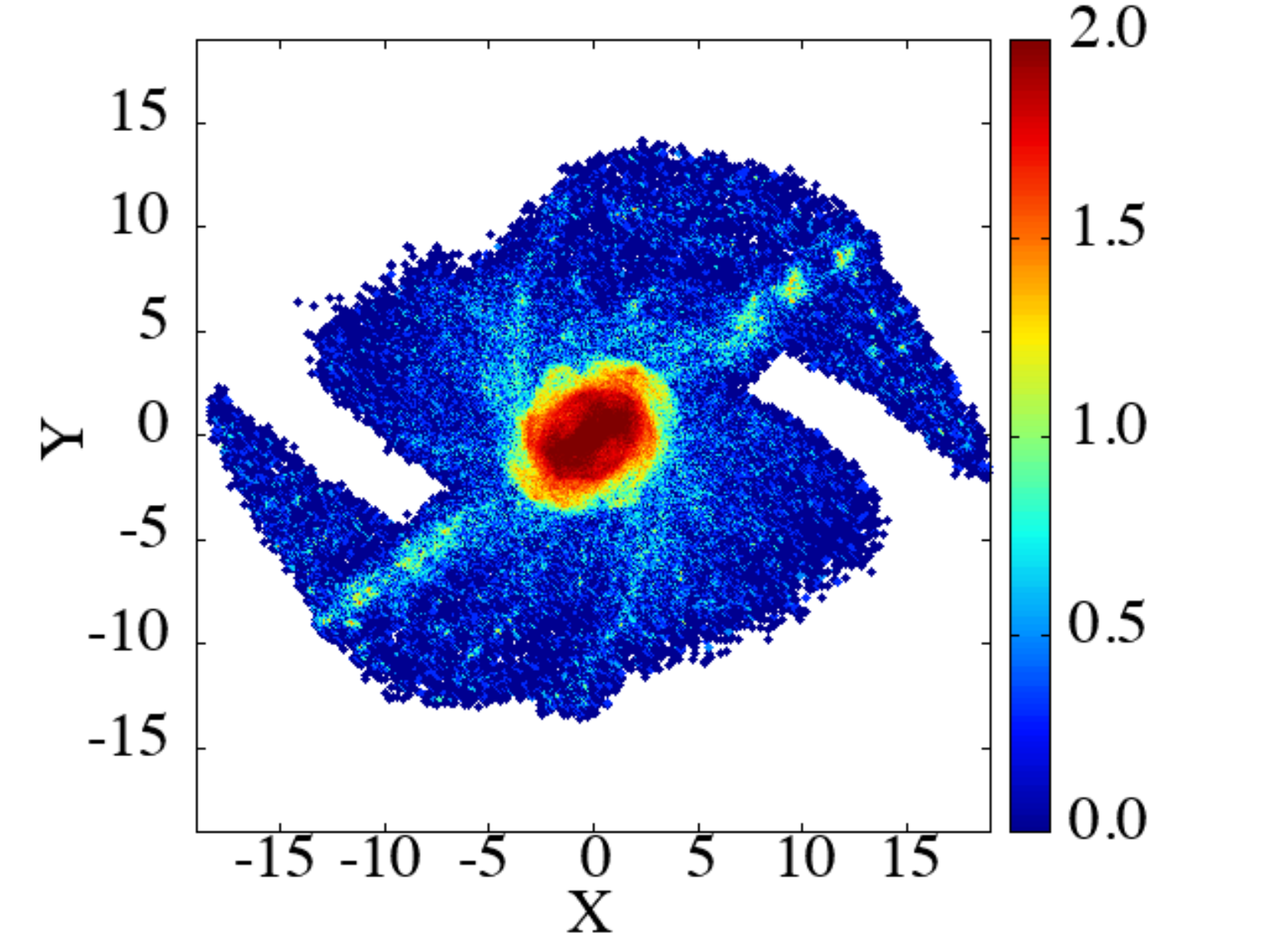}}  
\subfloat{\includegraphics[width =1.3in,height=1.1in]{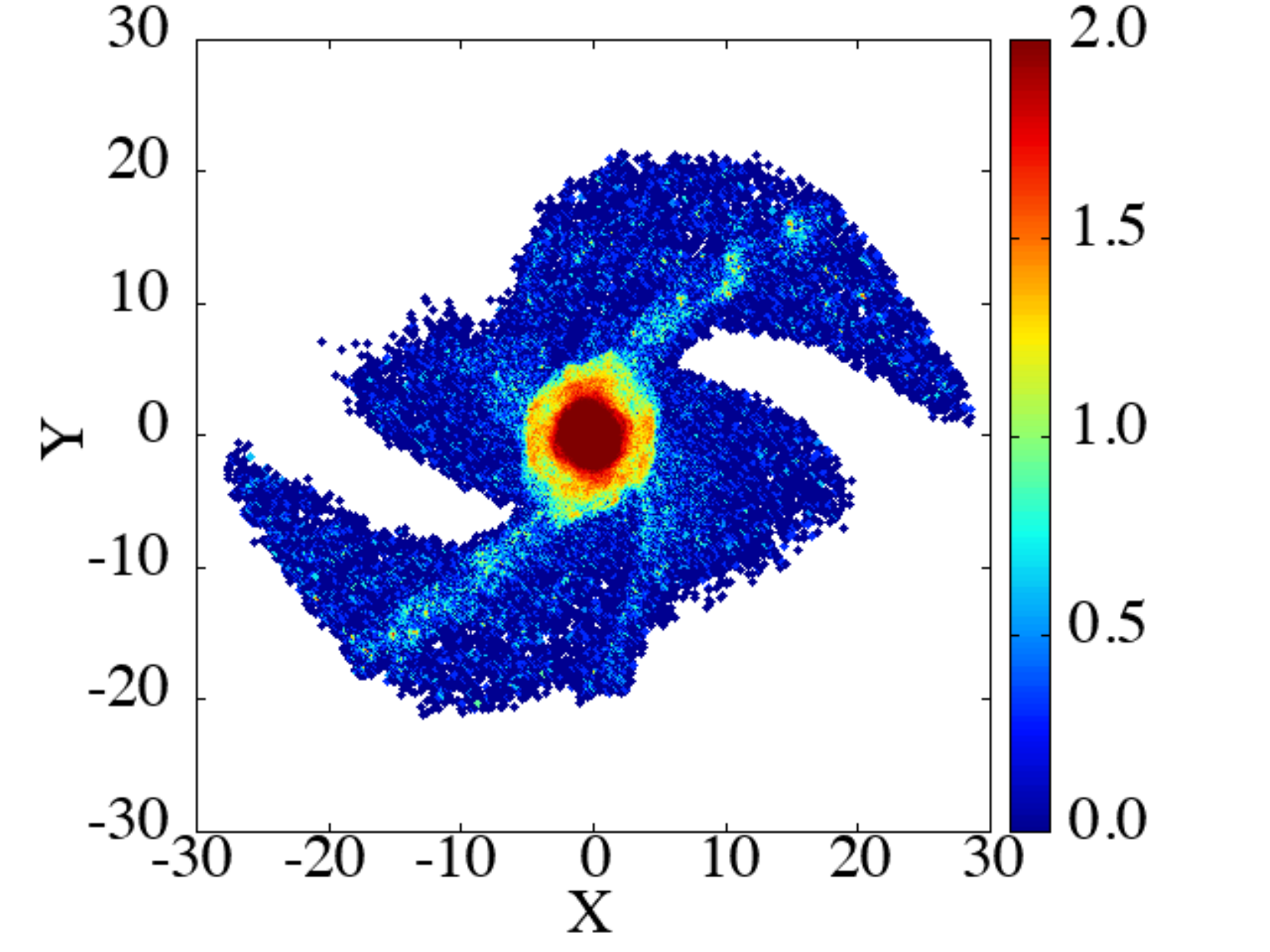}} \\ 
\subfloat{\includegraphics[width =1.3in,height=1.05in]{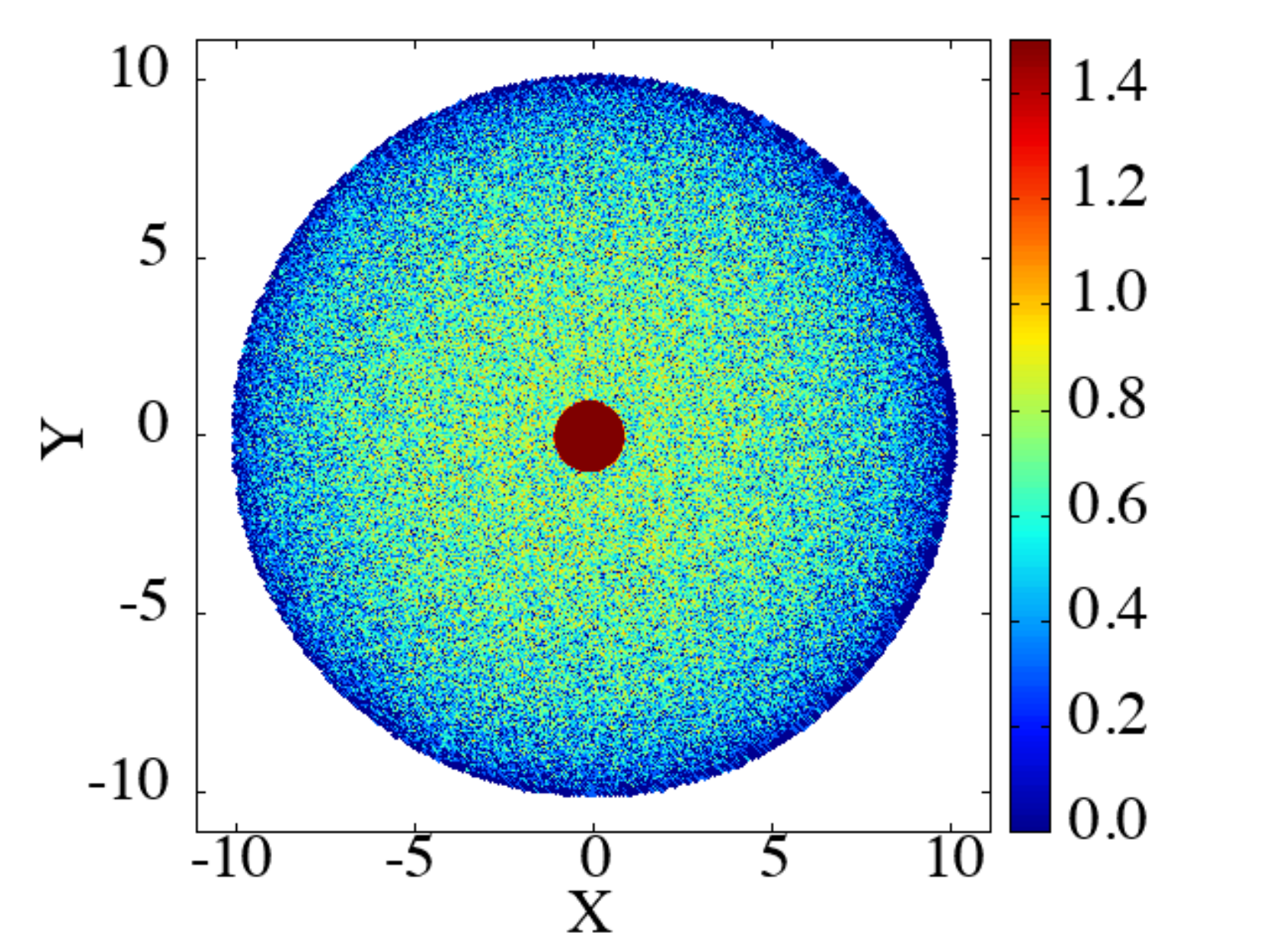}}  
\subfloat{\includegraphics[width =1.3in,height=1.05in]{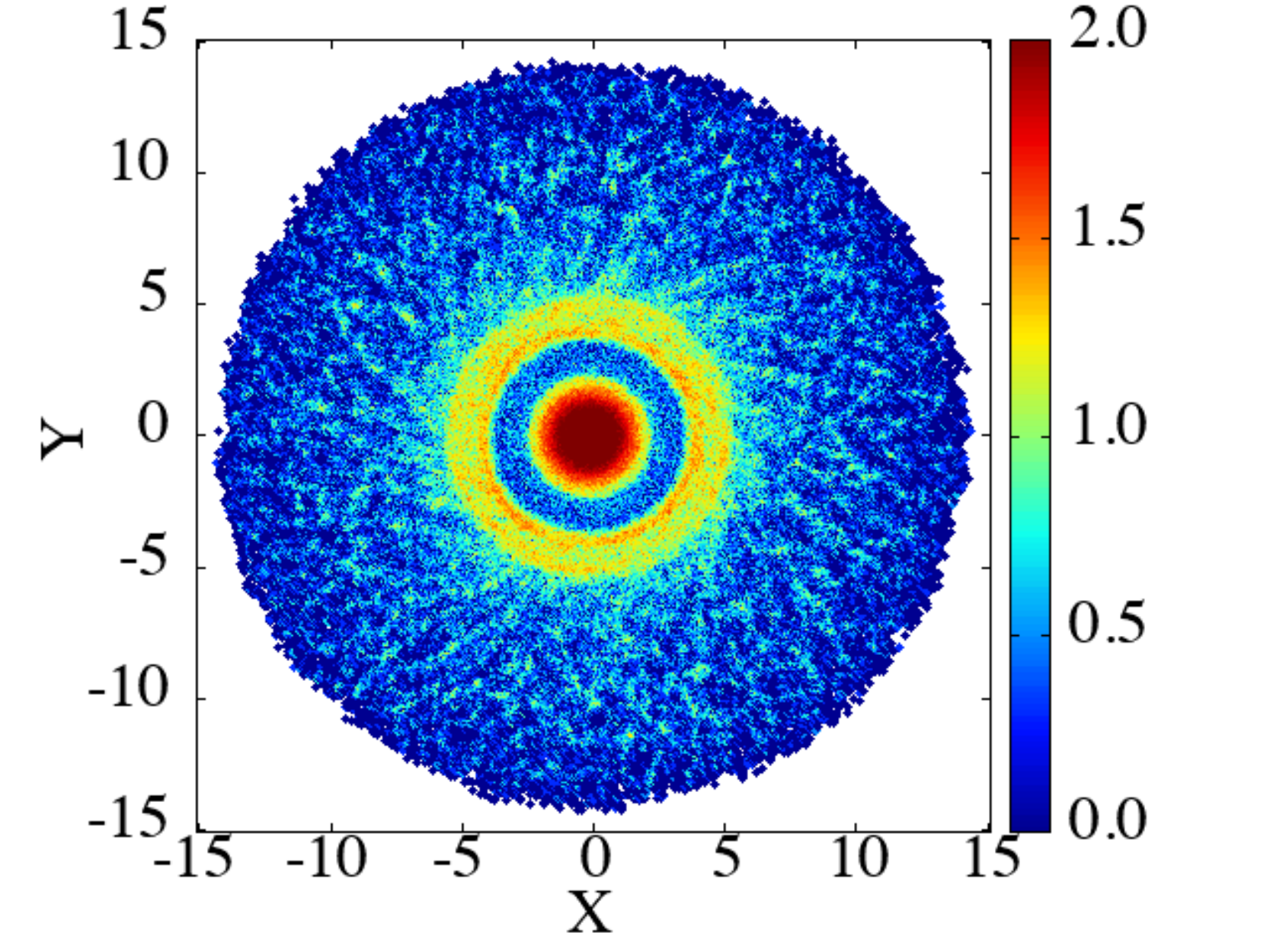}}  
\subfloat{\includegraphics[width =1.3in,height=1.05in]{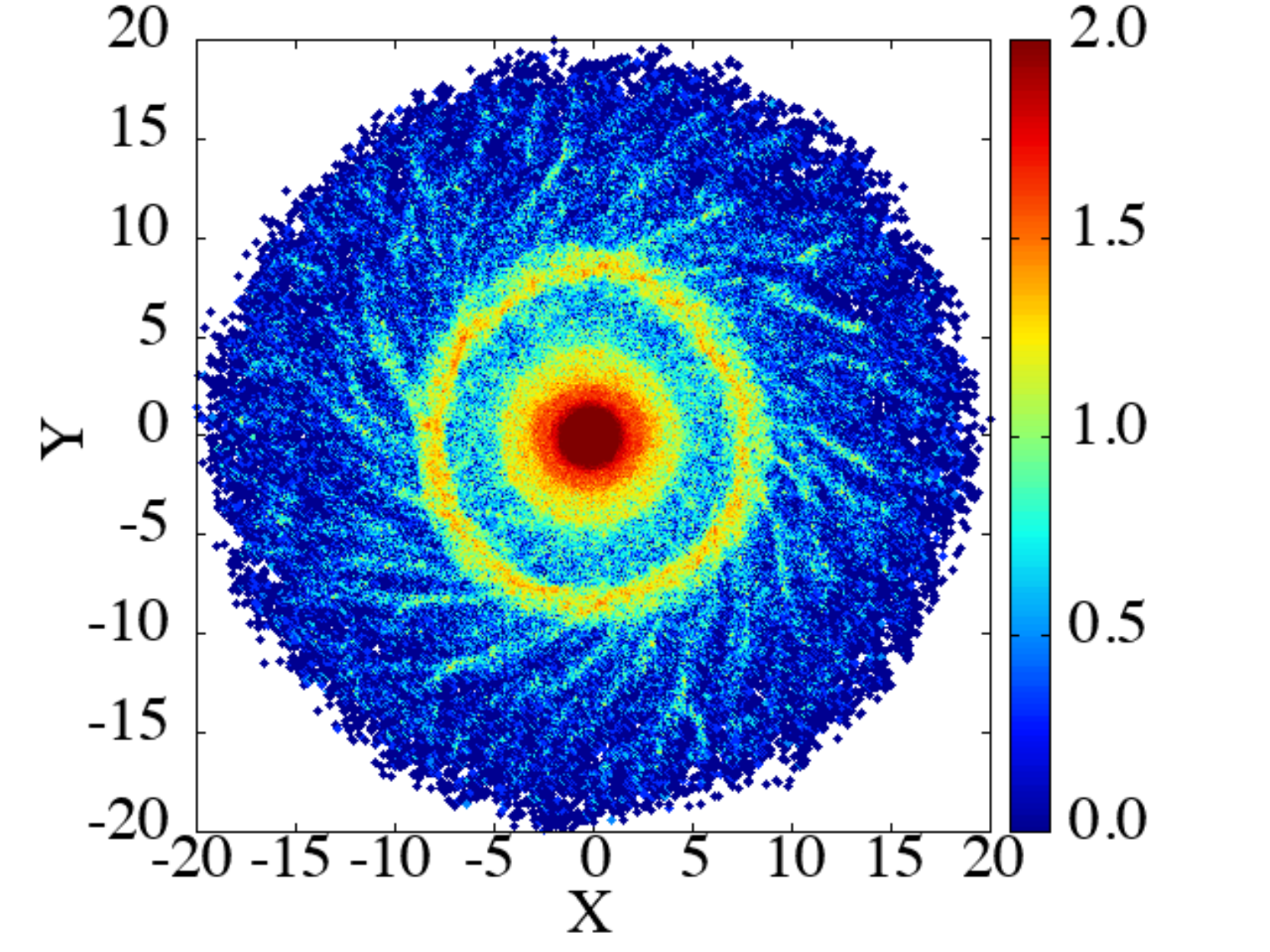}}  
\subfloat{\includegraphics[width =1.3in,height=1.05in]{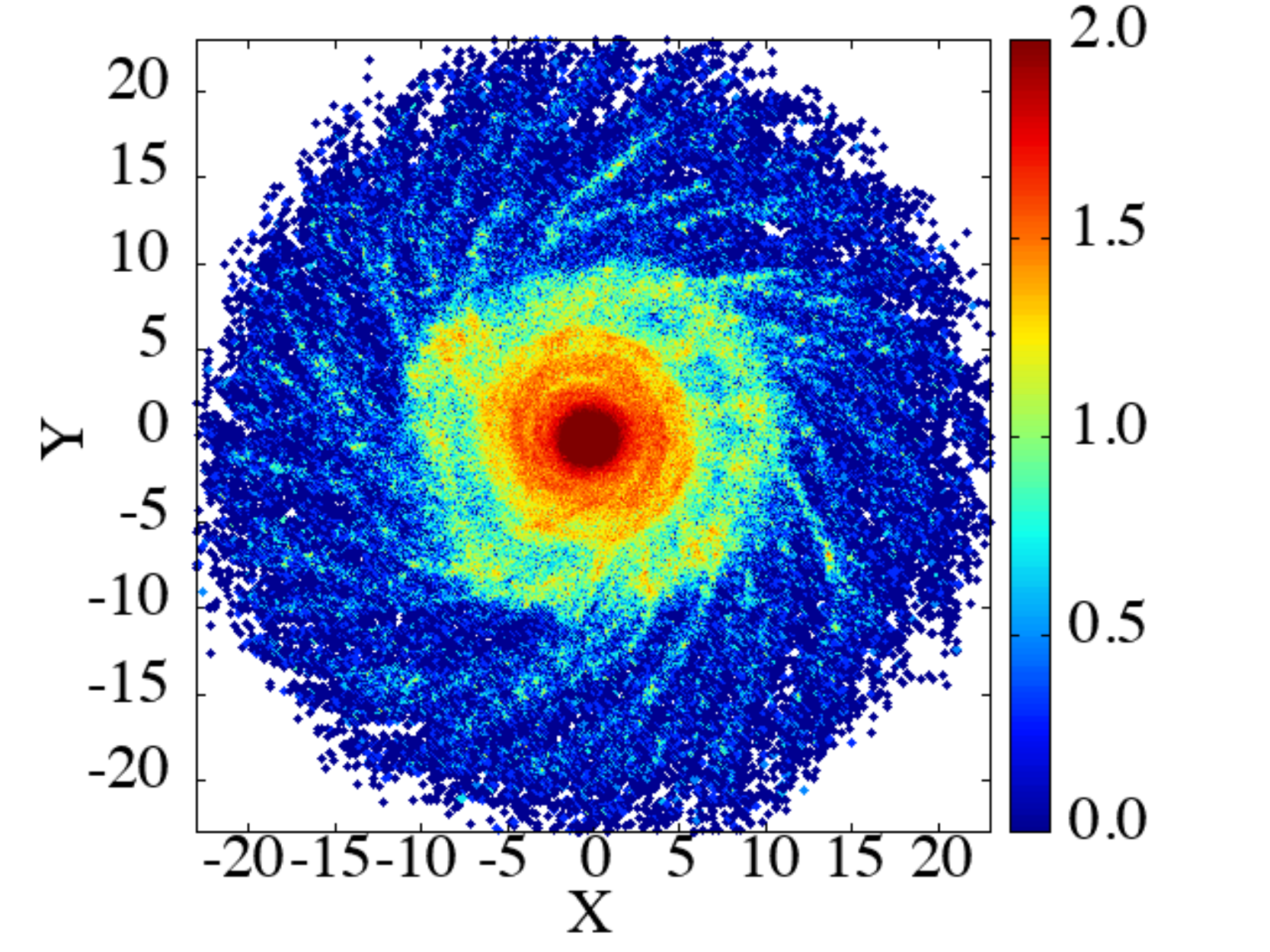}}  
\subfloat{\includegraphics[width =1.3in,height=1.05in]{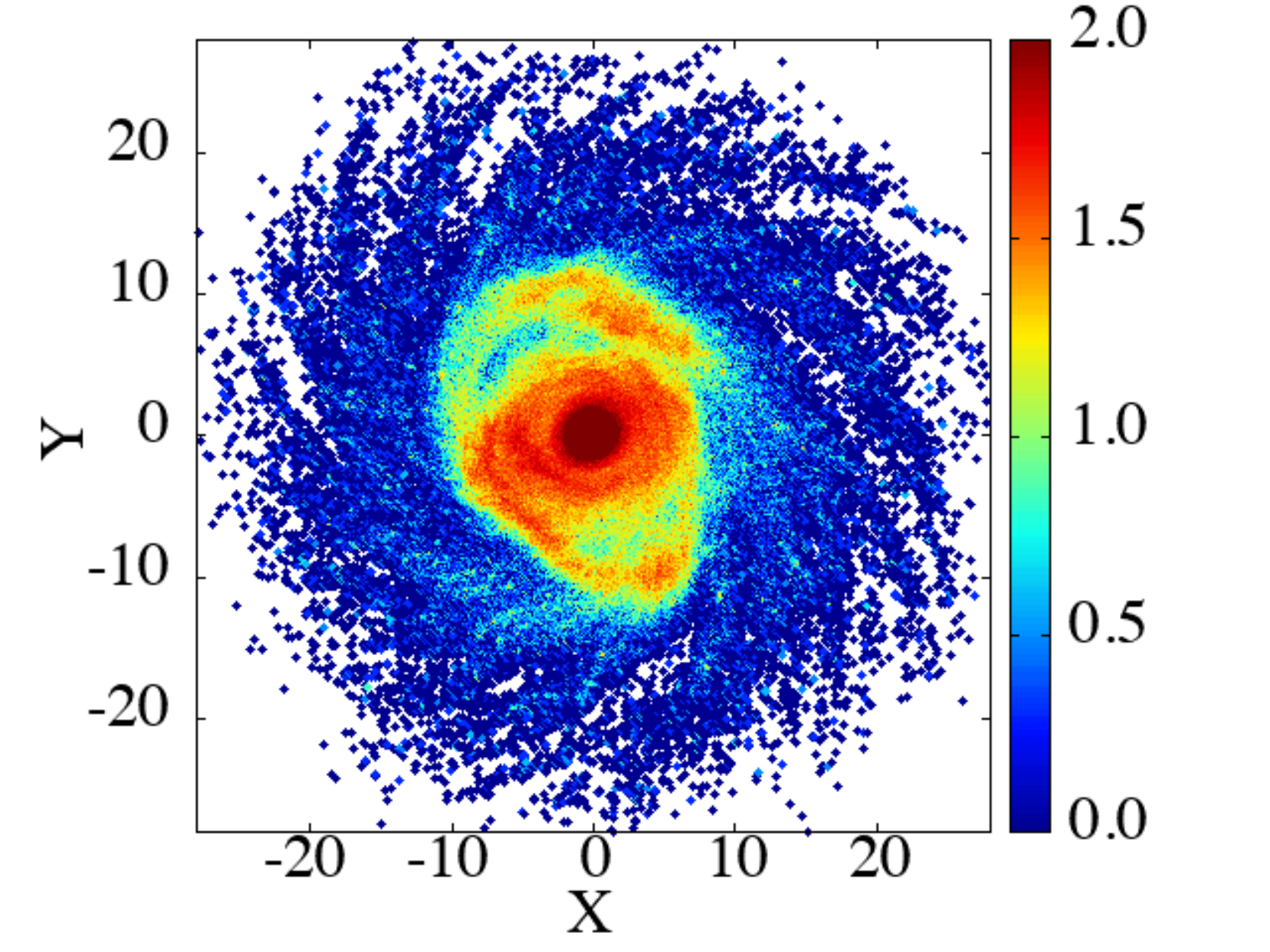}} \\  
\subfloat{\includegraphics[width =1.3in,height=1.1in]{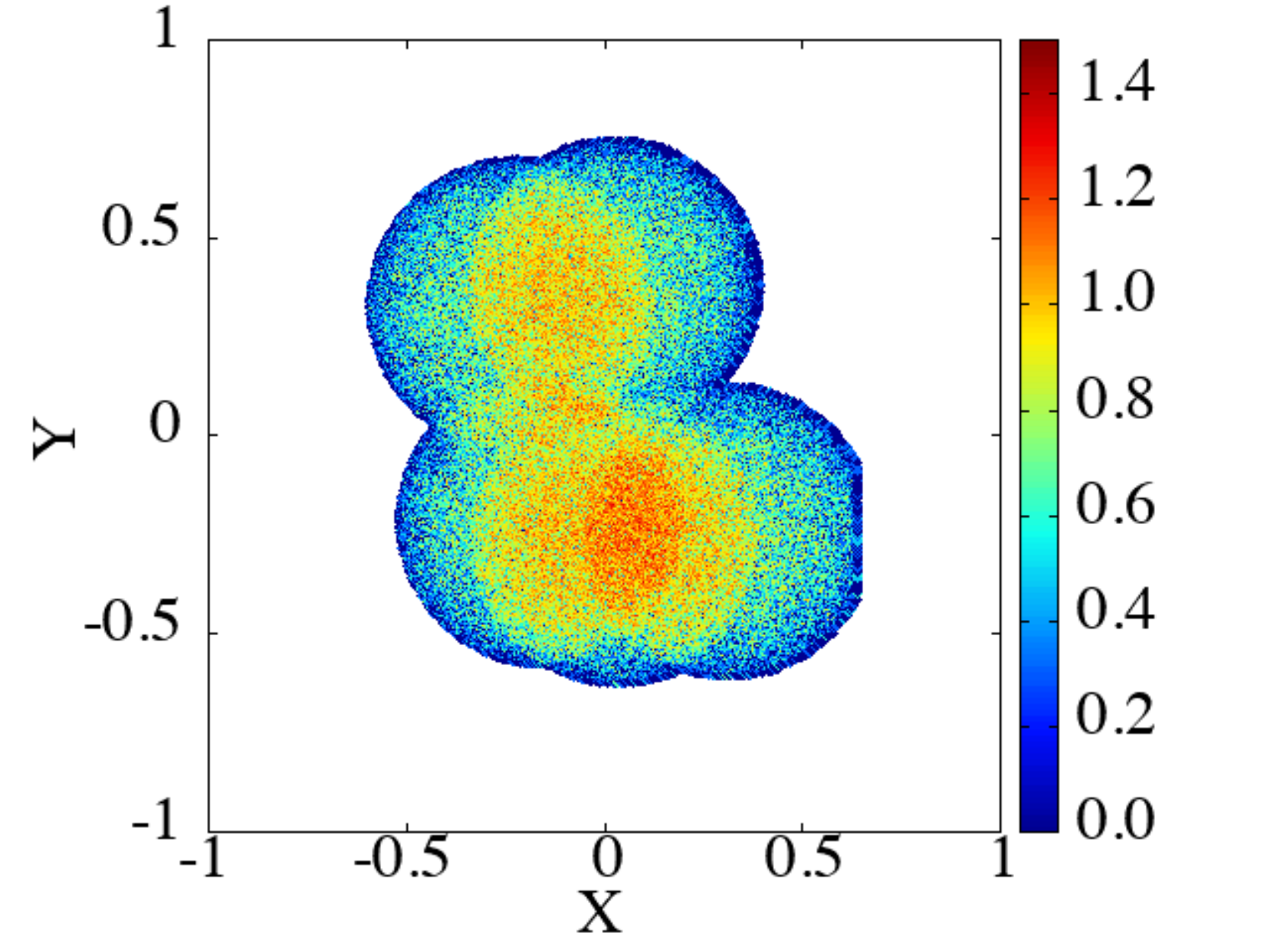}}  
\subfloat{\includegraphics[width =1.3in,height=1.1in]{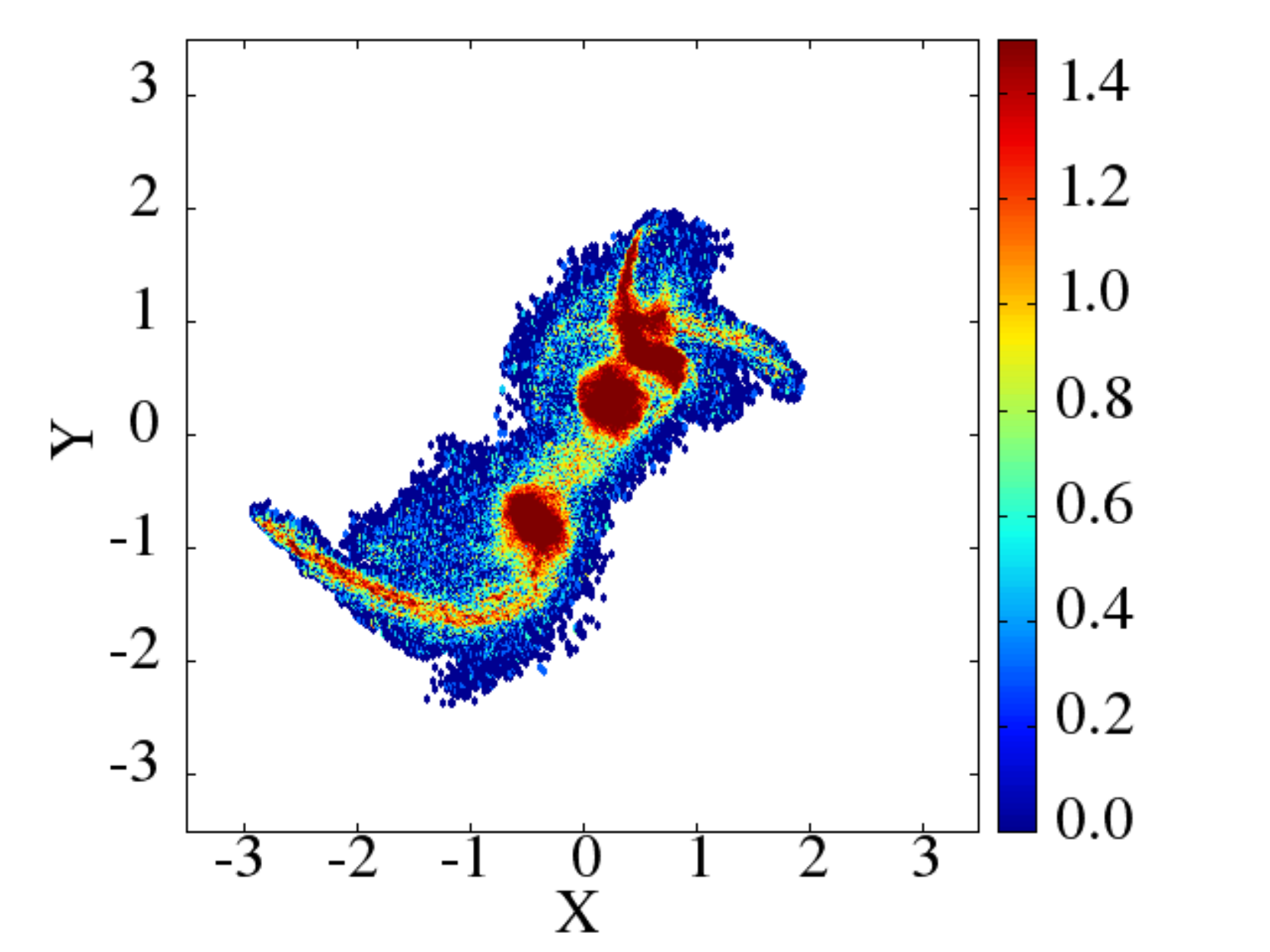}}  
\subfloat{\includegraphics[width =1.3in,height=1.1in]{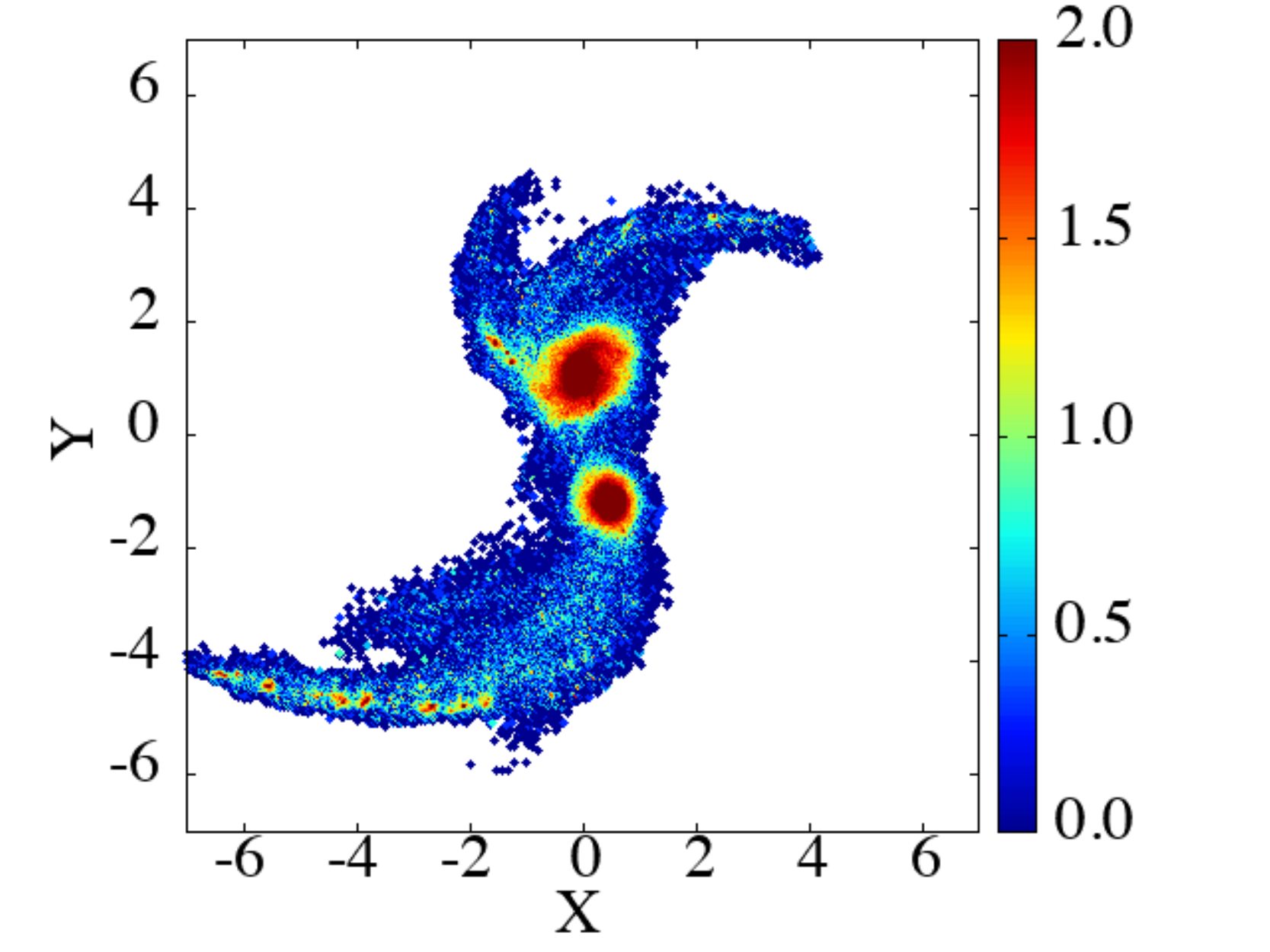}}  
\subfloat{\includegraphics[width =1.3in,height=1.1in]{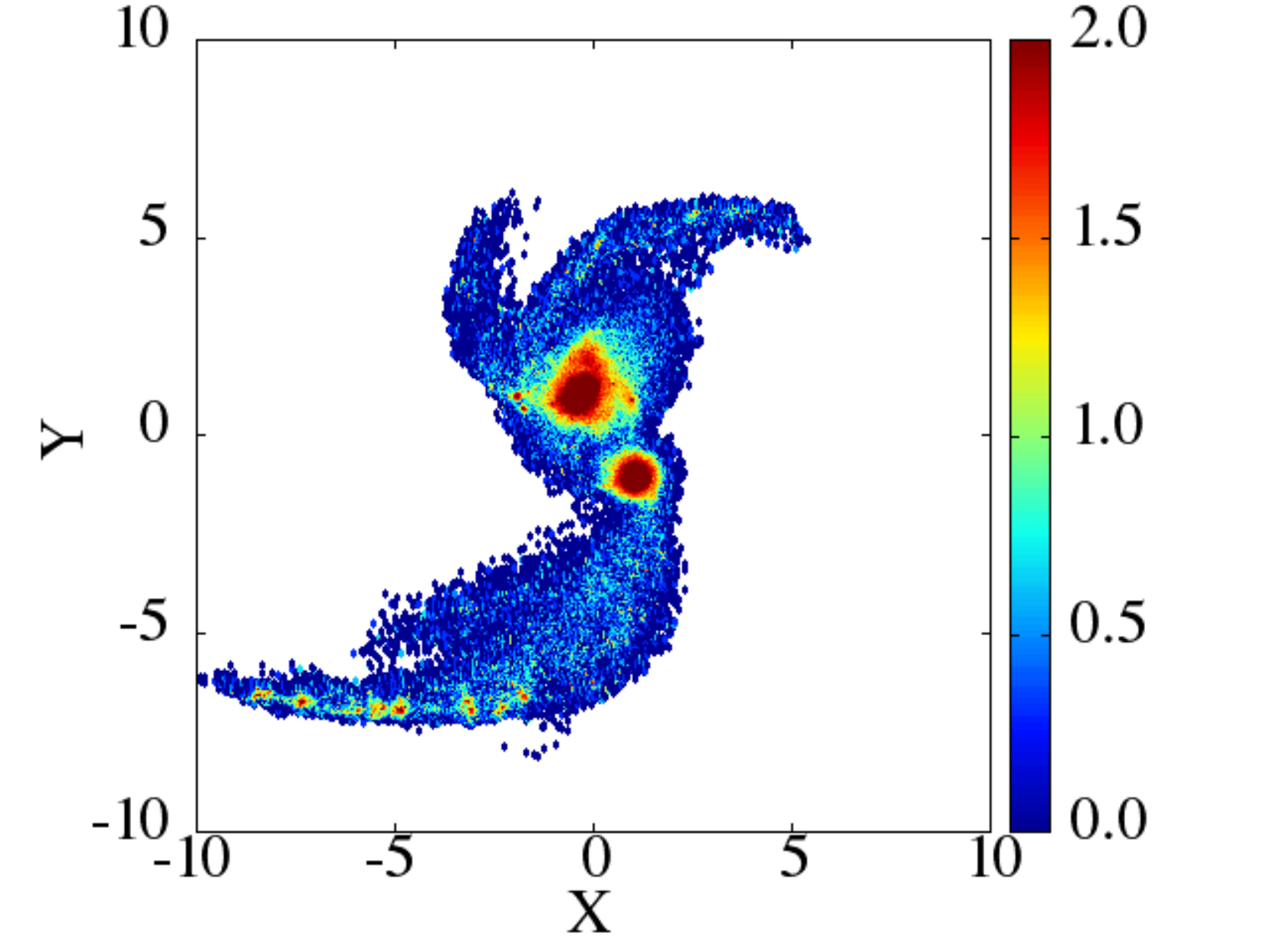}}  
\subfloat{\includegraphics[width =1.3in,height=1.1in]{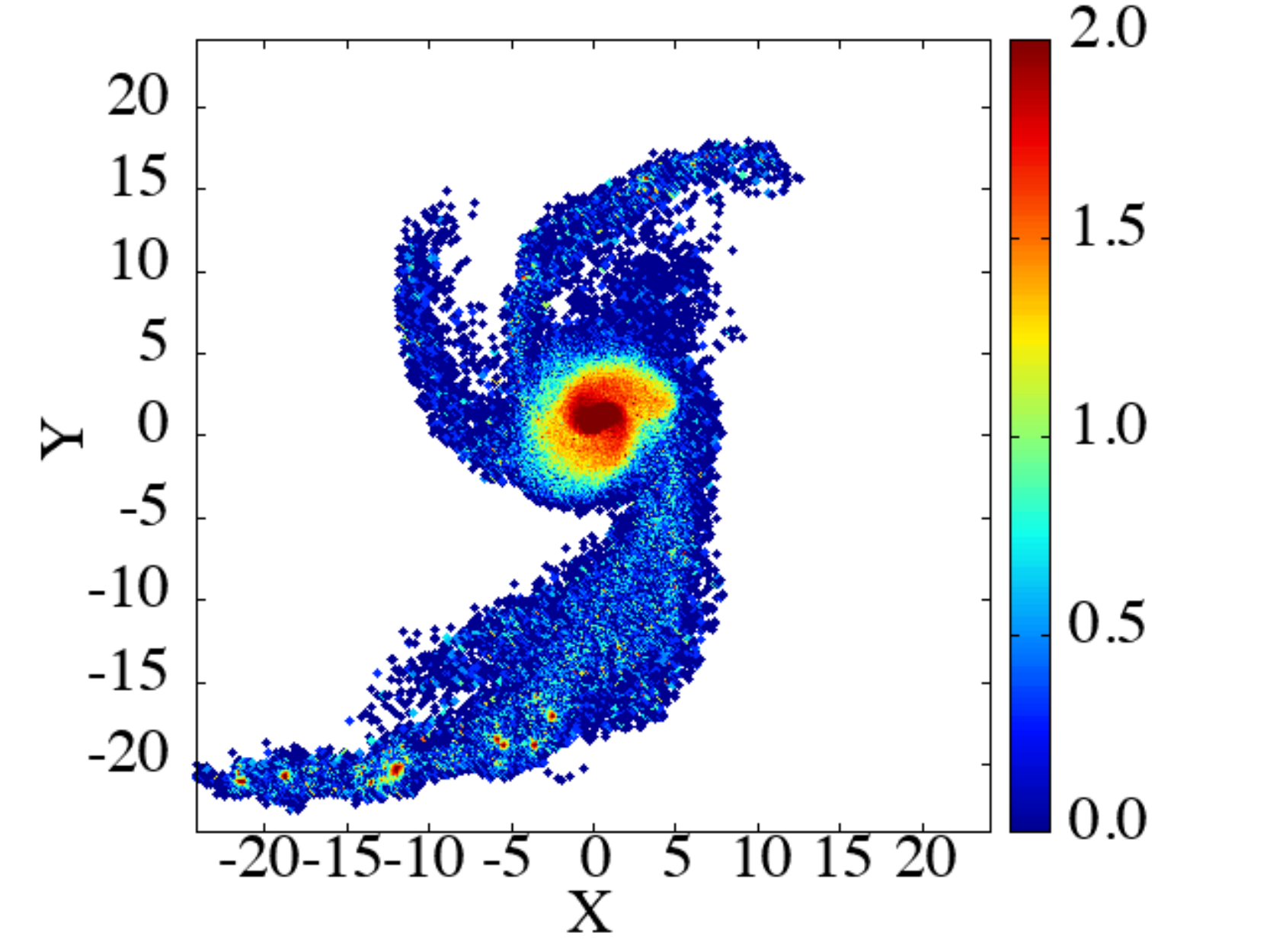}}\\  
\subfloat{\includegraphics[width =1.3in,height=1.1in]{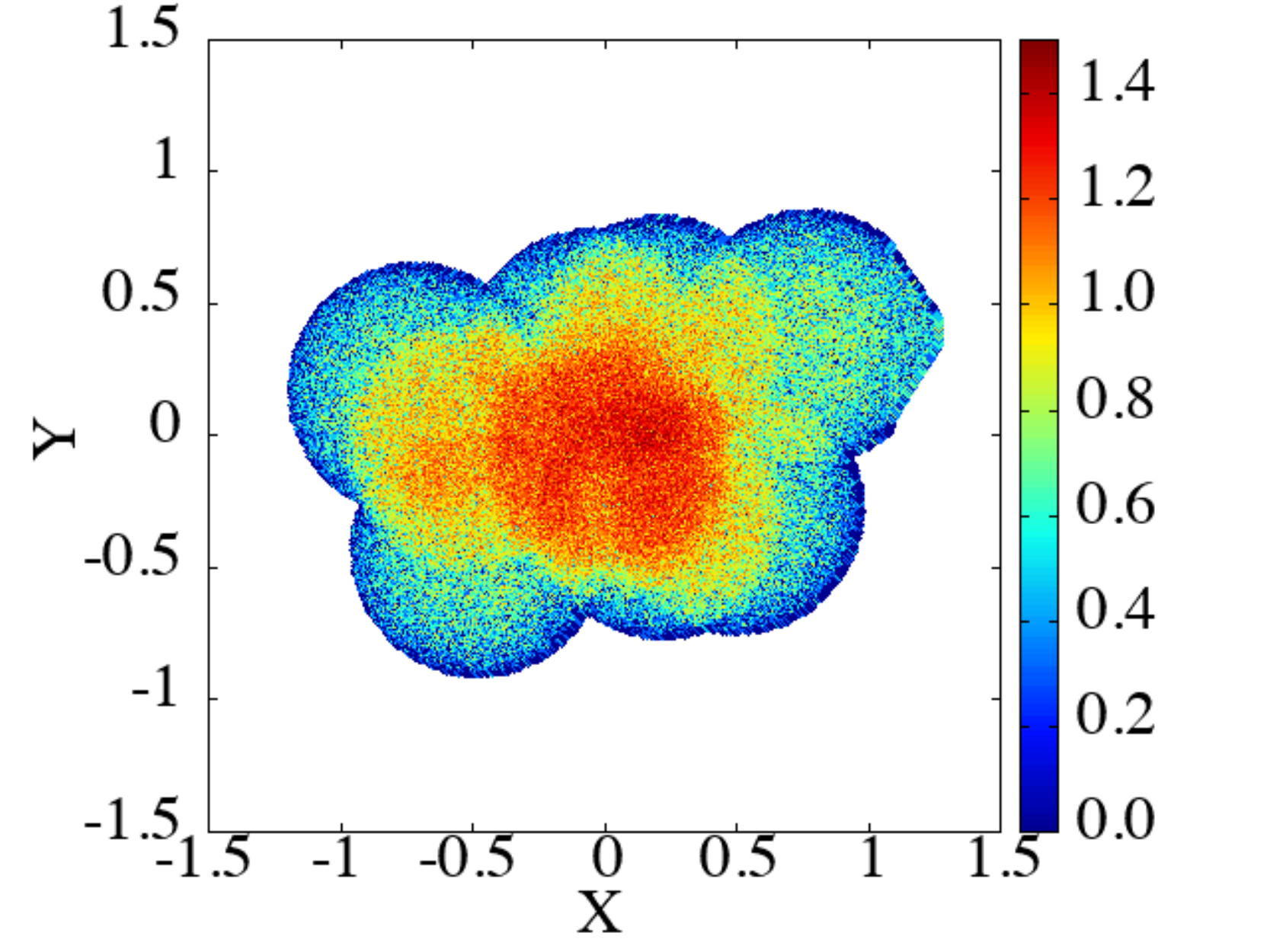}}  
\subfloat{\includegraphics[width =1.3in,height=1.1in]{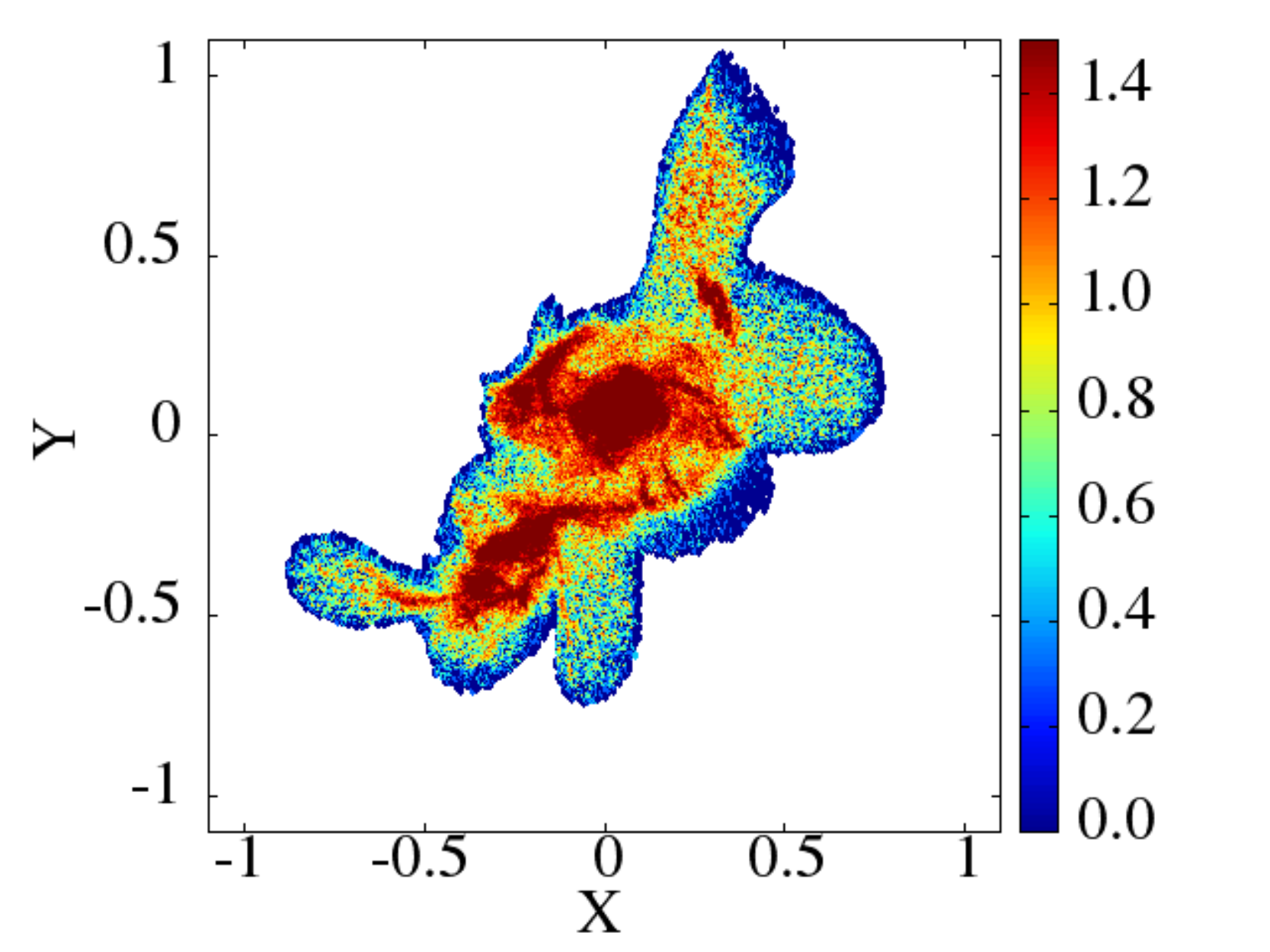}}  
\subfloat{\includegraphics[width =1.3in,height=1.1in]{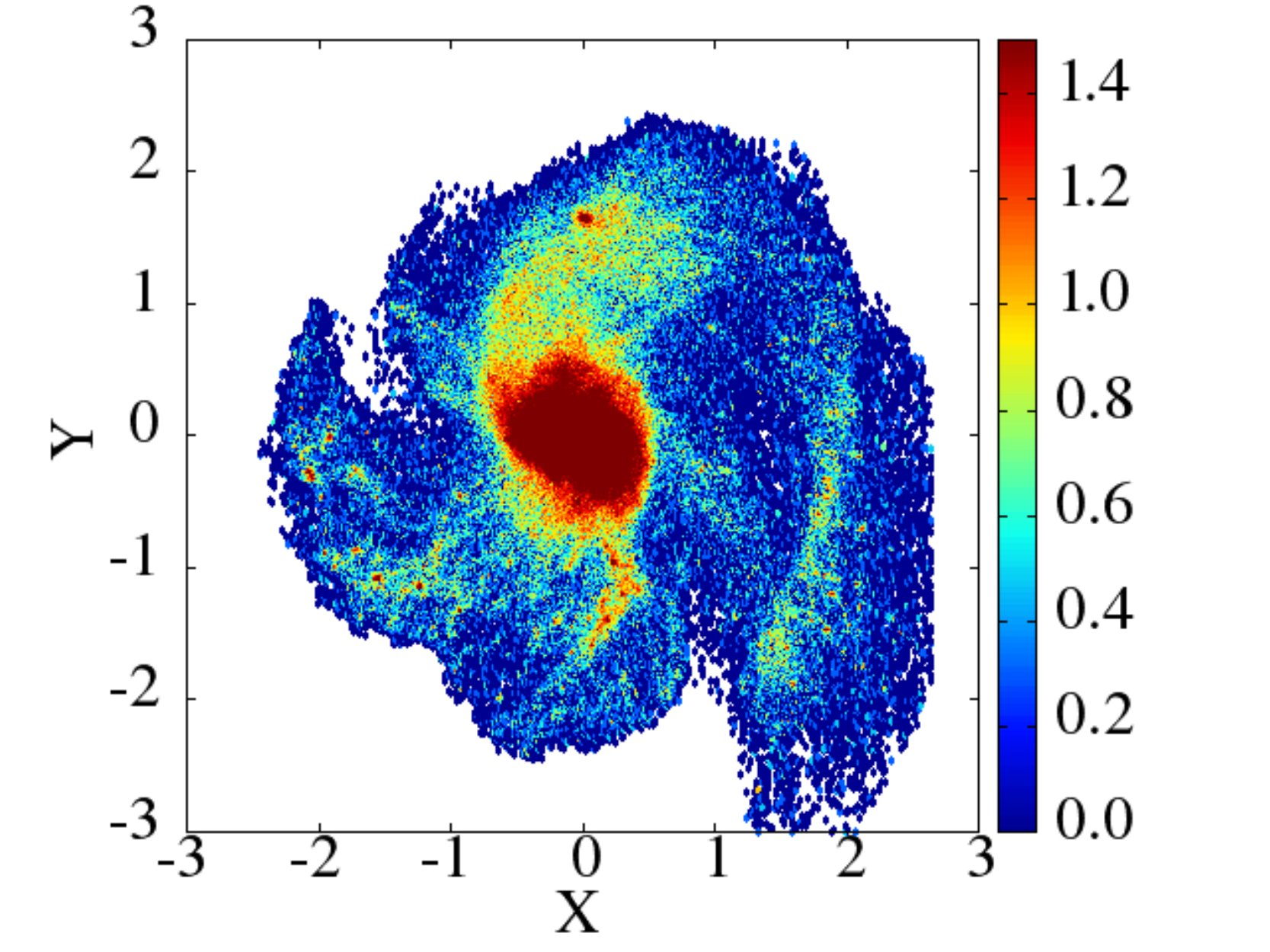}}  
\subfloat{\includegraphics[width =1.3in,height=1.1in]{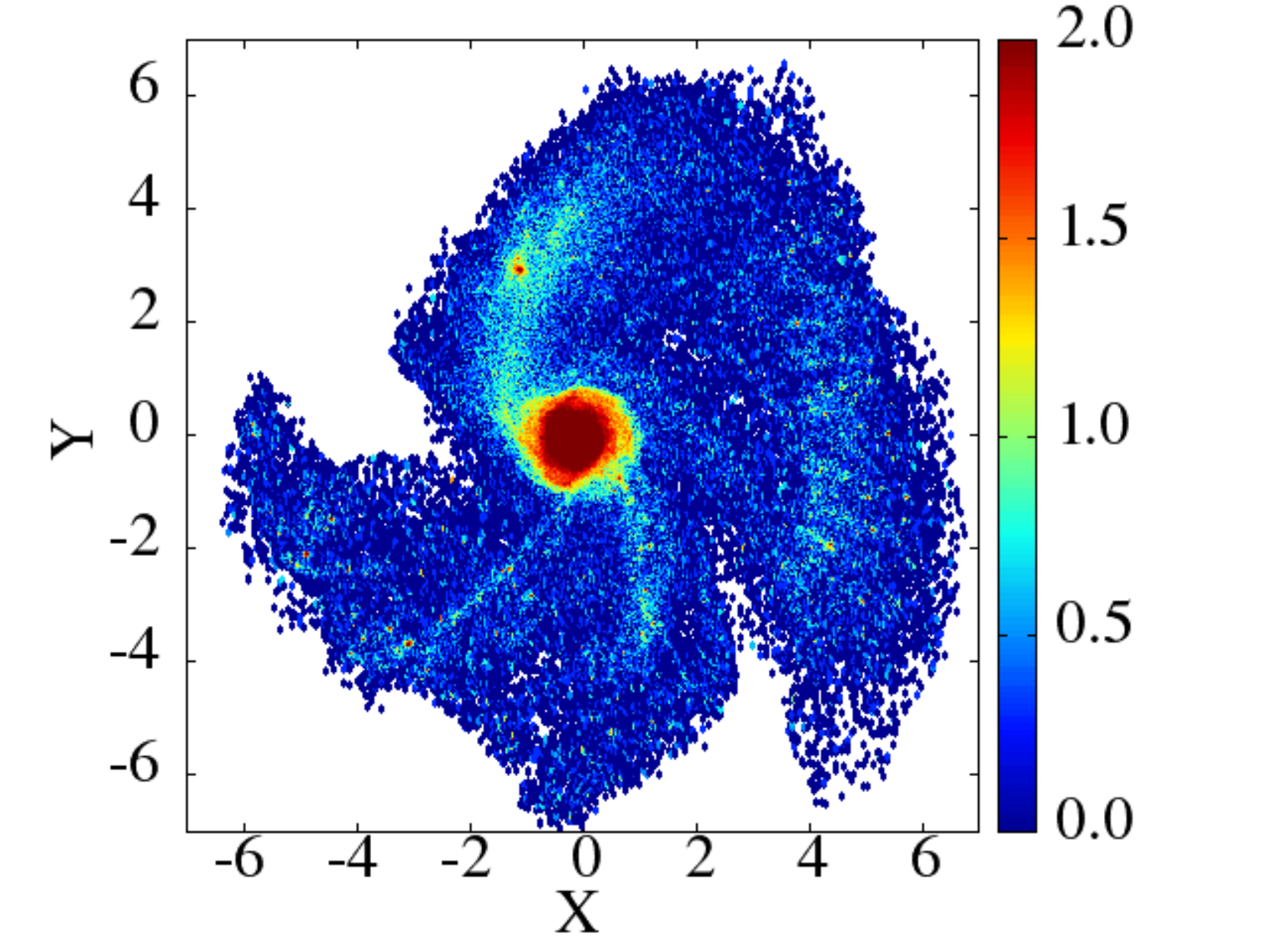}}  
\subfloat{\includegraphics[width =1.3in,height=1.1in]{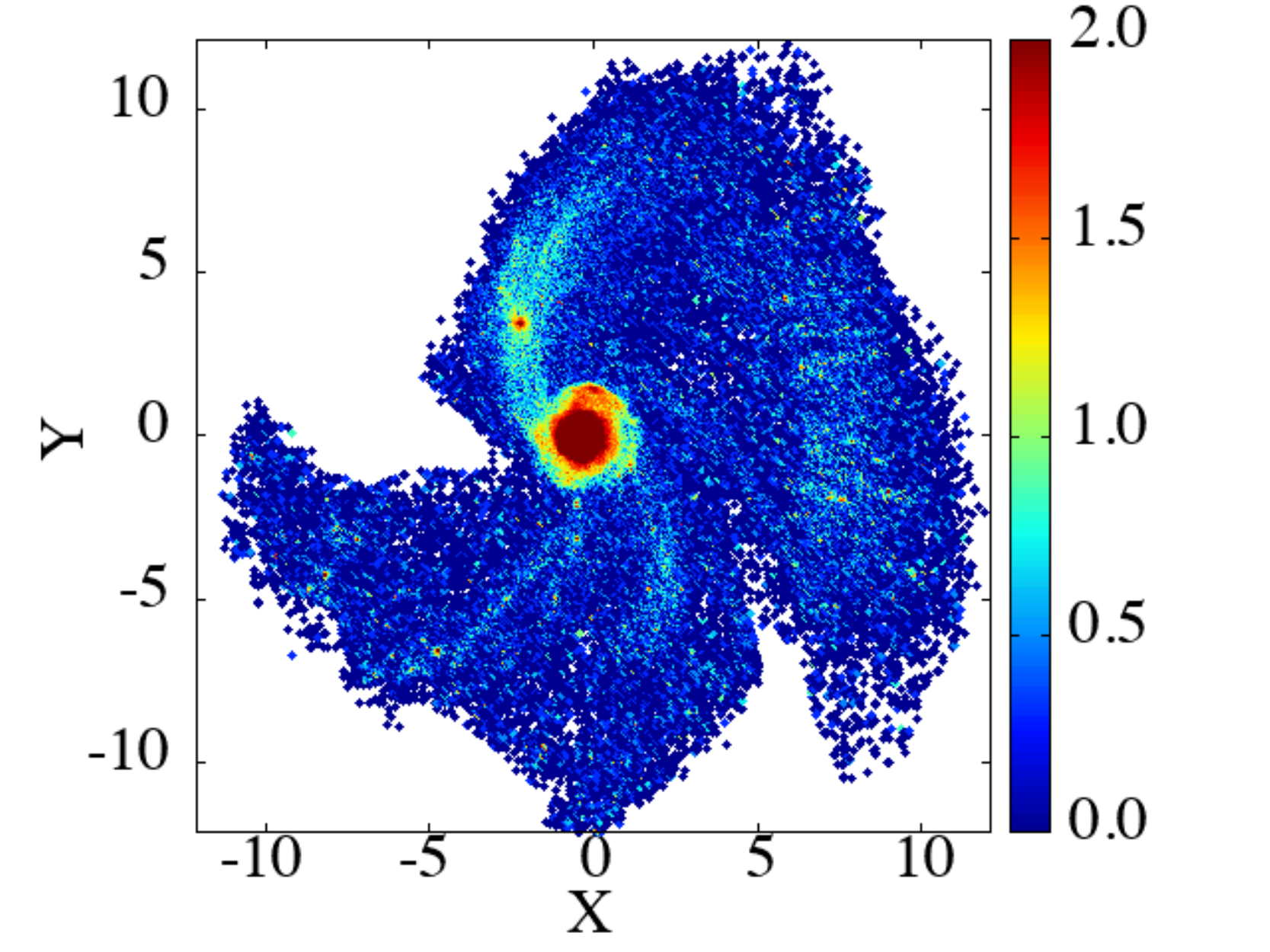}} 
\par\centering }
\caption{Density map of several snapshots of the different simulations
  at the indicated times. { Rows from top to bottom: A1, A2, B, B1a, B2, C1, C2.
  Columns from left to right: $t=0$, $t =2 \tau_{\rm max}$, 
  $t=5 \tau_{\rm max} $, $t=10 \tau_{\rm max}$, $t=20 \tau_{\rm max}$ where
  $\tau_{\rm max} $ is the time, determined approximately in each numerical simulation,
  at which the the gravitational potential reaches its maximum value, which corresponds to
  an estimate of when the system reaches its greatest contraction. The color code 
  is logarithmic in the  density. } The particle positions are
  projected on the plane orthogonal to the axis of initial rotation,
  and the direction of this initial rotation is {\it
    counter-clockwise} in these plots.}
\label{XY_evol}
\end{figure*}

\subsection{Non-stationary features at longer times}

We 
consider here the non-stationary features
of the mass distribution at longer times, of which the very 
non-trivial distinctive space and velocity structure is a 
result of the initial rotation of the IC. 
In space this can be seen very evidently in the 
(linear scale) snapshots of the evolving spatial configurations
projected on the plane orthogonal to the initial axis
of rotation of the IC, shown in 
Fig.\ref{XY_evol}.  
We focus first 
on how to quantify this non-stationary evolution which
shows up the features common to all the different IC,
and then discuss subsequently the genesis of the 
large variety of different forms which are evident. 

\begin{figure}
\vspace{1cm} { \par\centering
\includegraphics[width = 3.5in,height=3in]{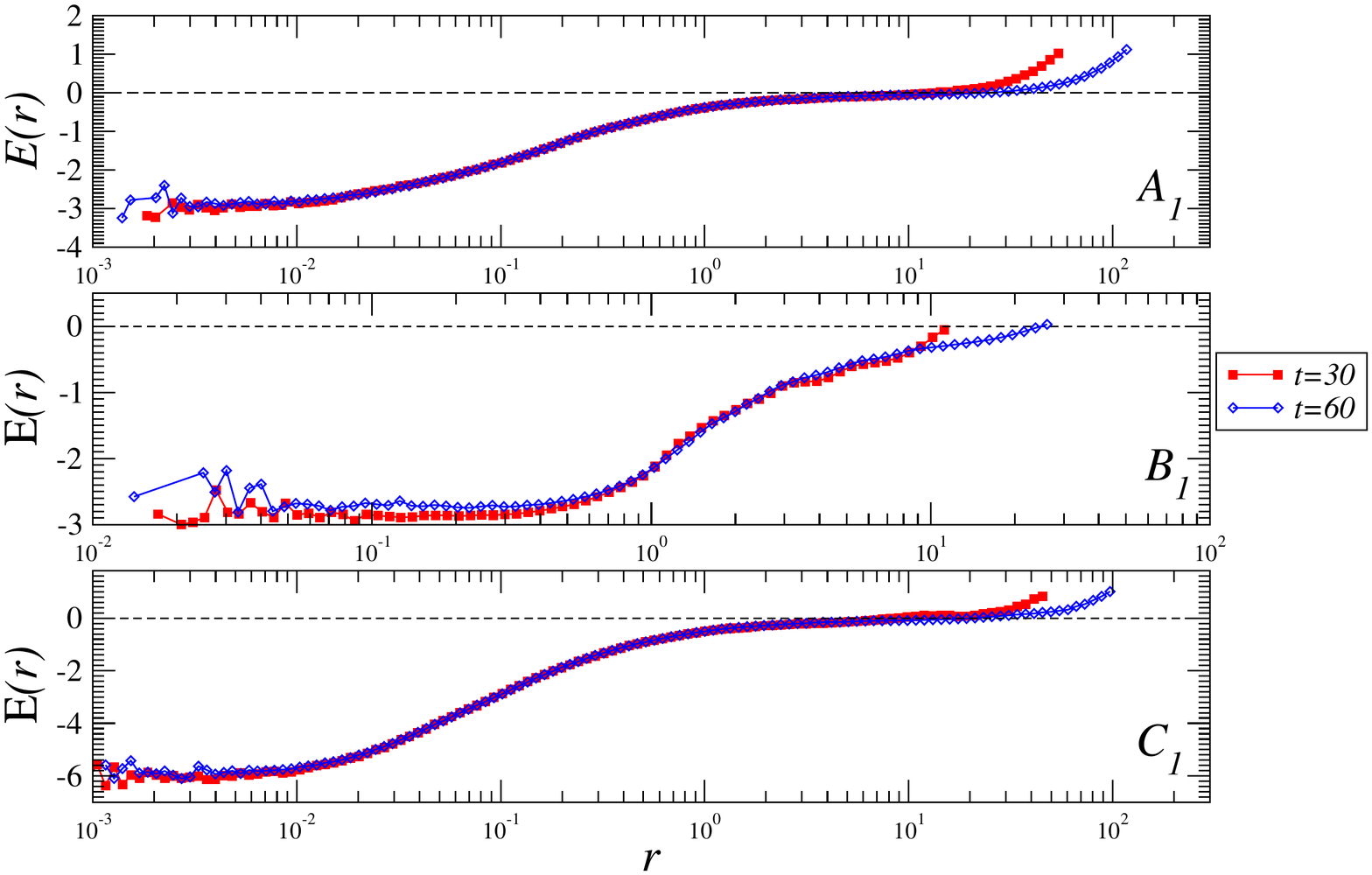} 
\par\centering }
\caption{Energy profile for the simulation A1 (upper panel), B1 (middle panel),
  and C1 (bottom panel) at different times (see labels). }
\label{er_A1B1C1}
\end{figure}

Fig. \ref{er_A1B1C1} shows the particle energies averaged in shells as
a function of radius, for the same three simulations as in
Figs. \ref{Rgt}-\ref{nr_A1B1C1}, and for two different times.  In all
cases, the non-stationarity is now clearly visible in the propagation to 
larger radii of the outermost mass, both loosely bound and unbound.  
This non-stationarity is essentially a consequence of the fact that the 
mass in this energy range has a
characteristic time for virialization with the rest of the mass which
diverges as $E  \rightarrow 0$ from below: to a first
approximation the potential they move is central and stationary, with
an associated Keplerian period $\tau \sim 1/(-E)^{3/2}$.

Fig. \ref{er_A1B1C1} shows clearly that the structure can be divided
in an inner stationary part and an outer non-stationary part.
Examination of the properties in velocity space, illustrated 
in  Figs.\ref{A1_vel} -\ref{C1_vel},  show that a further
refinement into three distinct regions is warranted:

\begin{figure} 
\subfloat[] {\includegraphics[width = 3.2in,height=2in]{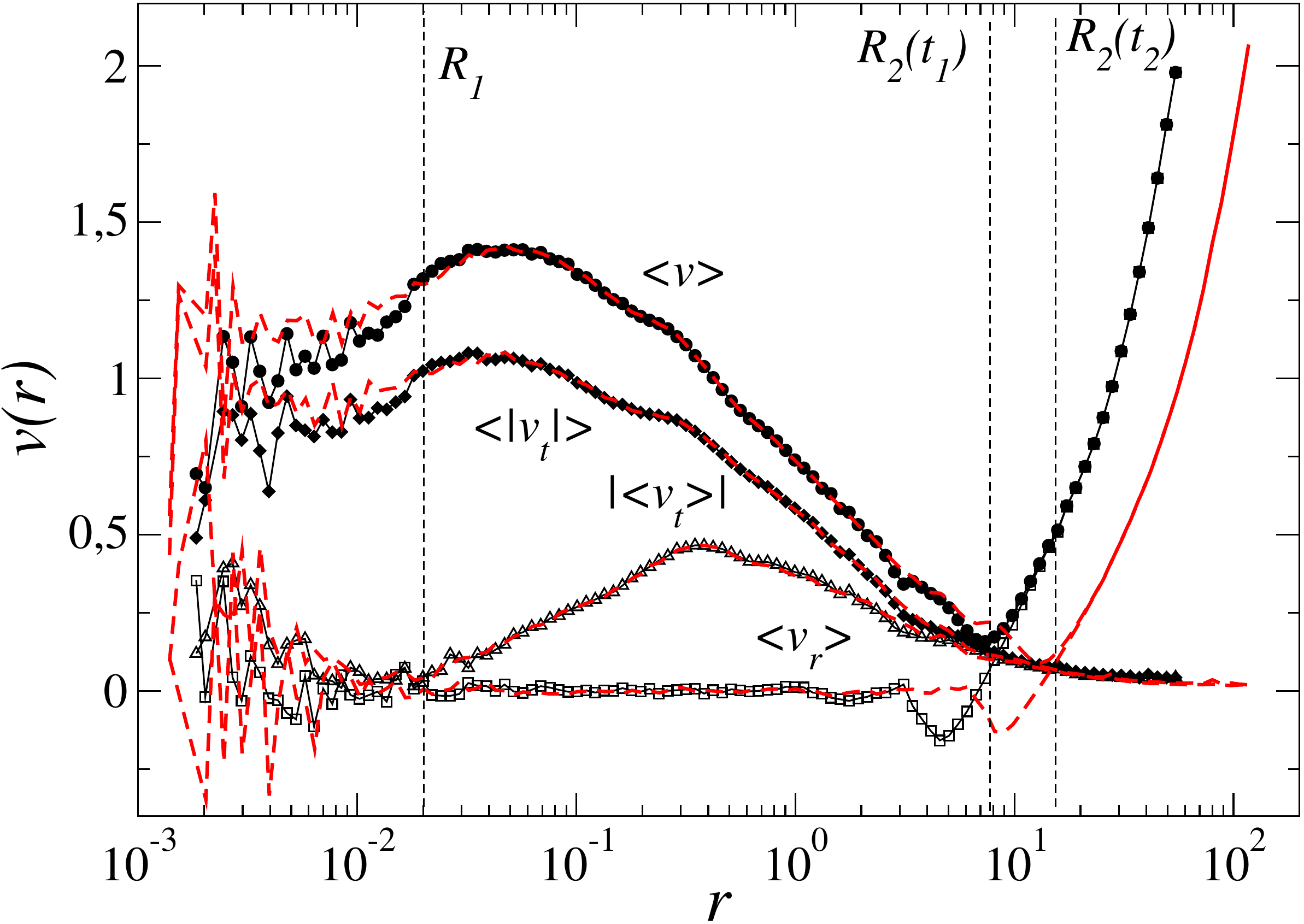}} \\ 
\subfloat[] {\includegraphics[width = 3.2in,height=2in]{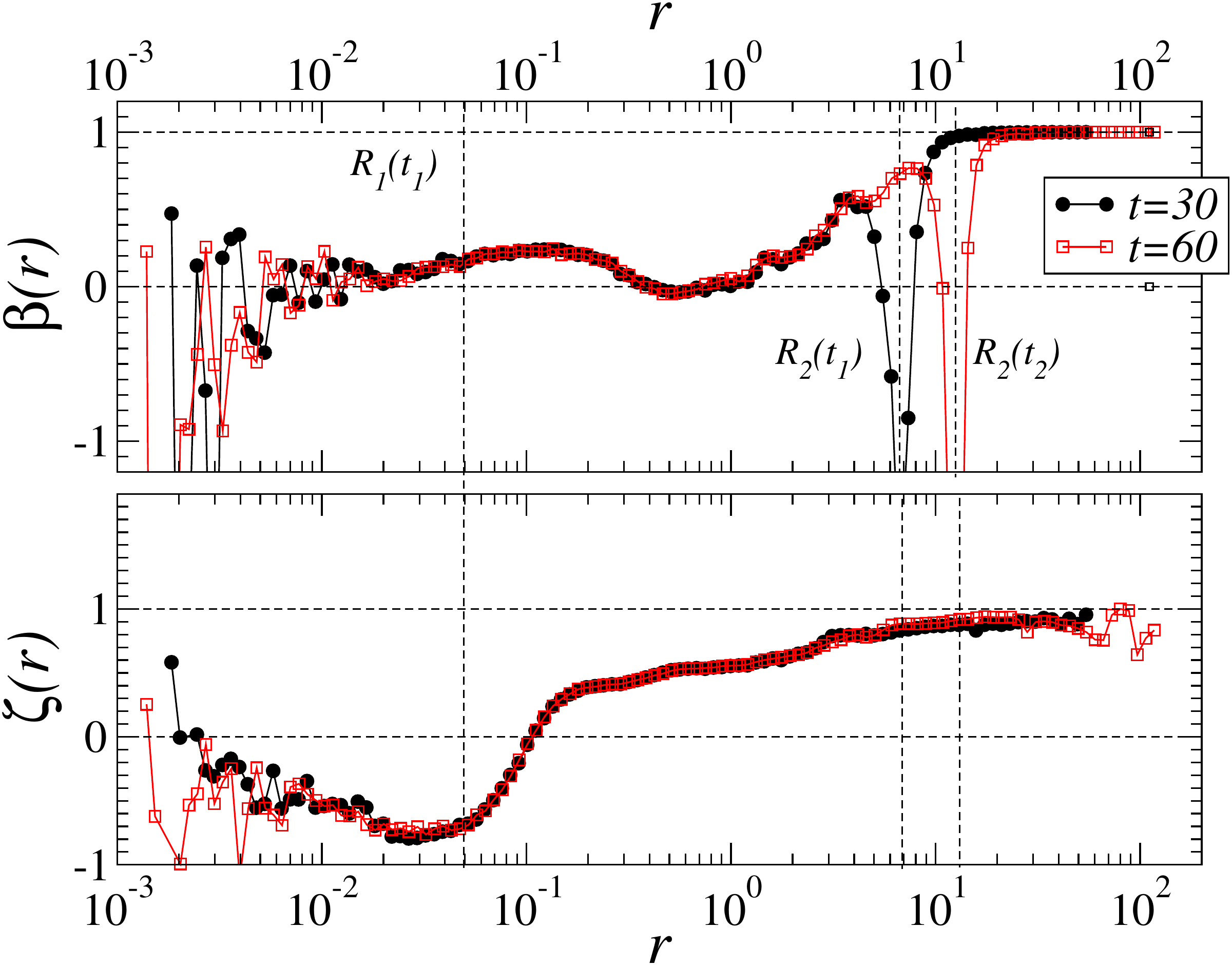}}  
\caption{Velocity field for the run A1. (a) Particles' velocity and its components averaged in radial
shells, as a function of radius, for two different times (symbols: $t_1=30$ and 
lines: $t_2=60$).
 (b) The functions $\beta(r)$ (upper panel) and $\zeta(r)$ (bottom panel) 
(see definitions in text) at two different indicated times.
 }
\label{A1_vel} 
\end{figure}
\begin{figure} 
\subfloat[] {\includegraphics[width = 3.2in,height=2in]{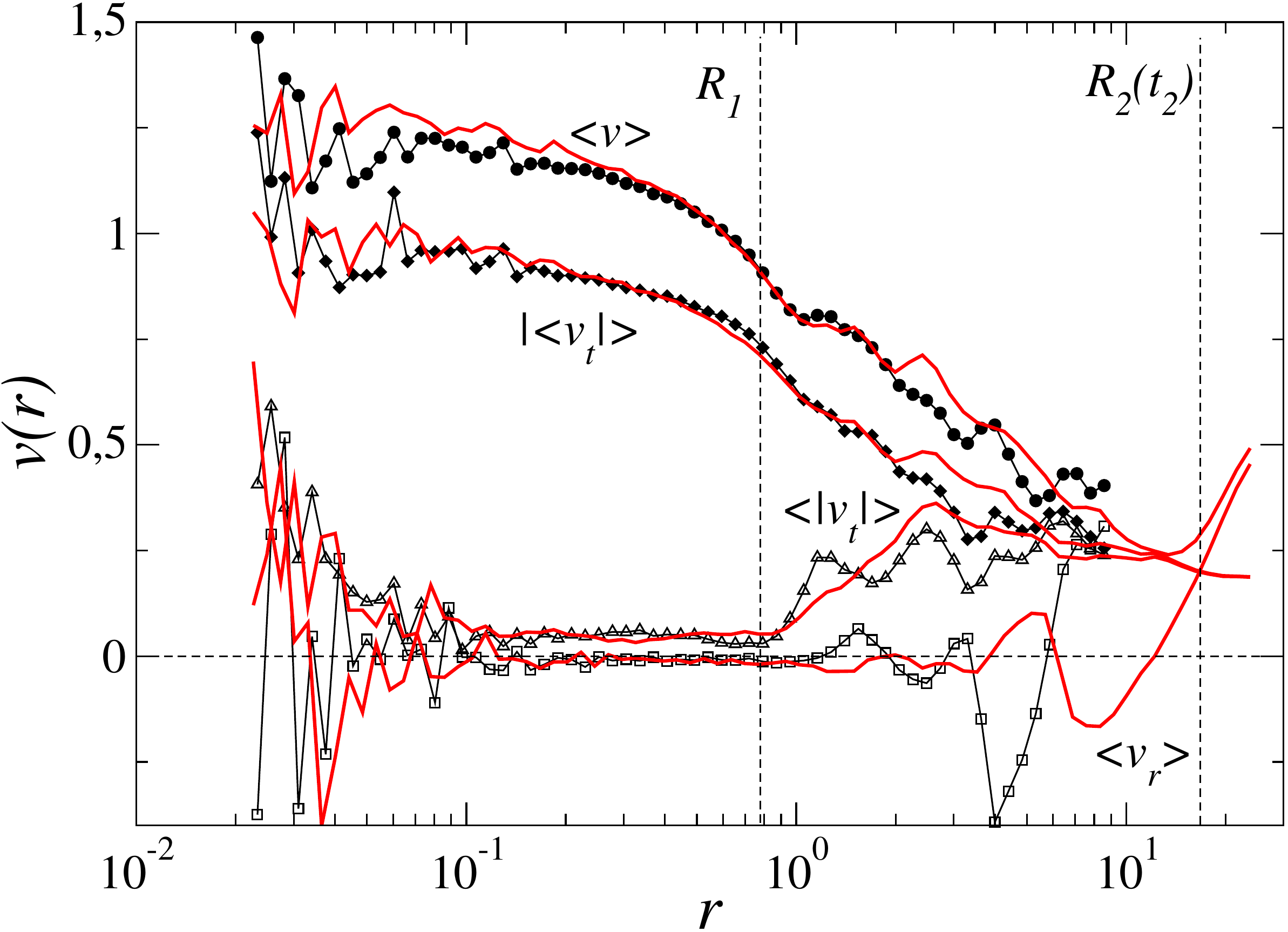}} \\ 
\subfloat[] {\includegraphics[width = 3.2in,height=2in]{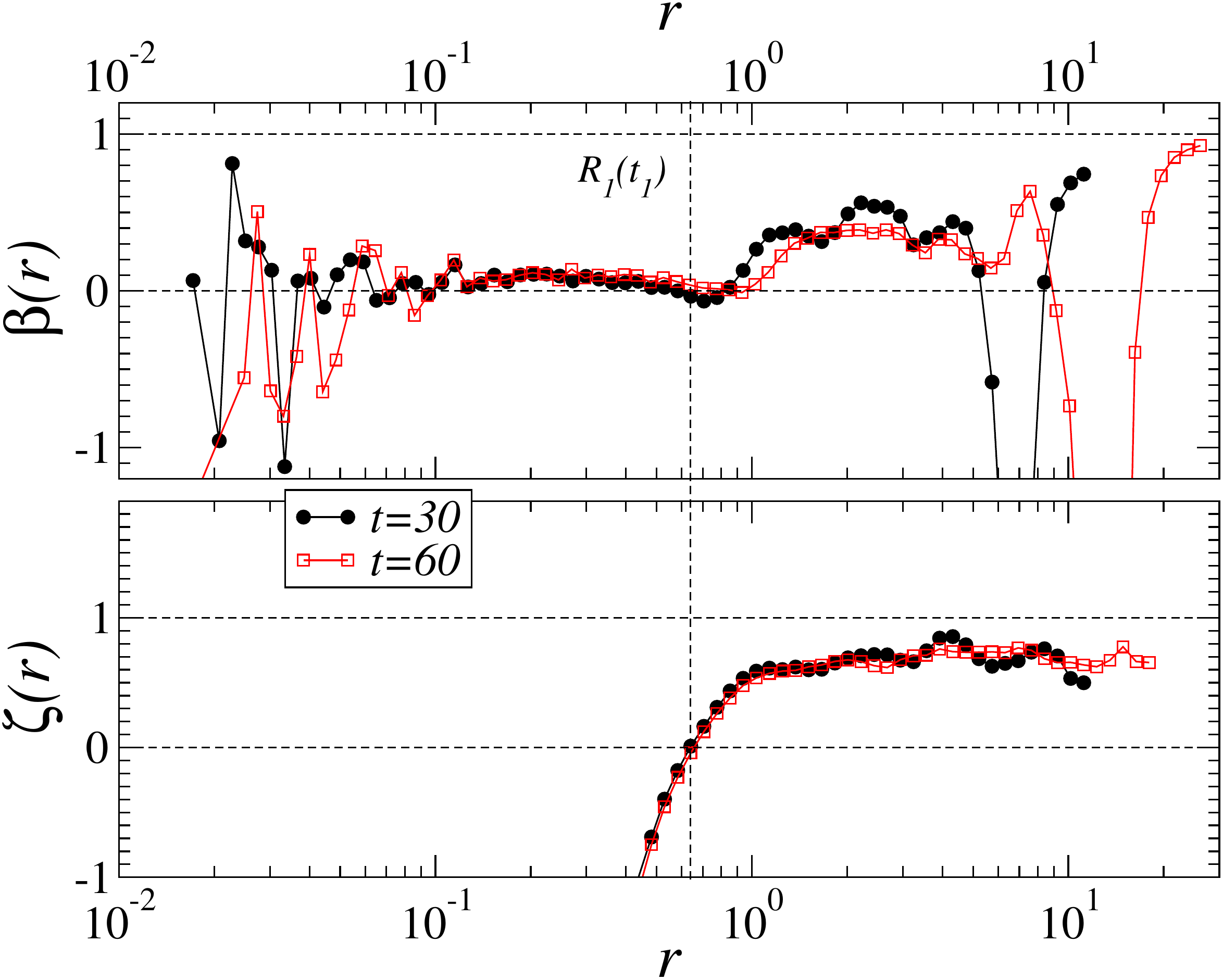}}  
\caption{As Fig.\ref{A1_vel} but for the run B1.} 
\label{B1_vel} 
\end{figure}

\begin{figure} 
\subfloat[] {\includegraphics[width = 3.2in,height=2in]{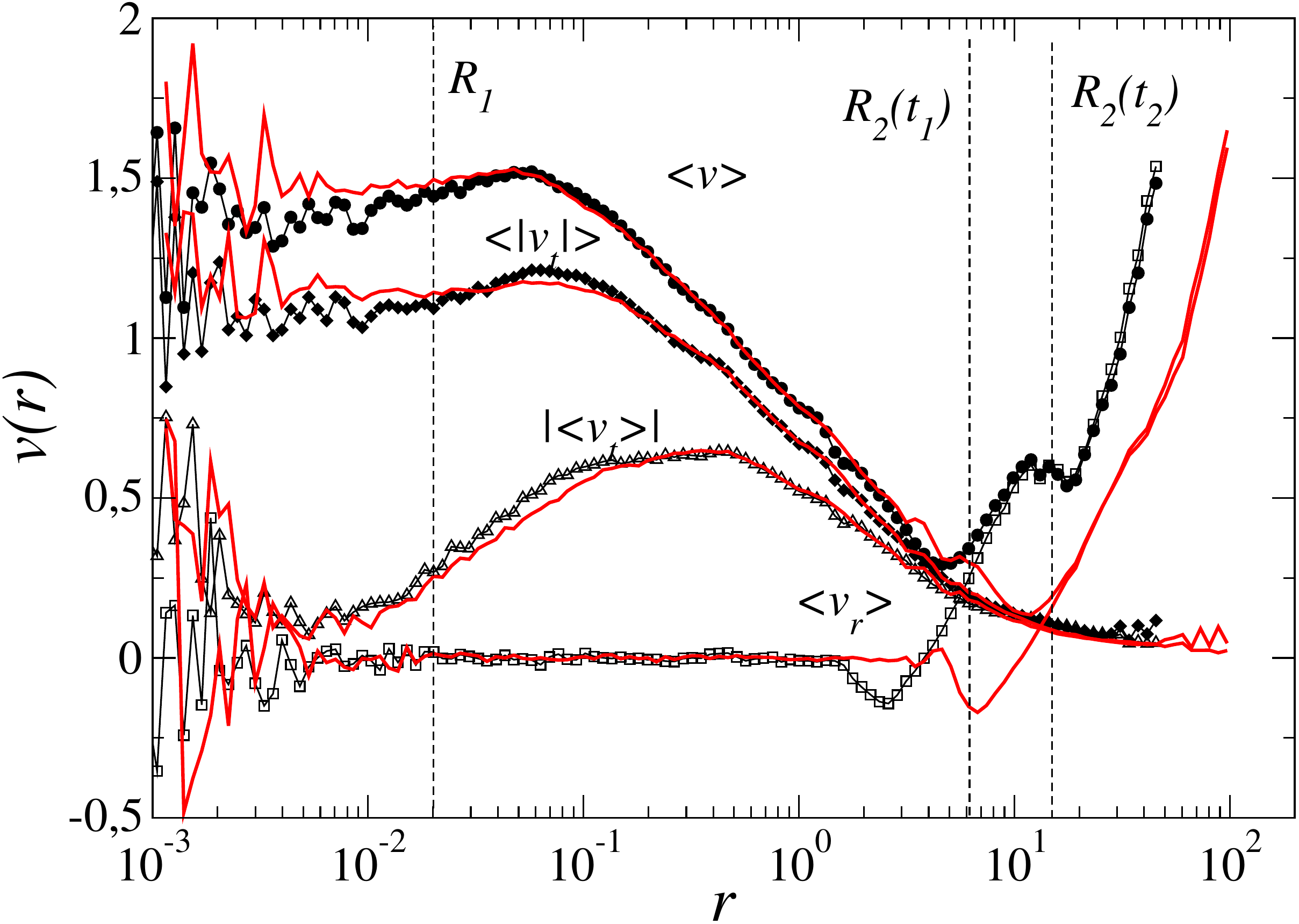}} \\ 
\subfloat[C1] {\includegraphics[width = 3.2in,height=2in]{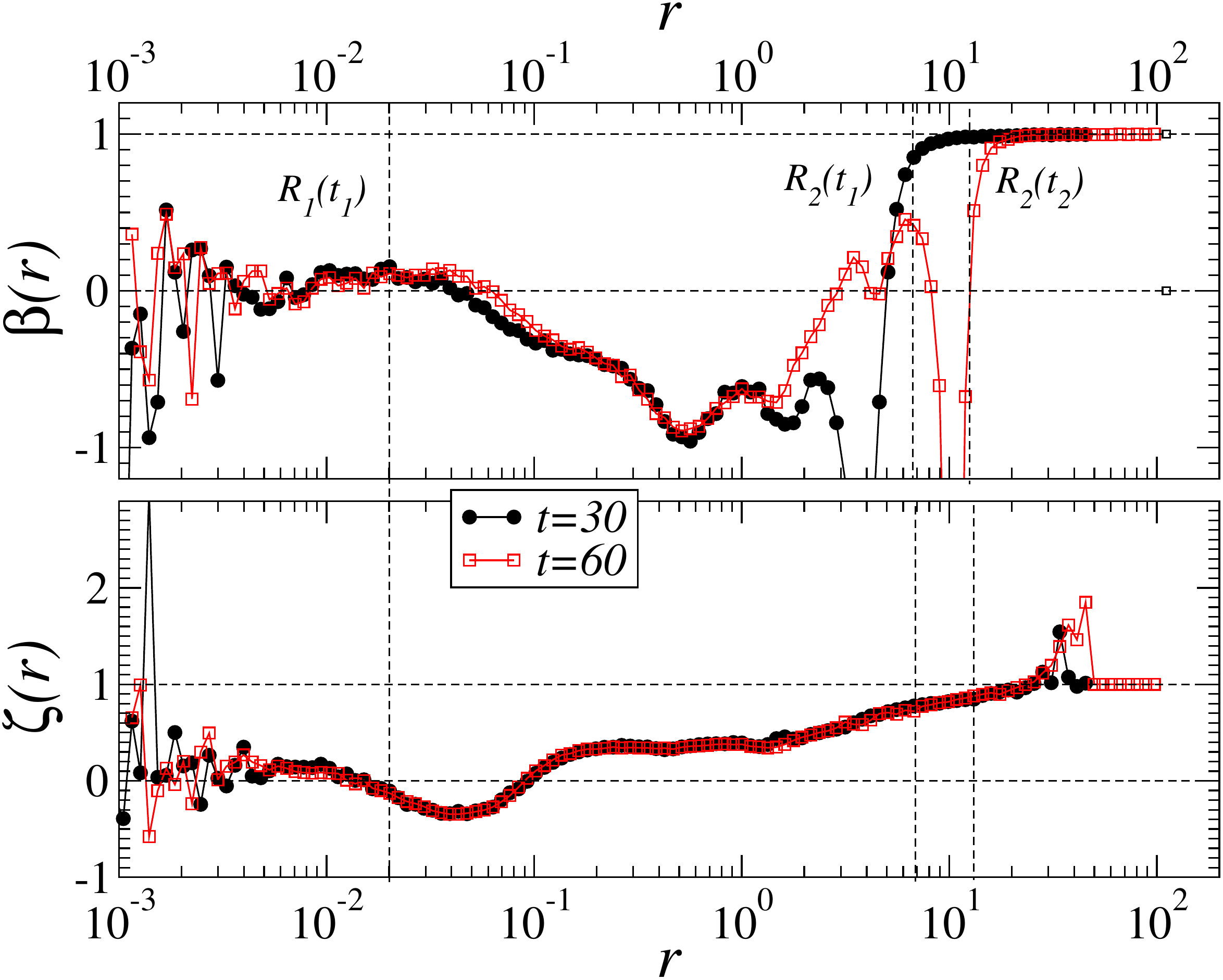}}  
\caption{As Fig.\ref{A1_vel} but for the run C1.} 
\label{C1_vel} 
\end{figure}

\begin{itemize}
\item For $r<R_1$, $\beta(r) \approx 0$, corresponding to an isotropic velocity
dispersion, there is neither net radial flow nor net rotation
(i.e. both $\langle v_r \rangle \approx 0$ and $|\langle \vec{v}_t \rangle| \approx 0$). 
This part of the distribution is the virialized core showing  the approximately flat density 
profile discussed above.

\item For $R_1 \le r \le R_2$, there is no net radial flow
(i.e.,  $\langle v_r \rangle \approx 0$) but there is a
significant net rotation. Indeed,  $|\langle \vec{v}_t \rangle|$
grows monotonically as a function of radius until it
reaches a value where it is comparable to 
$\langle |\vec{v}_t | \rangle$, and this remains so 
as both quantities slowly decline over the
radii up to $R_2$.  
This latter scale is defined such that $\langle v_r \rangle > |\langle \vec{v}_t \rangle|$
for $r> R_2$. 
Thus the rotational motion grows until it is close to a 
completely coherent one around a 
single axis. Correspondingly $\zeta$ is much smaller than unity as $\dot{v}_r$  
is small compared to $a_c$: this region corresponds to the flattened part of the distribution
where rotational motions dominate.
\item For $r>R_2$, net outward radial motion dominates (i.e.  $
  \langle v_r \rangle > \langle |\vec{v}_t | \rangle \approx 0$,
  $\beta \rightarrow 1$, $\zeta \rightarrow 1$ {(as $a_r \approx
    \dot{v}_r$)} and the sub-dominant transversal component of the
  velocity decays monotonically towards zero.  We note that in B1 
  this region is negligible as there  is almost no ejected mass.
 \end{itemize}

The scale $R_1$ does not evolve significantly with time, corresponding
to the stationarity of the region inside it. The scale $R_2$, which corresponds
approximately to the transition from bound to unbound mass, increases 
monotonically  with time.  Indeed, the mass distribution in both the outer 
part of the central  region, and in the entire outer region, is manifestly 
non-stationary on these  long time scales, and remains so for arbitrarily long times. 

{  Comparison of Fig.\ref{XY_evol} with Figs.\ref{A1_vel}-\ref{C1_vel}
shows a rather interesting property of this class of models: 
the more the shape of the structure formed after the collapse deviates from
axisymmetry, the larger is the radial velocity at large distance from its center.
In addition we stress that Figs.\ref{A1_vel}-\ref{C1_vel} show the 
average components of the velocity {  in} spherical shells, while the 
actual velocity fields are characterized by large anisotropies: most notably,
the amplitude of the radial velocity is correlated with the semi-major axis. 
These features, which will be studied in greater detail in a forthcoming work, 
must be considered when comparing  results of this class of simulations 
with observations (see Sect.\ref{conclusions}).}


\subsection{Emergence and evolution of spiral structure} 
 
Detailed study of the temporal evolution of the configurations
confirms that the mechanism for generation of the spiral-like
structure evident in Fig.\ref{XY_evol} is indeed the strong injection
of energy given to particles which pass through the center of the
system just after the time of the maximal compression.  This gives
these particles a significant radial component of their motion added
to the initial rotational motion.  The radial distance these particles
subsequently travel, once they are outside the core, because of
approximate conservation of angular momentum, is correlated with the
angle they move through: particles which have larger radial velocities
initially thus ``trail" behind particles with smaller radial
velocities.  For this process to generate a spiral-like structure, the
only necessary additional ingredient is that the distribution of the
directions of motion of these particles is anisotropic.  This is
indeed the case starting from these IC.  As we have discussed above,
as for cold non-rotating IC, the particles which gain energy are those
which lie furthest from the origin initially. In the case of an
ellipsoid, the radial motion arising from the energy injection is thus
preferentially correlated with the longest semi-principal axis.
Asymptotically, the motion of the particles with positive energy,
which are furthest out, becomes purely ballistic and radial.  As a
result the spiral-like structure is ``frozen" and stretches more and
more. Nevertheless this intrinsic non-axisymmetry of the disc is
typically not yet so marked even at the quite long times we show in
our plots.


We consider now in more detail the different forms arising from the
range of IC that we have selected.  We recall that the first two
simulations, A1 and A2, in Fig. \ref{XY_evol} show the evolution of
two prolate ellipsoids (see Table \ref{table-param}).  In both cases
the collapse is quite violent, with the system undergoing a very
strong contraction. This leads to the production of a significant
fraction (about 10 \%) of positive energy particles. The collapse of
most of the mass occurs along the direction of the initially shortest
semi-principal axis, while the subsequent marked expansion of the
system is along the direction of the longest semi-principal axis.

The simulations B1 shows a very similar morphology, but the
spiral-like arms are markedly more ``wound up" than in A1 and A2.
Nevertheless the basic process leading to the formation of this
structure is essentially the same, with an anisotropic energy
injection correlated with the axis along which particles arrive
latest.  This difference in the winding in particular is a direct
result of a much less violent collapse, in which only a relatively
small fraction of the total mass participates. As a result the radial
velocities injected into particles along the longest semi-principal
axis are smaller, and the particles thus travel a smaller radial
distance as they rotate with their initial rotational motion. The
outer shells initially collapse toward the center in a contraction
which is fastest along the shortest semi-axis. This contraction
enhances the initial anisotropy of the system which also leads to a
transient bar structure which we discuss in greater detail below.

In B1a, in which 
the core is a factor ten less extended than in B1,  
the evolution is very similar, except for the morphological 
details of the bar and spiral arms. 
{ This test shows that the precise morphology of the system formed after 
the collapse depends sensitively on the features of the IC.}

The simulation B2 likewise is characterized by a very anisotropic
collapse, but as it is an oblate ellipsoid, the contraction occurs
along the direction of the shortest semi-principal axis, and
approximate symmetry about this axis is maintained. The contraction
does not change the particle energy distribution very significantly,
but is enough to give to some particles a radial velocity component
directed outwards. In this case the particles which ``arrive late"
during the collapse are those initially close to the outer shells of
this plane of symmetry, and as they re-expand outward they give rise
to a quite different spiral-like structure with ``flocculent''
multiple arms. These appear to be seeded by the growth of the density
fluctuations in the system during its collapse phase: { 
for this reason we expect that these features depend
on the amplitude of the initial density fluctuations (see discussion 
in Sect.\ref{thermolimit}). In addition, a
transient ring structure emerges, which we discuss further below.}

In B1, B1a and B2 there is, because of the much more gentle collapse,
very little or no ejected mass (for B1, see Fig. \ref{B1_vel}) and
the motions are only predominantly radial at longer times, and only at
the very largest distances. Likewise the region where the coherent
rotational motion is dominant with respect to the radial motion is
much more extended than for the case of, e.g., the simulation A1.
Differently to the other cases, we observe in B2 that there appear to
be two distinct phases in which we see quite different spiral arms
emerge.

The simulations C1 and C2 show the typical behavior we observe in
this class of more inhomogeneous and anisotropic IC. Transient
structures similar to those in the other cases again emerge, and the
same basic physical mechanism is at play. The spiral arm structure
which forms after the collapse is clearly less axisymmetric,
reflecting the lower symmetry of the IC. Both the visual appearance
and the structure in velocity space (see Fig. \ref{C1_vel}) show that
the resultant structures are more similar to A1/A2 than B1/B2. This
reflects the fact that the collapse is indeed stronger in these cases.

\subsection{Formation of bar structures}

We have noted that the structures we observe are generically
non-axisymmetric (except in specific cases like B2 where the
axisymmetry of the IC survives the collapse phase better). Thus the
configuration which results is not just very flattened along the axis
parallel to the initially shortest semi-principal axis, but a coherent
anisotropic structure resembling a bar emerges in the plane defined by
the two longer semi-principal axes in the IC. This anisotropy is, as
one would anticipate, present not only in configuration space but also
in the velocity field.

\begin{figure}
{ \par\centering
  \subfloat{\includegraphics[width = 1.8in]{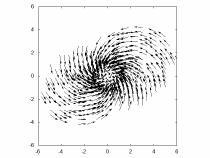}} 
  \subfloat{\includegraphics[width = 1.8in]{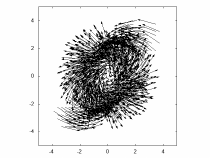}} \\ 
  \subfloat{\includegraphics[width = 1.8in]{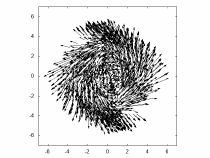}} 
  \subfloat{\includegraphics[width = 1.8in]{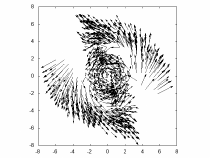}} 
  \par\centering }
  \caption{Spatial evolution of B1 (coarse grained on a grid) on the $x-y$ plane 
  of the particles that form the bar/arms structures at $t=25$.
 The times in the different panels is (from top right to bottom left): 5,10,15,20.
We have plotted  the coarse-grained  distribution (see text) with the corresponding 
velocity vector. }
  \label{D13_CG} 
\end{figure}

Fig.\ref{D13_CG}, showing several snapshots of B1 projected in the
$x-y$ plane, illustrates how these transient features form and
grow. The particle distribution has been coarse-grained onto a $32^3$
grid, and the average velocity determined in each cell.  We see that,
up $t \approx 15$ when the global collapse of the system occurs, the
velocities are essentially rotational, but progressively develop an
inward radial component, simply because of the collapse dynamics.
During the phase of maximal contraction, the particles originally
furthest out, i.e. along the initial semi-principal axis, gain a
radial velocity component directed outwards, but their total velocity
remains predominantly rotational.  The particles closer in initially,
on the other hand, have lower energy and remain more bound around the
central structure.  Thus the two arms which emerge clearly are formed
from two groups of particles which initially were located at opposite
ends of the longest principal semi-axis.  For longer time scales,
i.e. $t>50$, the spiral arms and the bar structure start to be washed
out by the radial component of the motion.

\subsection{Formation of ring structures}

\begin{figure}
\vspace{1cm} { \par\centering
\subfloat{\includegraphics[width = 3.8in, height= 4in]{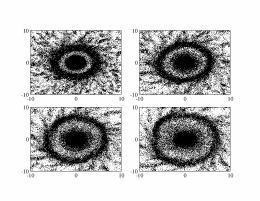}} \\
\subfloat{\includegraphics[width = 3.5in]{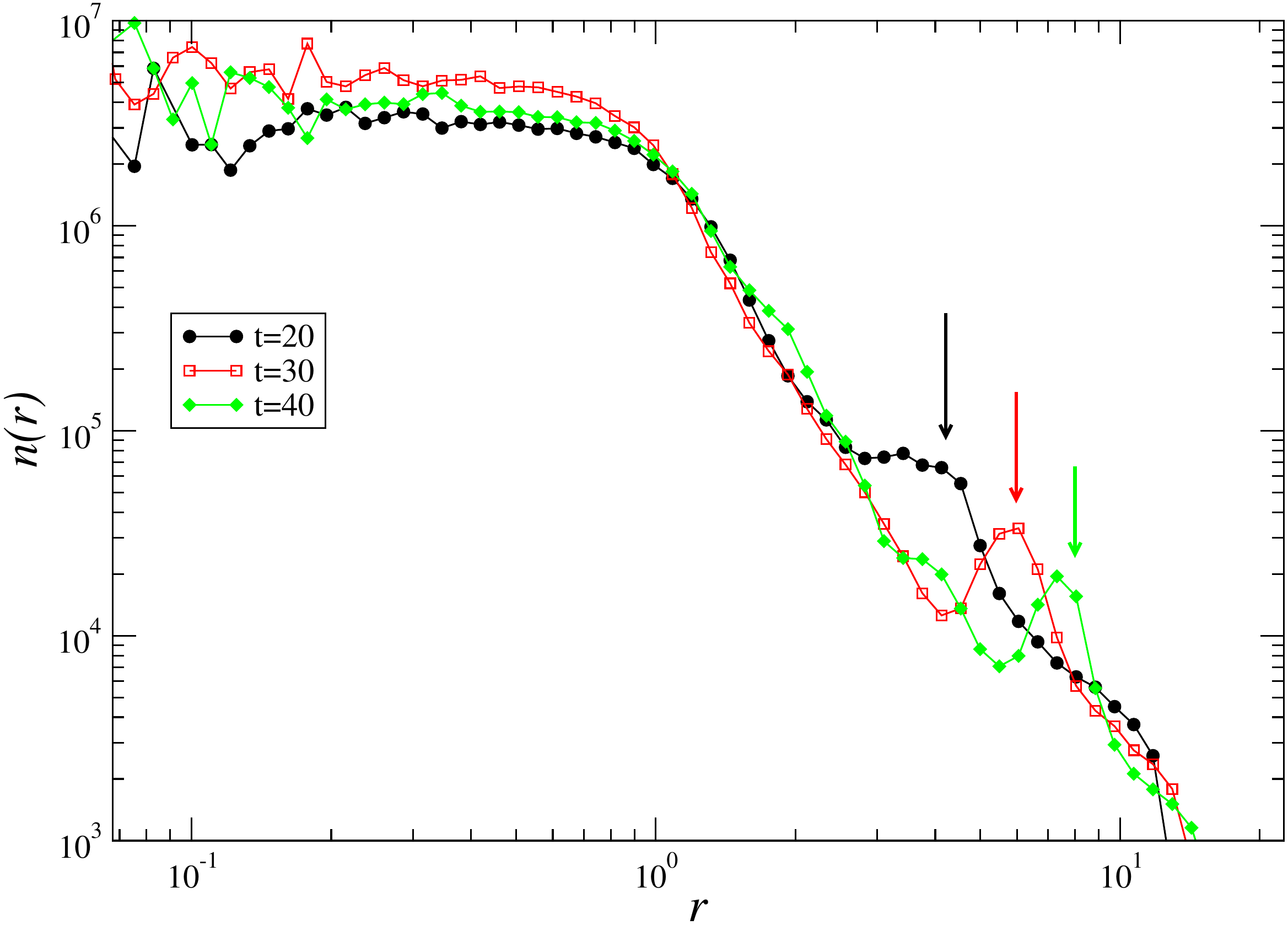}} 
\par\centering }
\caption{Upper figure: Zoom of the central part of  the simulation B2 (random sampled).
Bottom figure: evolution of the density profile of B2. 
The position of the expanding ring at different times
   (see labels) is indicated by an arrow.}
\label{B2_evol-ring}
\end{figure}

We have noted the formation of a time dependent ring-like structure in
simulation B2 at $t\approx 10$, in the plane corresponding to the two
largest initial principal semi-axes. As can be seen in
Fig.\ref{B2_evol-ring}, this is indeed a local density enhancement
which expands outwards in time. Investigation confirms that it is
generated by a fraction of particles moving outward at higher than
average radial velocity. These are particles which were initially in
or near the outermost radial shells in the plane of the oblate IC, and
which received a strong energy injection from the time dependent
potential generated by the collapse along the shortest axis. As these
particle carry also the initial velocities of the rotation about this
axis, the ring also rotates coherently. It persists up to the end of
our simulation at $t=50$.  As noted, we have varied the ratios
$a_1/a_3$ and $a_2/a_3$, for the more general case of a triaxial
ellipsoid, in a relatively wide range. We have found that, whenever we
are close to an oblate ellipsoid, such a {  ring-like structure}  is
formed.



\subsection{Shape parameters} 

As we have discussed the particles which gain most energy are those
which are initially furthest away from the center. For IC like the
ones we consider here, this leads to a very anisotropic distribution
for the loosely bound and ejected mass. The details of this
anisotropic distribution depend on the IC.
We recall that for cold prolate ellipsoidal IC without rotation these
particles are focused into two broad ``jets" in opposite directions
around the initially longest semi-principal axis
\citep{Benhaiem+SylosLabini_2015}.  This is the case simply because
the particles farthest from the center are indeed close to this axis
initially.  With additional coherent rotation, as considered here,
this jet-like structure, is, as we have seen, transformed into a
spiral-like structure as the particles propagate outward, and the axis
defined by the farthest out particles thus remains closely correlated
with the initially longest semi-principal axis.

\begin{figure}
\vspace{1cm} { \par\centering
\resizebox*{7cm}{7cm}{\includegraphics*{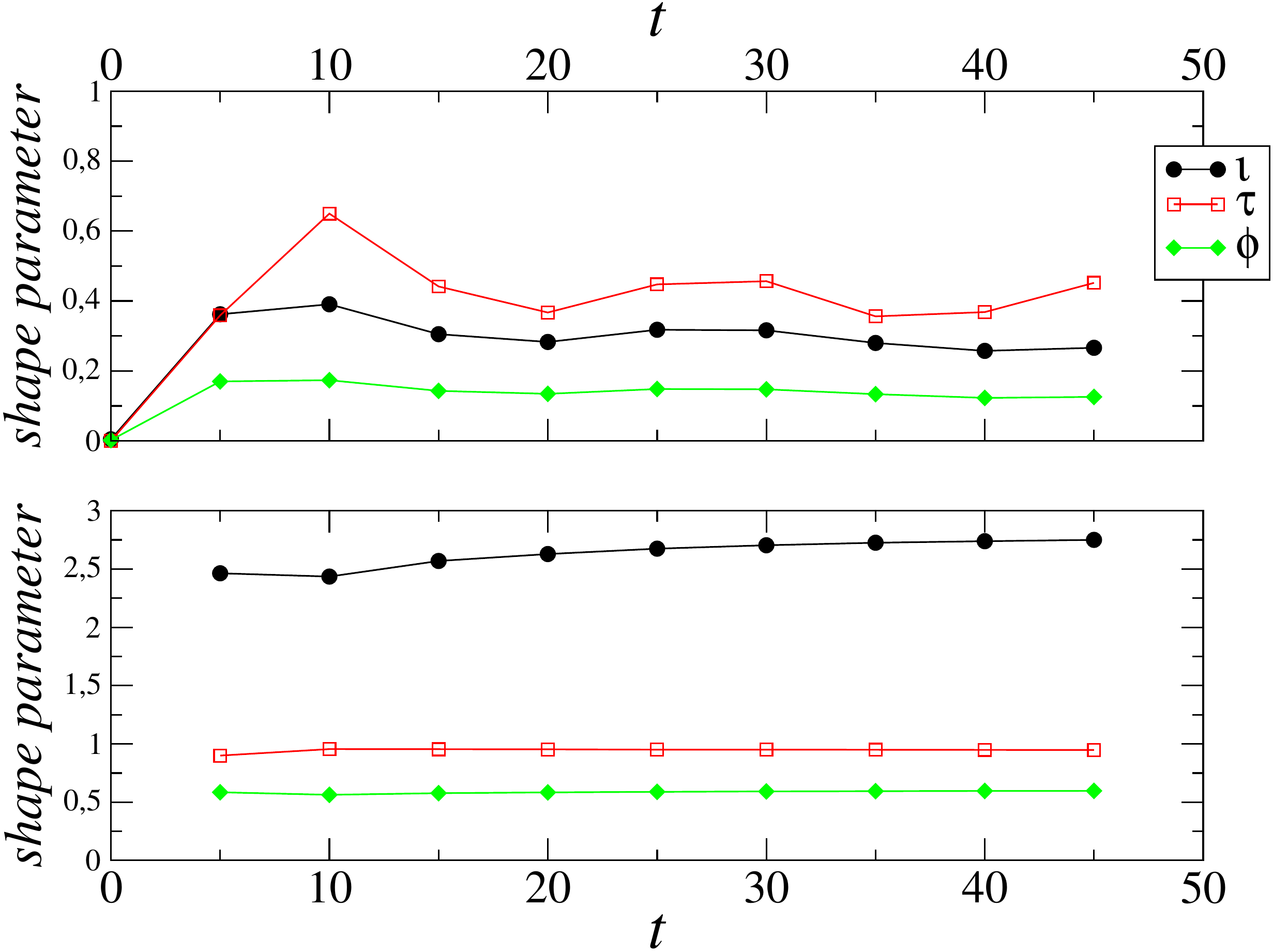}} 
\par\centering }
\caption{Shape parameters for the simulation A1: internal particles
  (upper panel), external particles (bottom panel).}
\label{S1_shape}
\end{figure}

In all our simulations the outer part of the structure is,
correspondingly, very flattened in the plane orthogonal to the
initially shortest semi-principal axis.  Fig. \ref{S1_shape} show the
evolution of the three shape parameters $\iota, \tau, \phi$ in
simulation A1, separately for the 'internal' particles constituting
the core of the structure (upper panels) and for the 'external'
particles constituting the spiral-like arms (bottom panels), where
this division is defined by the radius at which the measured averaged
radial density $n(r)$ discussed above reaches half its value in the
core.  In this case the core is a triaxial ellipsoid with,
respectively, $\iota \approx 0.3 \;, \phi \approx 0.1 \;, \tau \approx
0.4$ and $\iota \approx 0.7 \;, \phi \approx 0.2 \;, \tau \approx
0.5$.  On the other hand, the external particles are much more
flattened with $\iota \approx 3 \;, \phi \approx 0.5 \;, \tau \approx
0.5 $ in both cases.

A similar behavior is shown by the evolution of the flatness 
parameter for the simulations B1 and B2: in these cases
$\iota$ is computed by considering only the external, low energy, bound particles.
In simulation B1,
after the contraction phase during which $\iota$ is large and fluctuating,
it becomes of order one for $t>20$. In simulation B2, on the other hand, the 
flatness parameter remains of order one at all times without any 
significant variation. Simulations C1 and of C2 are in this respect
similar to A1/A2. 
%
%


\subsection{Sub-structures}

When the initial ellipsoidal deformation is large enough (e.g.,
$a_1/a_3 \approx 2$) the collapsing system may fragment into two or
more clumps which eventually merge long after the collapse : this is
the case of A2 (see Fig.\ref{XY_evol}). Clearly when the IC is made up
of clumps { with small amplitude fluctuations}, as in C1 and C2, this effect is enhanced.
On the other hand, during the collapse, initial density fluctuations
may evolve due to gravitational instability forming small aggregates
which, after the collapse, may correspond to sub-structures. These
sub-structures, which are typically not virialized as they are subject
to strong tidal fields, lie in the same plane defined by the
jets. Eventually they will fall into the largest virialized object and
be destroyed by the interaction with the core.  The formation of
sub-structures, anisotropically distributed on a planar configuration,
around the main virialized object appears to be a generic result of
the evolution from cold IC of this type
\citep{Benhaiem+SylosLabini_2015,Benhaiem+SylosLabini_2017}.

\subsection{Role of density fluctuations and the continuum limit}
\label{thermolimit}

{

Let us discuss further  the role of density fluctuations in the collapsing cloud
in an appropriate continuum limit defined by taking $ N \rightarrow \infty$. 
How this limit is taken must be specified, as it is not unique.
Indeed here there are (at least) two evident ways of taking such a 
limit, and the role of density fluctuations is different in each case.

First if we consider finite $N$ configurations as Poisson samplings of 
continuum configurations with fixed mass density $\rho_0$ i.e., without 
any intrinsic density fluctuations , we  can take the limit  $ N \rightarrow \infty$
together with the particle mass  $ m \rightarrow 0$, so that 
$\rho_0 = N \times m = const.$,. In this case the density fluctuations, that are 
proportional in amplitude  to $\delta \rho/ \rho_0 \propto N^{-1/2}$, also vanish. 
In this case all substructures generated by the growth of density fluctuations 
must disappear in the continuum limit. They can in this sense then 
be interpreted as finite $N$ effects. 

On the other hand,  we can also take the continuum limit in a different way,
by taking $ N \rightarrow \infty$ and $ m \rightarrow 0$ with $\rho_0 = N \times m = const.$ 
{\it and} with $\delta \rho/ \rho_0 = const.$ (or, more precisely, keeping the statistical properties of the latter fixed):
in this case  the internal fluctuations of the cloud grow in the same
way independently on $N$, and thus they give rise to the same substructures. 
Indeed, as it was shown in detail by \cite{Joyce+Marcos+SylosLabini_2009}
for the case of the spherical collapse model, while the whole system collapses
small density fluctuations inside the cloud grow and form substructures 
of increasing size with time. The collapse is halted when the size 
of non linear structures formed inside the collapsing cloud becomes 
of the same order of magnitude of the cloud itself (that meanwhile is collapsing). 
Thus in this process
there are two competing effects: the global monolithic collapse, which is a
top down process, and the 
bottom-up mechanism of structure formation. This latter mechanism is regulated 
by the properties of density fluctuations (in our case a simple Poisson distribution).

As final remark we note that the question of
how many particles are in
practice needed to simulate accurately the collisionless limit
(up  to  some  specified  time)
can only be answered in the
context of a given problem and it is in general a very difficult 
task to be sorted out. For instance, 
 it was recently shown  through the study of
the evolution a simple class of initial conditions (initially
spherical density profiles),  that  there is
a distinct
$N$ dependence associated with the presence of instabilities in the collisionless dynamics
that  arises because
the initial seeds for the instability themselves depend on
$N$  \cite{Benhaiem_etal_2018}: this dependence on
$N$
is very difficult to
find, as it manifests itself only in a very weak dependence
of the time of triggering of the instability, and
not, at sufficiently large
$N$, in the properties of the state to which
the instability drives the system.

The continuum limit for the dynamics of our finite $N$ systems is given by the Vlasov-Poisson system
and in principle could be simulated numerically directly. While much progress has been made on the numerical 
solution of these equation (see e.g. \cite{Sousbie-Colombi_2016}),
 such  an approach is, for the present,  feasible numerically only 
for systems with  high degrees of  symmetry, and not for those here in which the breaking notably of rotational 
symmetry plays a  crucial role.
}

\section{Models and observed structures}
\label{observations}

Both our initial conditions and the dynamics of the systems
we are studying are highly  idealized. In particular we expect
that in the formation of most astrophysical structures that
non-gravitational processes will play a crucial role 
(e.g., gas dynamics, star-formation and feedback from
it in galaxy formation).  Thus our models are intrinsically not
suitable to provide a detailed quantitative model for the formation of
astrophysical objects.  On the other hand, focusing on the specific
case of galaxy formation, we note that there is in fact little direct
constraint on initial conditions for it, as the fluctuations measured 
in the cosmic microwave background constrain strongly only
significantly larger scales.  Further the relative importance of gravity and other
forces in shaping galaxies is very uncertain.  As the structures we
have seen in our simulations bear a striking resemblance to spiral
galaxies, we believe it is worth looking more carefully at whether the
qualitative features of these structures are compatible or
incompatible with the observed qualitative properties of galaxies.

\subsection{Morphological features}

Firstly we note that there are several very common and non-trivial
features of spiral galaxies, which are accounted for in this kind of
model while they are problematic in the usual theoretical approaches,
in which spiral structures emerge through instability of an
equilibrium disc (see, e.g., \cite{Dobbs_Baba_2014, Binney_Tremine}).

Our models lead, as we have seen, very easily to two armed spiral
structure, which is observationally the most common kind.  Its
predominance has been considered puzzling and is difficult to account
for in the usual theoretical approaches.  Further the ``pitch angle"
$\alpha$, defined as the angle between the tangent to the arm and the
tangent to a circle at the same angle, gives values in the range
$10^\circ - 40^\circ$ in our simulations (at $t=50$) except for the
very cold cases like A1 in which the particles are ejected with very
high radial velocities (leading to an $\alpha$ approaching
$90^\circ$).  Pitch angles of this order are typical observationally,
while theoretical models predict much smaller angles and have great
difficulty accounting for those observed \cite{Dobbs_Baba_2014}. Further, as is almost
invariably the case observationally, the spiral arms formed in our
models are trailing, i.e. the outer tip points in the direction
opposite to rotation \citep{Binney_Tremine}.  As we have described
above, the spiral-like structure arises precisely because the
particles which are furthest out have, by angular momentum
conservation, smaller transverse velocities and thus lag behind in the
angle they rotate through up to a given time. This is again an
observational fact which does not have an apparent explanation in the
usual theoretical approaches.  Finally our mechanism produces, by
construction, structures which are non-axisymmetric and often the
central core is bar-like. Further, the spiral arms start at the end of
the bar. These properties likewise appear not only to be compatible
with observations (see, e.g., \cite{jog_2009,Dobbs_Baba_2014}), but to
potentially explain them in a very natural manner which eludes the
usual theoretical approaches.

{It is interesting to note that the mass of stars and gas in the 
spiral arms and in the outer part of the disc
in general  do not exceed the $\sim 20 \%$ of the luminous mass 
of the galaxy  (see, e.g., \cite{Sofue_2017}). 
In our simulations we also find that most of the mass is 
concentrated in the central bulge and that the fraction 
of particles with energy close to, or larger than, zero 
represent a small fraction, typically 
of the order of $10 \div 20 \%$, of the total mass.}


\subsection{Time-scales}

We have discussed in \cite{Benhaiem+Joyce+SylosLabini_2017} some
simple considerations about the compatibility of time and length
scales with real spiral galaxies, for the most idealized case of a
single ellipsoidal cloud. Making simple assumptions linking the final
size and velocity scale to those of real galaxies, the collapse
process which generated the structure must be assumed to occur on a
timescale of the order of $\sim$ 1 Gyr, that 
{   is the characteristic time scale of all out of equilibrium transient 
structures that are formed in our simulations, i.e. spiral arms, bars and 
ring-like structures.}
This is much shorter than the
age of the oldest stars in these galaxies ($\sim$ 10 Gyr) which is
usually assumed to correspond also to the age of such structures.
From the observational point of view, however, there is no definitive
evidence establishing the age of spiral arms; rather some observations
have suggested that spiral arms are not long-lived (see
e.g. \cite{Dobbs_Baba_2014, Binney_Tremine} and references therein).
{  Indeed the oldest stars and the galaxy  are  
formed by very different dynamical processes occurring on very different 
length scales (the size of the clouds where star formation occurs is  of the order of  $10^{-2}-10^{-1}$ kpc, while the size of a typical galaxy is $10-10^2$ kpc) 
and thus it is not at all evident that these two time scales must be of the same
order of magnitude.
}

The second family of IC we have studied here illustrate clearly that
this time-scale inferred from the space and velocity scales represents
only that of the violent collapse leading to this outer structure,
which could quite possibly be dissociated from that of the formation
of the central part of the galaxy, which could occur much before a
secondary collapse of surrounding matter giving rise to the disc and
extended ``halo" structure we have described.

{  In this respect it is perhaps useful to recall that the usual
assumption in modelling galaxies as dynamical equilibria 
(i.e. as QSS) } is intimately
linked to these considerations of time and length scales. Indeed if we
suppose a star orbits the galactic centers at a distance $R$, the
number of revolutions it has made in a time $T$ is
\be
\label{nrev} 
n_{rev} = \frac{T}{2 \pi R/v} \approx \, 30 \frac{T_{10} v_{200} }{R_{10}}
\ee
where $T_{10}$ is the time-scale in units of $10$ Gyr, 
$v_{200}$ is the velocity in units of $v=200$ km/sec, and
$R_{10}$ the radius in units of $10$ kpc. 

{  For stars 
(or other emitters) moving on closed Keplerian orbits, with a 
circular velocity $v \sim \sqrt{GM/r}$), this assumption 
(of ``stationarity")  appears to be reasonable only if $n_{rev} \gg 1$,  i.e., 
if these bodies have characteristic crossing times in the system considerably longer than its estimated lifetime. 
For smaller values they cannot have had the time to attain
orbits in which there is an equilibrium between centrifugal 
and centripetal acceleration. The precise value of the number of 
revolutions needed to reach a relaxed configuration cannot be  
 constrained  in a simple way theoretically because,
 as we have discussed above, it depends
 on the time in which the relaxation from an out of equilibrium 
 configuration to a QSS takes place. 
 However,  from a qualitative point of view, a reasonable 
 requirement is that $n_{rev} \gg 1$: only 
if  this condition is satisfied can the assumption of stationarity
possibly be justified. 

For a typical
disc galaxy of the size of the Milky Way, with
 a characteristic velocity $v_{200} \sim 1$
the number of revolutions is $n_{rev} \ge 10$ 
for  $R \le 30$ kpc only if $T_{10} \sim 1$, i.e. if the
age of the galaxy structure is of order of the
oldest stars.  If $T_{10} \sim 1$, i.e. if the
age of the galaxy structure is of order of the
oldest stars then objects in the inner part of the galaxy
(i.e., $R< 10$ kpc) had enough time to make 
$n_{rev} \ge 10^2$ while at larger distances
$n_{rev} \le 10$ and thus the assumption 
that emitters  move on closed Keplerian orbits
appears very difficult to justify for
the outermost regions of the galaxy. 
On the other hand, if the
age of the galaxy structure is  $T_{10} \sim 0.1$,  the assumption 
of stationarity clearly cannot be valid at larger radii
because $n_{rev} \ll 10$. 

The  out of equilibrium  scenario of our models for the outer
parts of a galaxy of the size of the Milky Way 
 is then, however, coherent with the observed 
velocity and distance scales.
}

\subsection{Velocities}

The compatibility of the velocity space structure of our transient
structures with observed properties of galaxies is much less
evident. Indeed a generic feature of the structures in our
simulations, arising from the nature of the mechanism, is that
velocities becomes predominantly radial at large radii. Extensive
observational study over decades, using different tracers, has placed
much constraint on such motions \citep{Sofue_2017}, and indicates that
motion in the outer parts of such galaxies is in fact very
predominantly rotational \citep{Kalberla_Dedes_2008, Sofue_2017},
although {a} significant radial motions have been detected in many
objects \citep{MLC_CGF_2016}.  As discussed in
\cite{Benhaiem+Joyce+SylosLabini_2017} for the results based on simple
ellipsoidal IC, it turns out that the naive expectation that such
motions are excluded by observations is not confirmed.  The reason is
that our velocity fields have a very particular spatial (anisotropic)
spatial structure which makes it difficult to distinguish them in
projection from rotating disc models. For the broader class of IC we
have explored here, the same considerations are valid, as there is a
similar kind of correlation between the velocities and the spatial
configuration.  Further, we have seen that different IC can produce a
less violent evolution than in the pure ellipsoidal model, leading to
less radial motion and a much more extended region in which there is
predominantly rotational motion. Thus the compatibility with
observations of galaxy kinematics depends also on the identification
of the time and length scales of the models with those of real
galaxies. { In our conclusion below we will comment about the radial motions
observed in our own Galactic disc, that are relevant to understand the possible
non stationary nature of the outer parts of the disc and of the spiral arms.}


\section{Dissipationless and dissipative disc formation}
\label{galform}

Let us now briefly discuss the difference between the models 
that we have presented, where the formation of a disc like flat structure is 
originated solely by a gravitational, and thus dissipationless, dynamics
and models in which instead the formation of a disc like structure is 
driven by dissipative effects. In standard models of galaxy formation the key element
in the formation of a disc galaxy, is the dissipation
associated with non-gravitational processes ---
gas dynamics, cooling, star formation, etc. { The
models}  used when gas dynamics is added to gravitational physics
consider the collapse of isolated and rotating clouds, like the one we studied here, but
solely with a spherical initial configuration. 
The central finding of our study is that disc-like configurations
 with transient spiral arms and with bars and/or rings in our
 simulations are formed by a purely dissipationless gravitational
dynamics if the initial conditions break spherical symmetry.
   There have been attempts  \cite{Katz_1991} 
   to study  the formation of quasi equilibrium configurations through a
   purely gravitational and dissipationless collapse dynamics in which
   the initial conditions are represented by isolated, spherically
   symmetric top hats in solid body rotation and in { Hubble flow.}
  These
   initial conditions differ from the ones we considered in this work by (i)
   the initial spherical shape and (ii) the small-scale fluctuations, 
   intended to model the fluctuations in standard ``cold dark
   matter cosmologies''. 
Starting from these initial conditions  the QSS formed are slowly rotating 
and are supported by an anisotropic velocity dispersion and closely  
resemble elliptical galaxies, and do not resemble at all spiral
galaxies.

The seminal work by \cite{Gott_1977} described
a scenario --- then developed in many other subsequent works --- in
which elliptical galaxies are products of a purely gravitational 
dissipationless collapse at high redshift, while spirals formed later with considerable
   dissipation. To simulate  such a scenario
dissipative gas dynamics was included in numerical 
   simulations  
     \cite{Katz+Gunn_1991} 
     in a system with the 
   same class of initial conditions as in \cite{Katz_1991}.
  It is precisely the dynamics of the gas in this two-component
  system, which leads then to structures resembling spiral galaxies: a
  thin disc made of gas and surrounded by purely gravitational 
  matter. Indeed since the gas
  can shock and dissipate energy it can develop a much flatter
  distribution than the dissipationless  matter. 
 By adding to the same initial conditions of \cite{Katz_1991}, 
 other non-gravitational effects, as star formation, supernova feedback
  \cite{Katz_1992}  and  metal enrichment due to supernovae
  \cite{Steinmetz+Mueller_1994,Steinmetz+Mueller_1995},
    the crucial mechanism for the formation of the disc and of 
 spiral arms is again played  dissipative non-gravitational 
 processes. These scenarios are thus completely different
  to how this structure emerges in our simulations, which 
  are pure gravitational and dissipationless.
  
\section{Discussion and conclusions} 
\label{conclusions}

We have described the results of numerical experiments exploring the
evolution under their self-gravity of non-spherical uniform and
non-uniform clouds with a coherent rigid-body rotation about their
shortest semi-principal axis. We have focused in particular on the
very rich spatial and velocity space organization of the outer parts
of these structures, which is a result of the combination of the
violent relaxation, which leads to a high energy tail in the energy
distribution, and the coherent rotation. Under very general conditions
spiral-like structure arises, while more or less evident bars and/or
rings appear depending on the properties of the initial conditions.
These outer parts of the structure are intrinsically non-stationary
and continue indefinitely to evolve in time. Although it will
disappear asymptotically, such structure is very long lived and, in
most cases, is still clearly defined at the longest times we simulate
to, of order $100$ to $200$ times the dynamical time (characterizing
the time for the formation, and characteristic time, of the virialized
core). The particles forming the spiral arms will escape from the system,
if they have positive energy, or will form an extremely dilute and anisotropic halo; eventually some of them, those having negative energy, will return back toward the core. On the other hand the largest fraction of the mass 
is bound in a triaxial system. 

It is perhaps relevant to remark why this simple path to producing
such structure in a self-gravitating system has, apparently, been
overlooked in the literature. Indeed the dynamics of self-gravitating
clouds of various forms, and with a wide variety of initial velocity
distributions, with and without initial angular momentum, has been
studied in {depth in} the literature, and it may seem surprising that
the phenomena we have discussed have not been noticed. We believe that
the explanation is probably linked to, on the one hand, the small
fraction of mass involved, and, on the other hand, the relatively long
times scales on which the system must be monitored. Indeed most
studies of this kind of system focus on the relatively short times on
which the system appears to virialize as indicated by global
parameters. Further most studies of this kind date back two to three
decades, and simulations in which the number of particles was
typically of $\sim 10^4$. In this case, the high energy particles
which are typically of order $5-10 \%$ are too few to resolve the
structures we have studied.

The spatial distributions of these transient structures are all
``spiral-like" , but even within the very circumscribed and idealized
set of IC we have considered, they show a wide variety of forms, from
ones qualitatively resembling grand design spiral galaxies, to
multi-armed and flocculent spiral galaxies and barred spiral galaxies.
The mechanism for producing these structures is completely different
in its physical principle to the mechanisms widely considered as
potentially explaining observed spiral structure. Indeed while such
mechanisms treat the spiral structure as a perturbative phenomenon ---
produced by the perturbation of an equilibrium rotating disc
\citep{Hohl_1971,Zang_Hohl_1978,Sellwood_1985,Sellwood_Moore_1999,Binney_Tremine,Fujii_etal_2011,Dobbs_Baba_2014}.
--- the mechanism at play in our simulations is intrinsically {\it far
  out-of-equilibrium}. While our model is too simple and idealized to
provide a quantitative model for real spiral galaxies, we have noted
that, in many respects, it apparently reproduces very naturally many
of their noted qualitative features.



{For what concerns the problem of cosmological galaxy formation, 
we note that the monolithic collapse discussed here is 
compatible with a top-down structure formation of the kind 
that occurs, e.g., in the so-called warm dark matter models. 
On the other hand, in models where dark matter is cold, structure
formation proceeds in a hierarchical bottom-up manner, so that 
galaxies are formed through an aggregation (i.e., merging) of smaller 
sub-structures. In this respect we note that when the initial conditions
break spherical symmetry and are inhomogeneous (as our models C1 and C2), 
before  the complete 
monolithic collapse of the whole cloud there are substructures  forming 
and  merging: in this scenario,  because all substructures
take part to the whole system collective dynamics, they can finally form 
coherent structures, like bars, rings and spiral arms, which are as large 
as the system itself. 
Thus, the scenario we have discussed is not 
in contradiction with the various observational evidences that merging was 
efficient in the early universe, but clearly a comprehensive theory 
of cosmological structure formation must be specified by 
the whole power-spectrum of density fluctuations: this does beyond 
the scope of the present work but will be addressed in forthcoming papers.}

We will explore in future work some of the questions opened up by our
results. One such question is of course whether the kind of initial
conditions we assume could be produced easily within a cosmological
framework.  As mentioned above, the problem of collapsing clouds has
  been wildly studied in the context of cosmological galaxy formation:
  however these studies were performed by taking a spherical
  overdensity while we used here more general shapes and
  a purely gravitational (and thus dissipationless) dynamics.
In addition, we note that while this seems to be excluded in typical
scenarios in which structure formation proceeds
hierarchically from very small scales (e.g., cold dark matter type scenarios),
 conditions like those we assume
might possibly be plausible in the case in which initial fluctuations
are highly suppressed below some large scale (e.g., as occurs in warm
dark matter type scenarios). A different { but complementary}
direction would be to explore the additional effects of
non-gravitational and dissipative physics, {like gas dynamics}, 
modeling the complex
processes inevitably at play in galaxy formation, and how they may or
may not modify formation and evolution of the structures we have
focused on here.

Finally it is interesting to mention that, while it has been known for
several decades that the disc of the Milky Way contains large-scale
non-axisymmetric features, the full knowledge of these asymmetric
structures and of their velocities fields is still lacking.
 The
recent Gaia DR2 maps \citep{Katz_etal_2018} have clearly shown that the
Milky Way is not, even to a first approximation, an axisymmetric system at 
equilibrium, but that it is characterized by streaming motions in all three 
velocity components. In particular it has confirmed the coherent radial motion in the
direction of the anti-center, earlier detected by \cite{MLC_CGF_2016},
up to 14 kpc. In addition, 
recent analyses of the radial velocity field in our Galaxy
by using different data-sets
\cite{MLC_Apogee,MLC_GAIA},  provide lots of new and corroborated information 
about the disk kinematics of our Galaxy: significant departures of
circularity in the mean orbits with radial galactocentric velocities, variations of 
rotation speed with position, asymmetries between Northern and Southern Galactic 
hemisphere and others
{  (note that 
the analysis of the velocity fields in external galaxies is model dependent 
and thus also the estimation of radial motions
\cite{SylosLabini_etal_2018})}.
These features of the full three-dimensional
velocity field seem to be compatible with the complex velocity fields
generated by the gravitational collapses we have discussed but a 
more detailed comparison of the models and observations is needed.
Such analysis will be reported in a forthcoming work.

 
%

\begin{acknowledgments}

This work was granted access to the HPC resources of The Institute for
scientific Computing and Simulation financed by Region Ile de France
and the project Equip@Meso (reference ANR-10-EQPX- 29-01) overseen by
the French National Research Agency (ANR) as part of the
Investissements d'Avenir program.
{ FSL thanks Martin  L{\'o}pez-Corredoira for very useful discussions and comments.}
\end{acknowledgments}. 


\begin{thebibliography}{63}%
\makeatletter
\providecommand \@ifxundefined [1]{%
 \@ifx{#1\undefined}
}%
\providecommand \@ifnum [1]{%
 \ifnum #1\expandafter \@firstoftwo
 \else \expandafter \@secondoftwo
 \fi
}%
\providecommand \@ifx [1]{%
 \ifx #1\expandafter \@firstoftwo
 \else \expandafter \@secondoftwo
 \fi
}%
\providecommand \natexlab [1]{#1}%
\providecommand \enquote  [1]{``#1''}%
\providecommand \bibnamefont  [1]{#1}%
\providecommand \bibfnamefont [1]{#1}%
\providecommand \citenamefont [1]{#1}%
\providecommand \href@noop [0]{\@secondoftwo}%
\providecommand \href [0]{\begingroup \@sanitize@url \@href}%
\providecommand \@href[1]{\@@startlink{#1}\@@href}%
\providecommand \@@href[1]{\endgroup#1\@@endlink}%
\providecommand \@sanitize@url [0]{\catcode `\\12\catcode `\$12\catcode
  `\&12\catcode `\#12\catcode `\^12\catcode `\_12\catcode `\%12\relax}%
\providecommand \@@startlink[1]{}%
\providecommand \@@endlink[0]{}%
\providecommand \url  [0]{\begingroup\@sanitize@url \@url }%
\providecommand \@url [1]{\endgroup\@href {#1}{\urlprefix }}%
\providecommand \urlprefix  [0]{URL }%
\providecommand \Eprint [0]{\href }%
\providecommand \doibase [0]{http://dx.doi.org/}%
\providecommand \selectlanguage [0]{\@gobble}%
\providecommand \bibinfo  [0]{\@secondoftwo}%
\providecommand \bibfield  [0]{\@secondoftwo}%
\providecommand \translation [1]{[#1]}%
\providecommand \BibitemOpen [0]{}%
\providecommand \bibitemStop [0]{}%
\providecommand \bibitemNoStop [0]{.\EOS\space}%
\providecommand \EOS [0]{\spacefactor3000\relax}%
\providecommand \BibitemShut  [1]{\csname bibitem#1\endcsname}%
\let\auto@bib@innerbib\@empty
\bibitem [{\citenamefont {Campa}\ \emph {et~al.}(2014)\citenamefont {Campa},
  \citenamefont {Dauxois}, \citenamefont {Fanelli},\ and\ \citenamefont
  {Ruffo}}]{Campa_etal_2014}%
  \BibitemOpen
  \bibfield  {author} {\bibinfo {author} {\bibfnamefont {A.}~\bibnamefont
  {Campa}}, \bibinfo {author} {\bibfnamefont {T.}~\bibnamefont {Dauxois}},
  \bibinfo {author} {\bibfnamefont {D.}~\bibnamefont {Fanelli}}, \ and\
  \bibinfo {author} {\bibfnamefont {S.}~\bibnamefont {Ruffo}},\ }\href@noop {}
  {\emph {\bibinfo {title} {Physics of Long-Range Interacting Systems}}}\
  (\bibinfo  {publisher} {Oxford},\ \bibinfo {year} {2014})\ \bibinfo {note}
  {oxford}\BibitemShut {NoStop}%
\bibitem [{\citenamefont {{Chavanis}}(2013)}]{Chavanis_2013}%
  \BibitemOpen
  \bibfield  {author} {\bibinfo {author} {\bibfnamefont {P.-H.}\ \bibnamefont
  {{Chavanis}}},\ }\href {\doibase 10.1051/0004-6361/201220607} {\bibfield
  {journal} {\bibinfo  {journal} {Astron.Astrophys.}\ }\textbf {\bibinfo
  {volume} {556}},\ \bibinfo {eid} {A93} (\bibinfo {year} {2013})},\ \Eprint
  {http://arxiv.org/abs/1210.5743} {arXiv:1210.5743} \BibitemShut {NoStop}%
\bibitem [{\citenamefont {Levin}\ \emph {et~al.}(2014)\citenamefont {Levin},
  \citenamefont {Pakter}, \citenamefont {Rizzato}, \citenamefont {Teles},\ and\
  \citenamefont {Benetti}}]{Levin_etal_2014}%
  \BibitemOpen
  \bibfield  {author} {\bibinfo {author} {\bibfnamefont {Y.}~\bibnamefont
  {Levin}}, \bibinfo {author} {\bibfnamefont {R.}~\bibnamefont {Pakter}},
  \bibinfo {author} {\bibfnamefont {F.~B.}\ \bibnamefont {Rizzato}}, \bibinfo
  {author} {\bibfnamefont {T.~N.}\ \bibnamefont {Teles}}, \ and\ \bibinfo
  {author} {\bibfnamefont {F.~P.}\ \bibnamefont {Benetti}},\ }\href {\doibase
  https://doi.org/10.1016/j.physrep.2013.10.001} {\bibfield  {journal}
  {\bibinfo  {journal} {Physics Reports}\ }\textbf {\bibinfo {volume} {535}},\
  \bibinfo {pages} {1 } (\bibinfo {year} {2014})},\ \bibinfo {note}
  {nonequilibrium statistical mechanics of systems with long-range
  interactions}\BibitemShut {NoStop}%
\bibitem [{\citenamefont {Padmanabhan}(1990)}]{Padmanabhan_1989}%
  \BibitemOpen
  \bibfield  {author} {\bibinfo {author} {\bibfnamefont {T.}~\bibnamefont
  {Padmanabhan}},\ }\href@noop {} {\bibfield  {journal} {\bibinfo  {journal}
  {Phys. Rept.}\ }\textbf {\bibinfo {volume} {188}},\ \bibinfo {pages} {285}
  (\bibinfo {year} {1990})}\BibitemShut {NoStop}%
\bibitem [{\citenamefont {Chavanis}\ \emph {et~al.}(2002)\citenamefont
  {Chavanis}, \citenamefont {Rosier},\ and\ \citenamefont
  {Sire}}]{Chavanis_2002}%
  \BibitemOpen
  \bibfield  {author} {\bibinfo {author} {\bibfnamefont {P.-H.}\ \bibnamefont
  {Chavanis}}, \bibinfo {author} {\bibfnamefont {C.}~\bibnamefont {Rosier}}, \
  and\ \bibinfo {author} {\bibfnamefont {C.}~\bibnamefont {Sire}},\ }\href@noop
  {} {\bibfield  {journal} {\bibinfo  {journal} {Phys. Rev. E}\ }\textbf
  {\bibinfo {volume} {66}},\ \bibinfo {pages} {036103} (\bibinfo {year}
  {2002})}\BibitemShut {NoStop}%
\bibitem [{\citenamefont {Dauxois}\ \emph {et~al.}(2002)\citenamefont
  {Dauxois}, \citenamefont {Ruffo}, \citenamefont {Arimondo},\ and\
  \citenamefont {Wilkens}}]{Dauxois_etal_2002}%
  \BibitemOpen
  \bibfield  {author} {\bibinfo {author} {\bibfnamefont {T.}~\bibnamefont
  {Dauxois}}, \bibinfo {author} {\bibfnamefont {S.}~\bibnamefont {Ruffo}},
  \bibinfo {author} {\bibfnamefont {E.}~\bibnamefont {Arimondo}}, \ and\
  \bibinfo {author} {\bibfnamefont {M.}~\bibnamefont {Wilkens}},\ }\href@noop
  {} {\emph {\bibinfo {title} {Dynamics and Thermodynamics of Systems with Long
  Range Interactions}}}\ (\bibinfo  {publisher} {Springer},\ \bibinfo {address}
  {Berlin},\ \bibinfo {year} {2002})\BibitemShut {NoStop}%
\bibitem [{\citenamefont {Yamaguchi}\ \emph {et~al.}(2004)\citenamefont
  {Yamaguchi}, \citenamefont {Barr\'e}, \citenamefont {Bouchet}, \citenamefont
  {Dauxois},\ and\ \citenamefont {Ruffo}}]{Yamaguchi_etal_2004}%
  \BibitemOpen
  \bibfield  {author} {\bibinfo {author} {\bibfnamefont {Y.~Y.}\ \bibnamefont
  {Yamaguchi}}, \bibinfo {author} {\bibfnamefont {J.}~\bibnamefont {Barr\'e}},
  \bibinfo {author} {\bibfnamefont {F.}~\bibnamefont {Bouchet}}, \bibinfo
  {author} {\bibfnamefont {T.}~\bibnamefont {Dauxois}}, \ and\ \bibinfo
  {author} {\bibfnamefont {S.}~\bibnamefont {Ruffo}},\ }\href@noop {}
  {\bibfield  {journal} {\bibinfo  {journal} {Physica A}\ }\textbf {\bibinfo
  {volume} {337}},\ \bibinfo {pages} {36} (\bibinfo {year} {2004})},\ \Eprint
  {http://arxiv.org/abs/cond-mat/0312480} {cond-mat/0312480} \BibitemShut
  {NoStop}%
\bibitem [{\citenamefont {{Campa}}\ \emph {et~al.}(2009)\citenamefont
  {{Campa}}, \citenamefont {{Dauxois}},\ and\ \citenamefont
  {{Ruffo}}}]{Campa_etal_2009}%
  \BibitemOpen
  \bibfield  {author} {\bibinfo {author} {\bibfnamefont {A.}~\bibnamefont
  {{Campa}}}, \bibinfo {author} {\bibfnamefont {T.}~\bibnamefont {{Dauxois}}},
  \ and\ \bibinfo {author} {\bibfnamefont {S.}~\bibnamefont {{Ruffo}}},\
  }\href@noop {} {\bibfield  {journal} {\bibinfo  {journal} {Phys. Reports}\
  }\textbf {\bibinfo {volume} {480}},\ \bibinfo {pages} {57} (\bibinfo {year}
  {2009})},\ \Eprint {http://arxiv.org/abs/0907.0323} {arXiv:0907.0323}
  \BibitemShut {NoStop}%
\bibitem [{\citenamefont {{Joyce}}\ and\ \citenamefont
  {{Worrakitpoonpon}}(2010)}]{Joyce+Worrakitpoonpon_2010}%
  \BibitemOpen
  \bibfield  {author} {\bibinfo {author} {\bibfnamefont {M.}~\bibnamefont
  {{Joyce}}}\ and\ \bibinfo {author} {\bibfnamefont {T.}~\bibnamefont
  {{Worrakitpoonpon}}},\ }\href {\doibase 10.1088/1742-5468/2010/10/P10012}
  {\bibfield  {journal} {\bibinfo  {journal} {Journal of Statistical Mechanics:
  Theory and Experiment}\ }\textbf {\bibinfo {volume} {10}},\ \bibinfo {pages}
  {12} (\bibinfo {year} {2010})},\ \Eprint {http://arxiv.org/abs/1004.2266}
  {arXiv:1004.2266 [cond-mat.stat-mech]} \BibitemShut {NoStop}%
\bibitem [{\citenamefont {Marcos}(2013)}]{Marcos_2013}%
  \BibitemOpen
  \bibfield  {author} {\bibinfo {author} {\bibfnamefont {B.}~\bibnamefont
  {Marcos}},\ }\href {\doibase 10.1103/PhysRevE.88.032112} {\bibfield
  {journal} {\bibinfo  {journal} {Phys. Rev. E}\ }\textbf {\bibinfo {volume}
  {88}},\ \bibinfo {pages} {032112} (\bibinfo {year} {2013})}\BibitemShut
  {NoStop}%
\bibitem [{\citenamefont {{Benetti}}\ \emph {et~al.}(2014)\citenamefont
  {{Benetti}}, \citenamefont {{Ribeiro-Teixeira}}, \citenamefont {{Pakter}},\
  and\ \citenamefont {{Levin}}}]{Benetti_etal_2014}%
  \BibitemOpen
  \bibfield  {author} {\bibinfo {author} {\bibfnamefont {F.~P.~C.}\
  \bibnamefont {{Benetti}}}, \bibinfo {author} {\bibfnamefont {A.~C.}\
  \bibnamefont {{Ribeiro-Teixeira}}}, \bibinfo {author} {\bibfnamefont
  {R.}~\bibnamefont {{Pakter}}}, \ and\ \bibinfo {author} {\bibfnamefont
  {Y.}~\bibnamefont {{Levin}}},\ }\href {\doibase
  10.1103/PhysRevLett.113.100602} {\bibfield  {journal} {\bibinfo  {journal}
  {Physical Review Letters}\ }\textbf {\bibinfo {volume} {113}},\ \bibinfo
  {eid} {100602} (\bibinfo {year} {2014})},\ \Eprint
  {http://arxiv.org/abs/1409.2060} {arXiv:1409.2060 [cond-mat.stat-mech]}
  \BibitemShut {NoStop}%
\bibitem [{\citenamefont {{Marcos}}\ \emph {et~al.}(2017)\citenamefont
  {{Marcos}}, \citenamefont {{Gabrielli}},\ and\ \citenamefont
  {{Joyce}}}]{Marcos_etal_2017}%
  \BibitemOpen
  \bibfield  {author} {\bibinfo {author} {\bibfnamefont {B.}~\bibnamefont
  {{Marcos}}}, \bibinfo {author} {\bibfnamefont {A.}~\bibnamefont
  {{Gabrielli}}}, \ and\ \bibinfo {author} {\bibfnamefont {M.}~\bibnamefont
  {{Joyce}}},\ }\href {\doibase 10.1103/PhysRevE.96.032102} {\bibfield
  {journal} {\bibinfo  {journal} {Phys.Rev.E}\ }\textbf {\bibinfo {volume}
  {96}},\ \bibinfo {eid} {032102} (\bibinfo {year} {2017})},\ \Eprint
  {http://arxiv.org/abs/1701.01865} {arXiv:1701.01865 [cond-mat.stat-mech]}
  \BibitemShut {NoStop}%
\bibitem [{\citenamefont {{Chandrasekhar}}(1943)}]{Chandrasekhar_1943}%
  \BibitemOpen
  \bibfield  {author} {\bibinfo {author} {\bibfnamefont {S.}~\bibnamefont
  {{Chandrasekhar}}},\ }\href {\doibase 10.1103/RevModPhys.15.1} {\bibfield
  {journal} {\bibinfo  {journal} {Reviews of Modern Physics}\ }\textbf
  {\bibinfo {volume} {15}},\ \bibinfo {pages} {1} (\bibinfo {year}
  {1943})}\BibitemShut {NoStop}%
\bibitem [{\citenamefont {Binney}\ and\ \citenamefont
  {Tremaine}(2008)}]{Binney_Tremine}%
  \BibitemOpen
  \bibfield  {author} {\bibinfo {author} {\bibfnamefont {J.}~\bibnamefont
  {Binney}}\ and\ \bibinfo {author} {\bibfnamefont {S.}~\bibnamefont
  {Tremaine}},\ }\href@noop {} {\emph {\bibinfo {title} {Galactic Dynamics}}}\
  (\bibinfo  {publisher} {Princeton University Press},\ \bibinfo {year}
  {2008})\BibitemShut {NoStop}%
\bibitem [{\citenamefont {{Dobbs}}\ and\ \citenamefont
  {{Baba}}(2014)}]{Dobbs_Baba_2014}%
  \BibitemOpen
  \bibfield  {author} {\bibinfo {author} {\bibfnamefont {C.}~\bibnamefont
  {{Dobbs}}}\ and\ \bibinfo {author} {\bibfnamefont {J.}~\bibnamefont
  {{Baba}}},\ }\href {\doibase 10.1017/pasa.2014.31} {\bibfield  {journal}
  {\bibinfo  {journal} {Pub.Astron.Soc.Austr.}\ }\textbf {\bibinfo {volume}
  {31}},\ \bibinfo {eid} {e035} (\bibinfo {year} {2014})}\BibitemShut {NoStop}%
\bibitem [{\citenamefont {{Gott}}(1977)}]{Gott_1977}%
  \BibitemOpen
  \bibfield  {author} {\bibinfo {author} {\bibfnamefont {J.~R.}\ \bibnamefont
  {{Gott}}, \bibfnamefont {III}},\ }\href {\doibase
  10.1146/annurev.aa.15.090177.001315} {\bibfield  {journal} {\bibinfo
  {journal} {Ann.Rev.Astron.Astrophys.}\ }\textbf {\bibinfo {volume} {15}},\
  \bibinfo {pages} {235} (\bibinfo {year} {1977})}\BibitemShut {NoStop}%
\bibitem [{\citenamefont {{Lake}}\ and\ \citenamefont
  {{Carlberg}}(1988)}]{Lake+Carlberg_1988}%
  \BibitemOpen
  \bibfield  {author} {\bibinfo {author} {\bibfnamefont {G.}~\bibnamefont
  {{Lake}}}\ and\ \bibinfo {author} {\bibfnamefont {R.~G.}\ \bibnamefont
  {{Carlberg}}},\ }\href {\doibase 10.1086/114908} {\bibfield  {journal}
  {\bibinfo  {journal} {Astron.J.}\ }\textbf {\bibinfo {volume} {96}},\
  \bibinfo {pages} {1581} (\bibinfo {year} {1988})}\BibitemShut {NoStop}%
\bibitem [{\citenamefont {{Katz}}(1991)}]{Katz_1991}%
  \BibitemOpen
  \bibfield  {author} {\bibinfo {author} {\bibfnamefont {N.}~\bibnamefont
  {{Katz}}},\ }\href {\doibase 10.1086/169696} {\bibfield  {journal} {\bibinfo
  {journal} {Astrophys.J.}\ }\textbf {\bibinfo {volume} {368}},\ \bibinfo
  {pages} {325} (\bibinfo {year} {1991})}\BibitemShut {NoStop}%
\bibitem [{\citenamefont {{Katz}}\ and\ \citenamefont
  {{Gunn}}(1991)}]{Katz+Gunn_1991}%
  \BibitemOpen
  \bibfield  {author} {\bibinfo {author} {\bibfnamefont {N.}~\bibnamefont
  {{Katz}}}\ and\ \bibinfo {author} {\bibfnamefont {J.~E.}\ \bibnamefont
  {{Gunn}}},\ }\href {\doibase 10.1086/170367} {\bibfield  {journal} {\bibinfo
  {journal} {Astrophys.J.}\ }\textbf {\bibinfo {volume} {377}},\ \bibinfo
  {pages} {365} (\bibinfo {year} {1991})}\BibitemShut {NoStop}%
\bibitem [{\citenamefont {{Katz}}(1992)}]{Katz_1992}%
  \BibitemOpen
  \bibfield  {author} {\bibinfo {author} {\bibfnamefont {N.}~\bibnamefont
  {{Katz}}},\ }\href {\doibase 10.1086/171366} {\bibfield  {journal} {\bibinfo
  {journal} {Astrophys.J.}\ }\textbf {\bibinfo {volume} {391}},\ \bibinfo
  {pages} {502} (\bibinfo {year} {1992})}\BibitemShut {NoStop}%
\bibitem [{\citenamefont {{Steinmetz}}\ and\ \citenamefont
  {{Mueller}}(1994)}]{Steinmetz+Mueller_1994}%
  \BibitemOpen
  \bibfield  {author} {\bibinfo {author} {\bibfnamefont {M.}~\bibnamefont
  {{Steinmetz}}}\ and\ \bibinfo {author} {\bibfnamefont {E.}~\bibnamefont
  {{Mueller}}},\ }\href@noop {} {\bibfield  {journal} {\bibinfo  {journal}
  {Astron.Astrophys.}\ }\textbf {\bibinfo {volume} {281}},\ \bibinfo {pages}
  {L97} (\bibinfo {year} {1994})},\ \Eprint
  {http://arxiv.org/abs/astro-ph/9312010} {astro-ph/9312010} \BibitemShut
  {NoStop}%
\bibitem [{\citenamefont {{Steinmetz}}\ and\ \citenamefont
  {{Muller}}(1995)}]{Steinmetz+Mueller_1995}%
  \BibitemOpen
  \bibfield  {author} {\bibinfo {author} {\bibfnamefont {M.}~\bibnamefont
  {{Steinmetz}}}\ and\ \bibinfo {author} {\bibfnamefont {E.}~\bibnamefont
  {{Muller}}},\ }\href {\doibase 10.1093/mnras/276.2.549} {\bibfield  {journal}
  {\bibinfo  {journal} {Mon. Not. R. Astron. Soc.}\ }\textbf {\bibinfo {volume}
  {276}},\ \bibinfo {pages} {549} (\bibinfo {year} {1995})},\ \Eprint
  {http://arxiv.org/abs/astro-ph/9407066} {astro-ph/9407066} \BibitemShut
  {NoStop}%
\bibitem [{\citenamefont {{Marinacci}}\ \emph {et~al.}(2014)\citenamefont
  {{Marinacci}}, \citenamefont {{Pakmor}},\ and\ \citenamefont
  {{Springel}}}]{Marinacci_2014}%
  \BibitemOpen
  \bibfield  {author} {\bibinfo {author} {\bibfnamefont {F.}~\bibnamefont
  {{Marinacci}}}, \bibinfo {author} {\bibfnamefont {R.}~\bibnamefont
  {{Pakmor}}}, \ and\ \bibinfo {author} {\bibfnamefont {V.}~\bibnamefont
  {{Springel}}},\ }\href {\doibase 10.1093/mnras/stt2003} {\bibfield  {journal}
  {\bibinfo  {journal} {Mon. Not. R. Astron. Soc.}\ }\textbf {\bibinfo {volume}
  {437}},\ \bibinfo {pages} {1750} (\bibinfo {year} {2014})},\ \Eprint
  {http://arxiv.org/abs/1305.5360} {arXiv:1305.5360} \BibitemShut {NoStop}%
\bibitem [{\citenamefont {{Naab}}\ and\ \citenamefont
  {{Ostriker}}(2017)}]{Naab_2017}%
  \BibitemOpen
  \bibfield  {author} {\bibinfo {author} {\bibfnamefont {T.}~\bibnamefont
  {{Naab}}}\ and\ \bibinfo {author} {\bibfnamefont {J.~P.}\ \bibnamefont
  {{Ostriker}}},\ }\href {\doibase 10.1146/annurev-astro-081913-040019}
  {\bibfield  {journal} {\bibinfo  {journal} {Ann.Rev.Astron.Astrophys.}\
  }\textbf {\bibinfo {volume} {55}},\ \bibinfo {pages} {59} (\bibinfo {year}
  {2017})},\ \Eprint {http://arxiv.org/abs/1612.06891} {arXiv:1612.06891}
  \BibitemShut {NoStop}%
\bibitem [{\citenamefont {{Henon}}(1973)}]{Henon_1973}%
  \BibitemOpen
  \bibfield  {author} {\bibinfo {author} {\bibfnamefont {M.}~\bibnamefont
  {{Henon}}},\ }\href@noop {} {\bibfield  {journal} {\bibinfo  {journal} {Ann.
  Astrophys.}\ }\textbf {\bibinfo {volume} {24}},\ \bibinfo {pages} {229}
  (\bibinfo {year} {1973})}\BibitemShut {NoStop}%
\bibitem [{\citenamefont {van Albada}(1982)}]{vanalbada_1982}%
  \BibitemOpen
  \bibfield  {author} {\bibinfo {author} {\bibfnamefont {T.}~\bibnamefont {van
  Albada}},\ }\href@noop {} {\bibfield  {journal} {\bibinfo  {journal} {Mon.
  Not. R. Astr. Soc.}\ }\textbf {\bibinfo {volume} {201}},\ \bibinfo {pages}
  {939} (\bibinfo {year} {1982})}\BibitemShut {NoStop}%
\bibitem [{\citenamefont {Aarseth}\ \emph {et~al.}(1988)\citenamefont
  {Aarseth}, \citenamefont {Lin},\ and\ \citenamefont
  {Papaloizou}}]{aarseth_etal_1988}%
  \BibitemOpen
  \bibfield  {author} {\bibinfo {author} {\bibfnamefont {S.}~\bibnamefont
  {Aarseth}}, \bibinfo {author} {\bibfnamefont {D.}~\bibnamefont {Lin}}, \ and\
  \bibinfo {author} {\bibfnamefont {J.}~\bibnamefont {Papaloizou}},\
  }\href@noop {} {\bibfield  {journal} {\bibinfo  {journal} {Astrophys. J.}\
  }\textbf {\bibinfo {volume} {324}},\ \bibinfo {pages} {288} (\bibinfo {year}
  {1988})}\BibitemShut {NoStop}%
\bibitem [{\citenamefont {David}\ and\ \citenamefont
  {Theuns}(1989)}]{david+theuns_1989}%
  \BibitemOpen
  \bibfield  {author} {\bibinfo {author} {\bibfnamefont {M.}~\bibnamefont
  {David}}\ and\ \bibinfo {author} {\bibfnamefont {T.}~\bibnamefont {Theuns}},\
  }\href@noop {} {\bibfield  {journal} {\bibinfo  {journal} {Mon. Not. R. Astr.
  Soc.}\ }\textbf {\bibinfo {volume} {240}},\ \bibinfo {pages} {957} (\bibinfo
  {year} {1989})}\BibitemShut {NoStop}%
\bibitem [{\citenamefont {Aguilar}\ and\ \citenamefont
  {Merritt}(1990)}]{aguilar+merritt_1990}%
  \BibitemOpen
  \bibfield  {author} {\bibinfo {author} {\bibfnamefont {L.}~\bibnamefont
  {Aguilar}}\ and\ \bibinfo {author} {\bibfnamefont {D.}~\bibnamefont
  {Merritt}},\ }\href@noop {} {\bibfield  {journal} {\bibinfo  {journal}
  {Astrophys. J.}\ }\textbf {\bibinfo {volume} {354}},\ \bibinfo {pages} {73}
  (\bibinfo {year} {1990})}\BibitemShut {NoStop}%
\bibitem [{\citenamefont {Theuns}\ and\ \citenamefont
  {David}(1990)}]{theuns+david_1990}%
  \BibitemOpen
  \bibfield  {author} {\bibinfo {author} {\bibfnamefont {T.}~\bibnamefont
  {Theuns}}\ and\ \bibinfo {author} {\bibfnamefont {M.}~\bibnamefont {David}},\
  }\href@noop {} {\bibfield  {journal} {\bibinfo  {journal} {Astrophys. Sp.
  Sci.}\ }\textbf {\bibinfo {volume} {170}},\ \bibinfo {pages} {276} (\bibinfo
  {year} {1990})}\BibitemShut {NoStop}%
\bibitem [{\citenamefont {Boily}\ \emph {et~al.}(2002)\citenamefont {Boily},
  \citenamefont {Athanassoula},\ and\ \citenamefont
  {Kroupa}}]{boily_etal_2002}%
  \BibitemOpen
  \bibfield  {author} {\bibinfo {author} {\bibfnamefont {C.}~\bibnamefont
  {Boily}}, \bibinfo {author} {\bibfnamefont {E.}~\bibnamefont {Athanassoula}},
  \ and\ \bibinfo {author} {\bibfnamefont {P.}~\bibnamefont {Kroupa}},\
  }\href@noop {} {\bibfield  {journal} {\bibinfo  {journal} {Mon. Not. R. Astr.
  Soc.}\ }\textbf {\bibinfo {volume} {332}},\ \bibinfo {pages} {971} (\bibinfo
  {year} {2002})}\BibitemShut {NoStop}%
\bibitem [{\citenamefont {{Barnes}}\ \emph {et~al.}(2009)\citenamefont
  {{Barnes}}, \citenamefont {{Lanzel}},\ and\ \citenamefont
  {{Williams}}}]{barnes_etal_2009}%
  \BibitemOpen
  \bibfield  {author} {\bibinfo {author} {\bibfnamefont {E.~I.}\ \bibnamefont
  {{Barnes}}}, \bibinfo {author} {\bibfnamefont {P.~A.}\ \bibnamefont
  {{Lanzel}}}, \ and\ \bibinfo {author} {\bibfnamefont {L.~L.~R.}\ \bibnamefont
  {{Williams}}},\ }\href {\doibase 10.1088/0004-637X/704/1/372} {\bibfield
  {journal} {\bibinfo  {journal} {Astrophys. J.}\ }\textbf {\bibinfo {volume}
  {704}},\ \bibinfo {pages} {372} (\bibinfo {year} {2009})},\ \Eprint
  {http://arxiv.org/abs/0908.3873} {arXiv:0908.3873 [astro-ph.CO]} \BibitemShut
  {NoStop}%
\bibitem [{\citenamefont {Joyce}\ \emph {et~al.}(2009)\citenamefont {Joyce},
  \citenamefont {Marcos},\ and\ \citenamefont
  {Sylos~Labini}}]{Joyce+Marcos+SylosLabini_2009}%
  \BibitemOpen
  \bibfield  {author} {\bibinfo {author} {\bibfnamefont {M.}~\bibnamefont
  {Joyce}}, \bibinfo {author} {\bibfnamefont {B.}~\bibnamefont {Marcos}}, \
  and\ \bibinfo {author} {\bibfnamefont {F.}~\bibnamefont {Sylos~Labini}},\
  }\href@noop {} {\bibfield  {journal} {\bibinfo  {journal} {Mon. Not. R.
  Astron. Soc.}\ }\textbf {\bibinfo {volume} {397}},\ \bibinfo {pages} {775}
  (\bibinfo {year} {2009})}\BibitemShut {NoStop}%
\bibitem [{\citenamefont {{Sylos Labini}}(2012)}]{syloslabini_2012}%
  \BibitemOpen
  \bibfield  {author} {\bibinfo {author} {\bibfnamefont {F.}~\bibnamefont
  {{Sylos Labini}}},\ }\href {\doibase 10.1111/j.1365-2966.2012.21019.x}
  {\bibfield  {journal} {\bibinfo  {journal} {Mon. Not. R. Astron. Soc.}\
  }\textbf {\bibinfo {volume} {423}},\ \bibinfo {pages} {1610} (\bibinfo {year}
  {2012})}\BibitemShut {NoStop}%
\bibitem [{\citenamefont {{Sylos Labini}}(2013)}]{syloslabini_2013}%
  \BibitemOpen
  \bibfield  {author} {\bibinfo {author} {\bibfnamefont {F.}~\bibnamefont
  {{Sylos Labini}}},\ }\href {\doibase 10.1093/mnras/sts365} {\bibfield
  {journal} {\bibinfo  {journal} {Mon. Not. R. Astron. Soc.}\ }\textbf
  {\bibinfo {volume} {429}},\ \bibinfo {pages} {679} (\bibinfo {year}
  {2013})}\BibitemShut {NoStop}%
\bibitem [{\citenamefont {{Benhaiem}}\ \emph {et~al.}(2017)\citenamefont
  {{Benhaiem}}, \citenamefont {{Joyce}},\ and\ \citenamefont {{Sylos
  Labini}}}]{Benhaiem+Joyce+SylosLabini_2017}%
  \BibitemOpen
  \bibfield  {author} {\bibinfo {author} {\bibfnamefont {D.}~\bibnamefont
  {{Benhaiem}}}, \bibinfo {author} {\bibfnamefont {M.}~\bibnamefont {{Joyce}}},
  \ and\ \bibinfo {author} {\bibfnamefont {F.}~\bibnamefont {{Sylos Labini}}},\
  }\href {\doibase 10.3847/1538-4357/aa96a7} {\bibfield  {journal} {\bibinfo
  {journal} {Astrophys.J.}\ }\textbf {\bibinfo {volume} {851}},\ \bibinfo {eid}
  {19} (\bibinfo {year} {2017})},\ \Eprint {http://arxiv.org/abs/1711.01913}
  {arXiv:1711.01913} \BibitemShut {NoStop}%
\bibitem [{\citenamefont {Springel}(2005)}]{Springel_2005}%
  \BibitemOpen
  \bibfield  {author} {\bibinfo {author} {\bibfnamefont {V.}~\bibnamefont
  {Springel}},\ }\href@noop {} {\bibfield  {journal} {\bibinfo  {journal} {Mon.
  Not. R. Ast. Soc.}\ }\textbf {\bibinfo {volume} {364}},\ \bibinfo {pages}
  {1105} (\bibinfo {year} {2005})}\BibitemShut {NoStop}%
\bibitem [{Note1()}]{Note1}%
  \BibitemOpen
  \bibinfo {note} {See
  https://wwwmpa.mpa-garching.mpg.de/gadget/users-guide.pdf.}\BibitemShut
  {Stop}%
\bibitem [{Note2()}]{Note2}%
  \BibitemOpen
  \bibinfo {note} {Specifically the parameter {\protect \tt ErrTolTheta} is
  $10^{-10}$ for HP compared to $0.7$ for LP, while {\protect \tt
  ErrTolForceAcc} is $10^{-10}$ for HP and $5\times 10^{-4}$ for
  LP.}\BibitemShut {Stop}%
\bibitem [{\citenamefont {{Peebles}}(1969)}]{Peebles_1969}%
  \BibitemOpen
  \bibfield  {author} {\bibinfo {author} {\bibfnamefont {P.~J.~E.}\
  \bibnamefont {{Peebles}}},\ }\href {\doibase 10.1086/149876} {\bibfield
  {journal} {\bibinfo  {journal} {Astrophys.J.}\ }\textbf {\bibinfo {volume}
  {155}},\ \bibinfo {pages} {393} (\bibinfo {year} {1969})}\BibitemShut
  {NoStop}%
\bibitem [{\citenamefont {{Peebles}}(1971)}]{Peebles_1971}%
  \BibitemOpen
  \bibfield  {author} {\bibinfo {author} {\bibfnamefont {P.~J.~E.}\
  \bibnamefont {{Peebles}}},\ }\href@noop {} {\bibfield  {journal} {\bibinfo
  {journal} {Astron.Astrophys}\ }\textbf {\bibinfo {volume} {11}},\ \bibinfo
  {pages} {377} (\bibinfo {year} {1971})}\BibitemShut {NoStop}%
\bibitem [{\citenamefont {{Barnes}}\ and\ \citenamefont
  {{Efstathiou}}(1987)}]{Barnes+Efstathiou_1987}%
  \BibitemOpen
  \bibfield  {author} {\bibinfo {author} {\bibfnamefont {J.}~\bibnamefont
  {{Barnes}}}\ and\ \bibinfo {author} {\bibfnamefont {G.}~\bibnamefont
  {{Efstathiou}}},\ }\href {\doibase 10.1086/165480} {\bibfield  {journal}
  {\bibinfo  {journal} {Astrophys.J.}\ }\textbf {\bibinfo {volume} {319}},\
  \bibinfo {pages} {575} (\bibinfo {year} {1987})}\BibitemShut {NoStop}%
\bibitem [{\citenamefont {{Bullock}}\ \emph {et~al.}(2001)\citenamefont
  {{Bullock}}, \citenamefont {{Dekel}}, \citenamefont {{Kolatt}}, \citenamefont
  {{Kravtsov}}, \citenamefont {{Klypin}}, \citenamefont {{Porciani}},\ and\
  \citenamefont {{Primack}}}]{Bullock_etal_2001}%
  \BibitemOpen
  \bibfield  {author} {\bibinfo {author} {\bibfnamefont {J.~S.}\ \bibnamefont
  {{Bullock}}}, \bibinfo {author} {\bibfnamefont {A.}~\bibnamefont {{Dekel}}},
  \bibinfo {author} {\bibfnamefont {T.~S.}\ \bibnamefont {{Kolatt}}}, \bibinfo
  {author} {\bibfnamefont {A.~V.}\ \bibnamefont {{Kravtsov}}}, \bibinfo
  {author} {\bibfnamefont {A.~A.}\ \bibnamefont {{Klypin}}}, \bibinfo {author}
  {\bibfnamefont {C.}~\bibnamefont {{Porciani}}}, \ and\ \bibinfo {author}
  {\bibfnamefont {J.~R.}\ \bibnamefont {{Primack}}},\ }\href {\doibase
  10.1086/321477} {\bibfield  {journal} {\bibinfo  {journal} {Astrophys.J.}\
  }\textbf {\bibinfo {volume} {555}},\ \bibinfo {pages} {240} (\bibinfo {year}
  {2001})},\ \Eprint {http://arxiv.org/abs/astro-ph/0011001}
  {arXiv:astro-ph/0011001 [astro-ph]} \BibitemShut {NoStop}%
\bibitem [{\citenamefont {{Benhaiem}}\ \emph {et~al.}(2018)\citenamefont
  {{Benhaiem}}, \citenamefont {{Joyce}}, \citenamefont {{Sylos Labini}},\ and\
  \citenamefont {{Worrakitpoonpon}}}]{Benhaiem_etal_2018}%
  \BibitemOpen
  \bibfield  {author} {\bibinfo {author} {\bibfnamefont {D.}~\bibnamefont
  {{Benhaiem}}}, \bibinfo {author} {\bibfnamefont {M.}~\bibnamefont {{Joyce}}},
  \bibinfo {author} {\bibfnamefont {F.}~\bibnamefont {{Sylos Labini}}}, \ and\
  \bibinfo {author} {\bibfnamefont {T.}~\bibnamefont {{Worrakitpoonpon}}},\
  }\href {\doibase 10.1093/mnras/stx2444} {\bibfield  {journal} {\bibinfo
  {journal} {Mon. Not. R. Astron. Soc.}\ }\textbf {\bibinfo {volume} {473}},\
  \bibinfo {pages} {2348} (\bibinfo {year} {2018})},\ \Eprint
  {http://arxiv.org/abs/1709.06657} {arXiv:1709.06657} \BibitemShut {NoStop}%
\bibitem [{\citenamefont {{Lin}}\ \emph {et~al.}(1965)\citenamefont {{Lin}},
  \citenamefont {{Mestel}},\ and\ \citenamefont {{Shu}}}]{Lin_Mestel_Shu_1965}%
  \BibitemOpen
  \bibfield  {author} {\bibinfo {author} {\bibfnamefont {C.~C.}\ \bibnamefont
  {{Lin}}}, \bibinfo {author} {\bibfnamefont {L.}~\bibnamefont {{Mestel}}}, \
  and\ \bibinfo {author} {\bibfnamefont {F.~H.}\ \bibnamefont {{Shu}}},\ }\href
  {\doibase 10.1086/148428} {\bibfield  {journal} {\bibinfo  {journal}
  {Astrophys. J.}\ }\textbf {\bibinfo {volume} {142}},\ \bibinfo {pages} {1431}
  (\bibinfo {year} {1965})}\BibitemShut {NoStop}%
\bibitem [{Note3()}]{Note3}%
  \BibitemOpen
  \bibinfo {note} {We take as center the particle which has the lowest
  gravitational potential. Note that we will use the same radial binning
  everywhere below when we compute averages.}\BibitemShut {Stop}%
\bibitem [{\citenamefont {{Benhaiem}}\ and\ \citenamefont {{Sylos
  Labini}}(2015)}]{Benhaiem+SylosLabini_2015}%
  \BibitemOpen
  \bibfield  {author} {\bibinfo {author} {\bibfnamefont {D.}~\bibnamefont
  {{Benhaiem}}}\ and\ \bibinfo {author} {\bibfnamefont {F.}~\bibnamefont
  {{Sylos Labini}}},\ }\href {\doibase 10.1093/mnras/stv075} {\bibfield
  {journal} {\bibinfo  {journal} {Mon.Not.R.Astron.Soc.}\ }\textbf {\bibinfo
  {volume} {448}},\ \bibinfo {pages} {2634} (\bibinfo {year}
  {2015})}\BibitemShut {NoStop}%
\bibitem [{\citenamefont {{Benhaiem}}\ and\ \citenamefont {{Sylos
  Labini}}(2017)}]{Benhaiem+SylosLabini_2017}%
  \BibitemOpen
  \bibfield  {author} {\bibinfo {author} {\bibfnamefont {D.}~\bibnamefont
  {{Benhaiem}}}\ and\ \bibinfo {author} {\bibfnamefont {F.}~\bibnamefont
  {{Sylos Labini}}},\ }\href {\doibase 10.1051/0004-6361/201628698} {\bibfield
  {journal} {\bibinfo  {journal} {Astron.Astrophys.}\ }\textbf {\bibinfo
  {volume} {598}},\ \bibinfo {eid} {A95} (\bibinfo {year} {2017})}\BibitemShut
  {NoStop}%
\bibitem [{\citenamefont {{Sousbie}}\ and\ \citenamefont
  {{Colombi}}(2016)}]{Sousbie-Colombi_2016}%
  \BibitemOpen
  \bibfield  {author} {\bibinfo {author} {\bibfnamefont {T.}~\bibnamefont
  {{Sousbie}}}\ and\ \bibinfo {author} {\bibfnamefont {S.}~\bibnamefont
  {{Colombi}}},\ }\href {\doibase 10.1016/j.jcp.2016.05.048} {\bibfield
  {journal} {\bibinfo  {journal} {Journal of Computational Physics}\ }\textbf
  {\bibinfo {volume} {321}},\ \bibinfo {pages} {644} (\bibinfo {year}
  {2016})},\ \Eprint {http://arxiv.org/abs/1509.07720} {arXiv:1509.07720
  [physics.comp-ph]} \BibitemShut {NoStop}%
\bibitem [{\citenamefont {{Jog}}\ and\ \citenamefont
  {{Combes}}(2009)}]{jog_2009}%
  \BibitemOpen
  \bibfield  {author} {\bibinfo {author} {\bibfnamefont {C.~J.}\ \bibnamefont
  {{Jog}}}\ and\ \bibinfo {author} {\bibfnamefont {F.}~\bibnamefont
  {{Combes}}},\ }\href {\doibase 10.1016/j.physrep.2008.12.002} {\bibfield
  {journal} {\bibinfo  {journal} {Phys.Rep.}\ }\textbf {\bibinfo {volume}
  {471}},\ \bibinfo {pages} {75} (\bibinfo {year} {2009})},\ \Eprint
  {http://arxiv.org/abs/0811.1101} {arXiv:0811.1101} \BibitemShut {NoStop}%
\bibitem [{\citenamefont {{Sofue}}(2017)}]{Sofue_2017}%
  \BibitemOpen
  \bibfield  {author} {\bibinfo {author} {\bibfnamefont {Y.}~\bibnamefont
  {{Sofue}}},\ }\href {\doibase 10.1093/pasj/psw103} {\bibfield  {journal}
  {\bibinfo  {journal} {Pub.Astron.Soc.Jap.}\ }\textbf {\bibinfo {volume}
  {69}},\ \bibinfo {eid} {R1} (\bibinfo {year} {2017})}\BibitemShut {NoStop}%
\bibitem [{\citenamefont {{Kalberla}}\ and\ \citenamefont
  {{Dedes}}(2008)}]{Kalberla_Dedes_2008}%
  \BibitemOpen
  \bibfield  {author} {\bibinfo {author} {\bibfnamefont {P.~M.~W.}\
  \bibnamefont {{Kalberla}}}\ and\ \bibinfo {author} {\bibfnamefont
  {L.}~\bibnamefont {{Dedes}}},\ }\href {\doibase 10.1051/0004-6361:20079240}
  {\bibfield  {journal} {\bibinfo  {journal} {Astron.Astrophys.}\ }\textbf
  {\bibinfo {volume} {487}},\ \bibinfo {pages} {951} (\bibinfo {year}
  {2008})}\BibitemShut {NoStop}%
\bibitem [{\citenamefont {{L{\'o}pez-Corredoira}}\ and\ \citenamefont
  {{Gonz{\'a}lez-Fern{\'a}ndez}}(2016)}]{MLC_CGF_2016}%
  \BibitemOpen
  \bibfield  {author} {\bibinfo {author} {\bibfnamefont {M.}~\bibnamefont
  {{L{\'o}pez-Corredoira}}}\ and\ \bibinfo {author} {\bibfnamefont
  {C.}~\bibnamefont {{Gonz{\'a}lez-Fern{\'a}ndez}}},\ }\href {\doibase
  10.3847/0004-6256/151/6/165} {\bibfield  {journal} {\bibinfo  {journal}
  {Astron.J.}\ }\textbf {\bibinfo {volume} {151}},\ \bibinfo {eid} {165}
  (\bibinfo {year} {2016})}\BibitemShut {NoStop}%
\bibitem [{\citenamefont {{Hohl}}(1971)}]{Hohl_1971}%
  \BibitemOpen
  \bibfield  {author} {\bibinfo {author} {\bibfnamefont {F.}~\bibnamefont
  {{Hohl}}},\ }\href {\doibase 10.1086/151091} {\bibfield  {journal} {\bibinfo
  {journal} {Astrophys. J.}\ }\textbf {\bibinfo {volume} {168}},\ \bibinfo
  {pages} {343} (\bibinfo {year} {1971})}\BibitemShut {NoStop}%
\bibitem [{\citenamefont {{Zang}}\ and\ \citenamefont
  {{Hohl}}(1978)}]{Zang_Hohl_1978}%
  \BibitemOpen
  \bibfield  {author} {\bibinfo {author} {\bibfnamefont {T.~A.}\ \bibnamefont
  {{Zang}}}\ and\ \bibinfo {author} {\bibfnamefont {F.}~\bibnamefont
  {{Hohl}}},\ }\href {\doibase 10.1086/156636} {\bibfield  {journal} {\bibinfo
  {journal} {Astrophys. J.}\ }\textbf {\bibinfo {volume} {226}},\ \bibinfo
  {pages} {521} (\bibinfo {year} {1978})}\BibitemShut {NoStop}%
\bibitem [{\citenamefont {{Sellwood}}(1985)}]{Sellwood_1985}%
  \BibitemOpen
  \bibfield  {author} {\bibinfo {author} {\bibfnamefont {J.~A.}\ \bibnamefont
  {{Sellwood}}},\ }\href {\doibase 10.1093/mnras/217.1.127} {\bibfield
  {journal} {\bibinfo  {journal} {Mon. Not. R. Astron. Soc.}\ }\textbf
  {\bibinfo {volume} {217}},\ \bibinfo {pages} {127} (\bibinfo {year}
  {1985})}\BibitemShut {NoStop}%
\bibitem [{\citenamefont {{Sellwood}}\ and\ \citenamefont
  {{Moore}}(1999)}]{Sellwood_Moore_1999}%
  \BibitemOpen
  \bibfield  {author} {\bibinfo {author} {\bibfnamefont {J.~A.}\ \bibnamefont
  {{Sellwood}}}\ and\ \bibinfo {author} {\bibfnamefont {E.~M.}\ \bibnamefont
  {{Moore}}},\ }\href {\doibase 10.1086/306557} {\bibfield  {journal} {\bibinfo
   {journal} {Astrophys. J.}\ }\textbf {\bibinfo {volume} {510}},\ \bibinfo
  {pages} {125} (\bibinfo {year} {1999})},\ \Eprint
  {http://arxiv.org/abs/astro-ph/9807010} {astro-ph/9807010} \BibitemShut
  {NoStop}%
\bibitem [{\citenamefont {{Fujii}}\ \emph {et~al.}(2011)\citenamefont
  {{Fujii}}, \citenamefont {{Baba}}, \citenamefont {{Saitoh}}, \citenamefont
  {{Makino}}, \citenamefont {{Kokubo}},\ and\ \citenamefont
  {{Wada}}}]{Fujii_etal_2011}%
  \BibitemOpen
  \bibfield  {author} {\bibinfo {author} {\bibfnamefont {M.~S.}\ \bibnamefont
  {{Fujii}}}, \bibinfo {author} {\bibfnamefont {J.}~\bibnamefont {{Baba}}},
  \bibinfo {author} {\bibfnamefont {T.~R.}\ \bibnamefont {{Saitoh}}}, \bibinfo
  {author} {\bibfnamefont {J.}~\bibnamefont {{Makino}}}, \bibinfo {author}
  {\bibfnamefont {E.}~\bibnamefont {{Kokubo}}}, \ and\ \bibinfo {author}
  {\bibfnamefont {K.}~\bibnamefont {{Wada}}},\ }\href {\doibase
  10.1088/0004-637X/730/2/109} {\bibfield  {journal} {\bibinfo  {journal}
  {Astrophys.J.}\ }\textbf {\bibinfo {volume} {730}},\ \bibinfo {eid} {109}
  (\bibinfo {year} {2011})},\ \Eprint {http://arxiv.org/abs/1006.1228}
  {arXiv:1006.1228} \BibitemShut {NoStop}%
\bibitem [{\citenamefont {{Gaia Collaboration}}\ \emph
  {et~al.}(2018)\citenamefont {{Gaia Collaboration}}, \citenamefont {{Katz}},
  \citenamefont {{Antoja}}, \citenamefont {{Romero-G{\'o}mez}}, \citenamefont
  {{Drimmel}}, \citenamefont {{Reyl{\'e}}}, \citenamefont {{Seabroke}},
  \citenamefont {{Soubiran}}, \citenamefont {{Babusiaux}}, \citenamefont {{Di
  Matteo}},\ and\ \citenamefont {et~al.}}]{Katz_etal_2018}%
  \BibitemOpen
  \bibfield  {author} {\bibinfo {author} {\bibnamefont {{Gaia Collaboration}}},
  \bibinfo {author} {\bibfnamefont {D.}~\bibnamefont {{Katz}}}, \bibinfo
  {author} {\bibfnamefont {T.}~\bibnamefont {{Antoja}}}, \bibinfo {author}
  {\bibfnamefont {M.}~\bibnamefont {{Romero-G{\'o}mez}}}, \bibinfo {author}
  {\bibfnamefont {R.}~\bibnamefont {{Drimmel}}}, \bibinfo {author}
  {\bibfnamefont {C.}~\bibnamefont {{Reyl{\'e}}}}, \bibinfo {author}
  {\bibfnamefont {G.~M.}\ \bibnamefont {{Seabroke}}}, \bibinfo {author}
  {\bibfnamefont {C.}~\bibnamefont {{Soubiran}}}, \bibinfo {author}
  {\bibfnamefont {C.}~\bibnamefont {{Babusiaux}}}, \bibinfo {author}
  {\bibfnamefont {P.}~\bibnamefont {{Di Matteo}}}, \ and\ \bibinfo {author}
  {\bibnamefont {et~al.}},\ }\href {\doibase 10.1051/0004-6361/201832865}
  {\bibfield  {journal} {\bibinfo  {journal} {Astron.Astrophys.}\ }\textbf
  {\bibinfo {volume} {616}},\ \bibinfo {eid} {A11} (\bibinfo {year} {2018})},\
  \Eprint {http://arxiv.org/abs/1804.09380} {arXiv:1804.09380} \BibitemShut
  {NoStop}%
\bibitem [{\citenamefont {{L{\'o}pez-Corredoira}}\ \emph
  {et~al.}(2019)\citenamefont {{L{\'o}pez-Corredoira}}, \citenamefont {{Sylos
  Labini}}, \citenamefont {{Kalberla}},\ and\ \citenamefont {{Allende
  Prieto}}}]{MLC_Apogee}%
  \BibitemOpen
  \bibfield  {author} {\bibinfo {author} {\bibfnamefont {M.}~\bibnamefont
  {{L{\'o}pez-Corredoira}}}, \bibinfo {author} {\bibfnamefont {F.}~\bibnamefont
  {{Sylos Labini}}}, \bibinfo {author} {\bibfnamefont {P.~M.~W.}\ \bibnamefont
  {{Kalberla}}}, \ and\ \bibinfo {author} {\bibfnamefont {C.}~\bibnamefont
  {{Allende Prieto}}},\ }\href {\doibase 10.3847/1538-3881/aaf3b3} {\bibfield
  {journal} {\bibinfo  {journal} {Astron.J.}\ }\textbf {\bibinfo {volume}
  {157}},\ \bibinfo {eid} {26} (\bibinfo {year} {2019})},\ \Eprint
  {http://arxiv.org/abs/1901.01300} {arXiv:1901.01300 [astro-ph.GA]}
  \BibitemShut {NoStop}%
\bibitem [{\citenamefont {{L{\'o}pez-Corredoira}}\ and\ \citenamefont {{Sylos
  Labini}}(2019)}]{MLC_GAIA}%
  \BibitemOpen
  \bibfield  {author} {\bibinfo {author} {\bibfnamefont {M.}~\bibnamefont
  {{L{\'o}pez-Corredoira}}}\ and\ \bibinfo {author} {\bibfnamefont
  {F.}~\bibnamefont {{Sylos Labini}}},\ }\href@noop {} {\bibfield  {journal}
  {\bibinfo  {journal} {Astron.Astrophys. in the press}\ } (\bibinfo {year}
  {2019})},\ \Eprint {http://arxiv.org/abs/1810.13436.} {arXiv:1810.13436.
  [astro-ph.GA]} \BibitemShut {NoStop}%
\bibitem [{\citenamefont {{Sylos Labini}}\ \emph {et~al.}(2019)\citenamefont
  {{Sylos Labini}}, \citenamefont {{Benhaiem}}, \citenamefont {{Comeron}},\
  and\ \citenamefont {{Lopez-Corredoira}}}]{SylosLabini_etal_2018}%
  \BibitemOpen
  \bibfield  {author} {\bibinfo {author} {\bibfnamefont {F.}~\bibnamefont
  {{Sylos Labini}}}, \bibinfo {author} {\bibfnamefont {D.}~\bibnamefont
  {{Benhaiem}}}, \bibinfo {author} {\bibfnamefont {S.}~\bibnamefont
  {{Comeron}}}, \ and\ \bibinfo {author} {\bibfnamefont {M.}~\bibnamefont
  {{Lopez-Corredoira}}},\ }\href@noop {} {\bibfield  {journal} {\bibinfo
  {journal} {Astron.Astrohys., in the press}\ } (\bibinfo {year} {2019})},\
  \Eprint {http://arxiv.org/abs/1812.01447} {arXiv:1812.01447 [astro-ph.GA]}
  \BibitemShut {NoStop}%
\end{thebibliography}
%

\end{document}